\input preprint.sty                                                 %
\input epsf                                                      %
\def\sgn{{\mathop{\rm sgn}\nolimits}{}}                                %
\def\const{{\rm const}\, {}}                                           %
\def\lesssim{\lap}                                                     %
\def\gtrsim{\gap}                                                      %
\def\scri{{\cal I}}                                                    %
\newcount\EEK                                                          %
\EEK=0                                                                 %
\def\eek{\global\advance\EEK by 1\eqno(\the\EEK )}                     %
\newcount\FEET                                                         %
\FEET=0                                                                %
\def\fnote#1{\global\advance\FEET by 1\footnote{${}^{\the\FEET}$}{#1}} %
\def\ignore#1{\relax}                                                  %
\pptstyle
\title{Do black holes radiate?}

\author{Adam D Helfer}

\address{Department of Mathematics, Mathematical Sciences Building, 
University of Missouri, Columbia, Missouri 65211, U.S.A.}

\jl{0}

\beginabstract
The prediction that black holes radiate due to quantum effects is often
considered one of the most secure in quantum field theory in curved
space--time.  Yet this prediction rests on two dubious assumptions:  that
ordinary physics may be applied to vacuum fluctuations at energy scales
increasing exponentially without bound; and that quantum--gravitational effects
may be neglected.  Various suggestions have been put forward to address these
issues:  that they might be explained away by lessons from sonic black hole
models; that the prediction is indeed successfully reproduced by quantum
gravity; that the success of the link provided by the prediction between black
holes and thermodynamics justifies the prediction.

\cabs
This paper explains the nature of the difficulties, and reviews the proposals
that have been put forward to deal with them.   None of the proposals put
forward can so far be considered to be really successful, and simple
dimensional arguments show that quantum--gravitational effects might well alter
the evaporation process outlined by Hawking.   Thus a definitive theoretical
treatment will require an understanding of quantum gravity in at least some
regimes.  Until then, no compelling theoretical case for or against radiation
by black holes is likely to be made.

\cabs
The possibility that non--radiating ``mini'' black holes exist should be taken
seriously; such holes could be part of the dark matter in the Universe.
Attempts to place observational limits on the number of ``mini'' black holes
(independent of the assumption that they radiate) would be most welcome.
\endabstract


\date

\noindent\bf Contents \rm

1. Introduction 

2. The predictions

3. The Hawking process

\qquad 3.1 The space--time

\qquad 3.2 Null infinity, the black hole, and time--asymmetry

\qquad 3.3 The mapping of surfaces of constant phase

\qquad 3.4 Propagation of the quantum field

\qquad 3.5 The quantum state and the two--point functions

\qquad\qquad 3.5.1 The two--point function in Minkowski space

\qquad\qquad 3.5.2 The two--point function in the Hawking model

\qquad\qquad 3.5.3 Particles

\qquad 3.6 Stress--energy

\qquad 3.7 The analysis of Fredenhagen and Haag

\qquad 3.8 Almost--black holes

4. The trans--Planckian problem

\qquad 4.1 Localization of the problem

\qquad\qquad 4.1.1 The family (CH)

\qquad\qquad 4.1.2 The family (SO)

\qquad 4.2 Discussion

5. Connection with the Unruh process

\qquad 5.1 The Unruh process

\qquad 5.2 The argument from the equivalence principle

\qquad 5.3 The Bisognano--Wichmann theorem

6. Lessons from moving--mirror models

\qquad 6.1 The trajectories

\qquad 6.2 The energy budget

7. Connections with thermodynamics

\qquad 7.1 Classical black--hole thermodynamics

\qquad 7.2 General relativity and the second law

\qquad 7.3 The generalized second law

\qquad 7.4 The generalized second law and the Hawking process

\qquad\qquad 7.4.1 Quantum--gravitational issues

\qquad\qquad 7.4.2 Definition of entropy

\qquad\qquad 7.4.3 Rate of entropy production

\qquad 7.5 The Geroch--Wheeler process revisited

\vfill\eject

\noindent\bf Contents \rm (continued)

8. Nonstandard propagation

\qquad 8.1 Jacobson's cut--off model

\qquad 8.2 The model of Corley and Jacobson

\qquad\qquad 8.2.1 The space--times

\qquad\qquad 8.2.2 The dispersive propagation

\qquad\qquad 8.2.3 Quantization

\qquad 8.3 The model of Brout, Massar, Parentani and Spindel

\qquad 8.4 Summary and discussion

9.  't Hooft's $S$--matrix black hole theory

\qquad 9.1 Overview of the program

\qquad\qquad 9.1.1 The spectrum

\qquad\qquad 9.1.2 The holographic principle

\qquad\qquad 9.1.3 The trans--Planckian problem

\qquad\qquad 9.1.4 The $S$--matrix theory

\qquad\qquad 9.1.5 Purity and unitarity

\qquad 9.2 Summary

10. Evidence from theories of quantum gravity

\qquad 10.1 Dilatonic black holes

\qquad 10.2 TTFKAS

\qquad 10.3 Ashtekar's approach

\qquad 10.4 Euclidean quantum gravity

11. Quantum character of space--time

\qquad 11.1 Validity of the semiclassical approximation

\qquad 11.2 Quantization of black--hole area

\qquad 11.3 Quantum measurement issues

12. Experimental prospects

\qquad 12.1 What sorts of black holes might exist?

\qquad 12.2 What would a Hawking--radiating black hole really look
like?

\qquad 12.3 Summary

13. Conclusions

Appendix.  Text passages supporting the conclusions

\vfill\eject

\section{Introduction}

In 1974 and 1975, Stephen Hawking published his analysis of the
effects of gravitational collapse on quantum fields, and predicted
that black holes are not in fact black, but radiate thermally and
eventually explode.  Whether black holes turn out to radiate or not,
it would be hard to overstate the significance of these papers.
Hawking had found one of those key physical systems which at once
bring vexing foundational issues to a point, are accessible to
analytic techniques, and suggest deep connections between
disparate areas of physics.  

Work stimulated by Hawking's led to clarifications of what it means to detect
or create particles, and even of what it means to define a quantum field
theory.  Perhaps more excitingly, Hawking's was a concrete proposal that
quantum effects might qualitatively alter the character of general relativity,
turning black holes into sources rather than perfect sinks.   This mechanism,
involving in essential ways as it did quantum theory and general relativity,
suggested that the system might be a point of entry to the problem of
quantizing gravity. And it suggested that the connections between black holes
and thermodynamics, which until that time had appeared to most workers to be
formal (see however Bekenstein 1973) were real and might be understood
quantum--theoretically. 

Given the profound nature of the issues addressed, it is perhaps not surprising
that some disagreement and controversy exists over exactly what has been
achieved.  Thus some authors write that the arguments for thermal radiation are
almost certainly secure (Carlip 2001),  others that there are problems but that
those can almost certainly be fixed (Wald 2001), or that there are problems
which probably can be fixed (Jacobson 1990), or that there are problems which
hopefully can be fixed (Parentani 2001), or that there are problems which may
not be fixable (Belinski 1995).   According to some, the prediction depends on
speculations about ultra--high energy physics (Gibbons 1977); according to
others, this has been shown not to be the case (Visser 2001).  Most workers
assume \it a priori \rm that quantum--gravitational effects will be negligible,
since there are no locally Planck--scale curvatures (except perhaps for holes
of around the Planck mass), but others have argued that quantum--gravitational
effects could completely alter or even obviate the thermal radiation
(Bekenstein and Mukhanov 1995, Ashtekar 1998).  There are those who feel that
the case for Hawking radiation  is so compelling that it should be a touchstone
for theories of quantum gravity: we should grade the plausibility of theories
of quantum gravity according to whether they reproduce Hawking's predictions
(Carlip 2001) --- which should be contrasted with grading the plausibility of
the predictions according to whether they are sustained by theories of quantum
gravity.

How confident should we be in the prediction of thermal radiation from black
holes?  Is the evidence really good enough that we can pass or reject theories
of quantum gravity (or of links between thermodynamics and quantum fields in
curved space--time) on the basis of it?  If the evidence is not that good, are
we potentially selecting against the correct theory of quantum gravity (or the
correct understanding of the link with thermodynamics) by only considering
theories compatible with the predictions? To what extent does the proposed
mechanism really probe quantum gravity?  

It is these questions which are the subject of the present review.  So this
paper is not a pedagogical treatment of a well--understood area, but a
critical review of the current state of a difficult issue.  The emphasis will be
on a clear presentation of the physical issues and their interplay.
 
The main conclusions may be summarized as follows:\fnote{The appendix
lists the places in the text where these assertions are justified.}

(a) None of the derivations that have been given of the prediction of radiation
from black holes is convincing.  All involve, at some point, 
speculations of what physics is like at scales which are not merely orders of
magnitude  beyond any that have so far been investigated experimentally ($\sim
10^3$ GeV), but at and \it increasing beyond \rm the Planck
scale ($\sim 10^{19}$ GeV), where essentially quantum--gravitational effects
are expected to be dominant.  (In Hawking's treatment, this increase occurs \it
exponentially quickly.\rm ) Some of these speculations    may be plausible, but
none can be considered reliable.\fnote{It should be remarked that very similar
issues arise in contemporary inflationary cosmology.  See e.g. Brandenberger
2002 and references therein.}

(b) There are equally plausible speculations about physics at such scales which
result in no radiation at all, or in non--thermal spectra.

(c) The various derivations which have been put forward are not all mutually
consistent.  Thus, even among the derivations
which do give rise to thermal radiation, there is no single accepted
physical mechanism.

(d) Quantum--gravitational corrections are very plausibly of a size
to completely alter or even obviate the prediction of thermal radiation.

(e) A number of the arguments put forward in support of the Hawking
mechanism are not really direct evidence for the existence of thermal 
radiation,
but rather are arguments for interpreting black holes' areas as entropies.

(f) The proposed mechanism, at least as conventionally understood,
relies precisely on the assumption that quantum--gravitational effects can be
neglected, and so no deep test of quantum gravity can emerge from it.

\smallskip\noindent
It should be emphasized that the problems uncovered here are entirely
physical, not mathematical.  While there are some technical mathematical
concerns with details of Hawking's computation, we do not anticipate any real
difficulty in resolving these (cf. Fredenhagen and Haag 1990).  The
issues are whether the physical assumptions underlying the mathematics
are correct, and whether the correct physical lessons are being
drawn from the calculations.

The significance of these conclusions is not negative, however, but
positive and exciting.  They show that the precise assumptions we make about
certain aspects of quantum gravity and high--energy physics do have an effect
on Hawking's predictions.  So the challenge to any theory of quantum gravity is
(not necessarily to reproduce Hawking's predictions, but) to provide a detailed
physical picture of what happens to quantum fields in a region where a black
hole forms.  Different theories will lead to different pictures, more or less
plausible and with differing implications.  Even at the theoretical level, the
requirement of being able to produce a plausible and self--consistent model is
a terrific constraint.  And of course the sensitivity to assumptions means
that, should experimental data become available, we can discriminate between
different theories.  

It should also be noted that the issues that occur in the black--hole case ---
the appearance of trans--Planckian modes, and the question of
quantum--gravitational corrections -- are also of interest in contemporary
inflationary cosmology.  (See e.g. Brandenberger 2002.)

A few historical comments may help to orient the reader.

The trans--Planckian problem and the question of quantum--gravitational
corrections were apparently recognized almost immediately on the publication of
Hawking's work; for example, they are raised in Gibbons's (1977) paper. 
However, for some reason (which is to me puzzling), these issues did not get
much attention for a long time.   They were Unruh's main motivation for
introducing the idea of sonic black holes (Unruh 1981); he clearly
considered the issues serious, but other workers did not pursue them. Jacobson
(1990, 1991) gave what is as far as I know the first explicit statement of the
problem since Gibbons's and Unruh's papers.  

Jacobson's (1991)  paper was very influential in one sector of the relativity
community (it made the LANL ``top cite'' list).  He carefully explained the
trans--Planckian problem, and then suggested a template for a scheme to
``save'' the Hawking prediction without appealing to trans--Planckian physics. 
Then several papers by Jacobson and Unruh developed an idea that if new physics
intervened and the propagation of field modes at high frequencies had a certain
dispersive character, one could recover Hawking's prediction (by, however, a
mechanism essentially different from that envisaged by Hawking).  This
mechanism still required trans--Planckian wave--numbers.  

At this point, a chronological account becomes of limited value, because, while
the papers of Jacobson and Unruh did attract attention, there was no unanimity
about what lessons were to be drawn from them.  Had the trans--Planckian problem
been solved?  If not, how serious was the situation?  Even if the issue was
still open and serious, surely the connections which had been established
between the Hawking effect and other physics must be good circumstantial
evidence that the prediction of black--hole evaporation is in fact correct?  
Physics (and particularly quantum field
theory) has some spectacular examples of the right answers gotten for
the wrong reasons.  Perhaps that is the case here?


That is more or less where we are today.  In recent years, a few papers have
been appearing with cautionary remarks that there are unresolved difficulties
with the mechanism proposed by Hawking.  
Yet most workers seem uncertain about just how serious those difficulties are. 
The aim of this paper is to lay out the status of our understanding so that
they may judge for themselves.  
For earlier perspectives, see Brout et al
(1995a), Jacobson (1999).

The organization of the paper is this.

Section 2 is a brief overview of Hawking's predictions, before explaining the
theory behind them.  The emphasis is on understanding the scales of the
predicted effects.    Section 3 is a review of the mechanism as it is usually
understood.  The emphasis is on making clear the nature of the physical
assumptions, rather than on details of the computations.   Section 4 discusses
the trans--Planckian problem (the occurrence of arbitrarily high energy
scales), and the sense in which it is localized.  

Sections 5--10 cover the arguments that have been put forward in
support of the Hawking process.  These include an attempt to derive it
from a combination of the Unruh effect and the principle of equivalence,
connections between it and moving--mirror models, arguments that it
can be derived from or essentially links general relativity and
thermodynamics, attempts to avoid the trans--Planckian problem by
introducing non--standard rules for the propagation of the quantum
field, and arguments in support of the Hawking mechanism from quantum
theories of gravity.  't Hooft's program for analyzing the
quantum structure of black holes is discussed in section 9.

Section 11 takes up quantum--gravitational issues from another
direction.  Rather than examining the consequences of particular
theories of quantum gravity (strings, loops, etc.), it considers the
general physical consequences which might be expected to arise from
any sort of quantum character of space--time.  These include quantum
limitations on measurement, discreteness of eigenvalues, and so on.

Section 12 discusses the experimental prospects for resolving the
question, and section 13 is a brief summary of the conclusions.
An appendix lists the points in the text supporting the main conclusions.

\smallskip

\it Conventions.  \rm  In most places in this paper, factors of $c$,
$\hbar$, $G$ and $k$ (Boltzmann's constant) are given explicitly, but
in a few places, where it would be too cumbersome, factors of $c$ are
omitted.  As has become conventional in this area, a factor of $k^{-1}$ 
is absorbed in the definition of entropies, so that they are pure
numbers.
The conventions for general relativity are those of Penrose
and Rindler (1984--6), and for quantum field theory are those of Schweber
(1961).  The metric signature is $+{}-{}-{}-$.  The Planck length, Planck
time, Planck energy and Planck mass are
$$\eqalign{l_{\rm Pl}&=(G\hbar /c^3)^{1/2}\simeq 1.6\cdot 10^{-33}\;{\rm cm}
  \; ,\cr t_{\rm Pl}&=(G\hbar /c^5)^{1/2}\simeq 5.4\cdot 10^{-44}\;
{\rm s}\; ,\cr
  E_{\rm Pl}&=(\hbar c^5/G)^{1/2}\simeq 1.2\cdot 10^{19}\; {\rm GeV}\;
,\cr
  m_{\rm Pl}&=(\hbar c/G)^{1/2}\simeq 2.2\cdot 10^{-5}\; {\rm g}\; 
.\cr}
$$

\section{The predictions}

This section gives a brief overview of Hawking's predictions
in the case of linear quantum fields, and without
discussing their derivations.
The aim is to acquaint the reader with the physical scales involved.  We shall
see that the Hawking effect would be (in most circumstances, and also in a
certain scale--invariant sense) extremely tiny. 
One will need to think carefully about \it all \rm possible physical effects at
or above these scales in order to have confidence that black holes radiate.

Hawking's predictions apply to isolated black holes which formed by
gravitational collapse but have settled down to (macroscopically)
stationary states.  The assumption that the holes formed by collapse
(rather than were created with the Universe) is essential, as will be
discussed in section 3.4.  According to the ``black hole uniqueness''
theorems, each isolated, stationary black hole which occurs in Nature
should be characterized by its mass $M$, angular momentum $J$ and
electric charge $Q$; its exterior will be a Kerr--Newman solution of
Einstein's equation.  Since we are interested in a critical assessment
of the theory, we consider for the most part the simplest case, an
uncharged, spherically symmetric hole (so $Q=0$ and $J=0$), whose
exterior is Schwarzschild.  The difficulties are already apparent in
this case.

A key quantity in black--hole physics is the \it surface gravity \rm
$\kappa$, which may be defined for example as the acceleration
measured by red--shifts of light rays passing close to the horizon
(Helfer 2001b).
For a Schwarzschild black hole, one has
$$\kappa =c^4/(4GM)\, .\eek$$
Since the Schwarzschild radius is $R_{\rm Sch}=2GM/c^2$, one can think
of $c/\kappa =2R_{\rm Sch}/c$ as the light--crossing time of the
hole.  This will be an important time scale in later sections.
Classically, this is the time scale for the final approach to the
black--hole state (which occurs exponentially quickly).
In the Hawking process, this is the time--scale for an exponential
blue--shift in the frequencies of the particular vacuum modes which
give rise to the Hawking quanta at any time.  Numerically, one has
$$c/\kappa \simeq 2.0\cdot 10^{-5}(M/M_\odot )\; {\rm s}\, ,\eek$$
where $M_\odot$ is the mass of the Sun.  Thus for solar--mass or
smaller black holes the time scale is very short by ordinary
standards.  Even for a super--massive black hole of size $\sim 10^8
M_\odot$, the time scale would only be about half an hour.

According to Hawking, a free massless field will radiate at a
temperature
$$T_{\rm H}=(\hbar /2\pi ck)\kappa\, .\eek$$
(Of course, we do not know of any free fields in Nature --- all known
fields interact.  The question of how Hawking's predictions might be
modified by interactions (and mass) will be taken up in Section 12.2.)
For a Schwarzschild hole, one has
$$\eqalign{ T_{\rm H}&=\hbar c^3/(8\pi GMk)\cr
  &\simeq 6.2\cdot 10^{-8}\; (M_\odot /M)\; {\rm K}\, .\cr}\eek$$
Thus radiation from a solar--mass black hole would be exceedingly cold
--- about $5\cdot 10^7$ times colder than the cosmic microwave background.
Larger black holes would be colder still.  This gives one a sense of
just how easily Hawking radiation can be lost in other, apparently
small, effects.  Of course, smaller black holes would have higher
temperatures, and there is some possibility that ``mini'' black holes
might exist and that Hawking radiation from them might be detected.
A ``mini''black hole of mass $\sim 10^{15}$ g would have $T_{\rm H}\sim
10^{11}$~K. 

The luminosity of the black hole can be estimated from the
Stefan--Boltzmann law if one has a measure of the effective radiating
area.  Precise calculations of this are lengthy, but their results are
that the area is $\alpha A$, where $\alpha$ is a numerical factor of
order unity and $A$ is the area of the hole.  Thus for a Schwarzschild
hole the luminosity due to a given massless field is
$$L=(\pi ^2k^4/60\hbar ^3c^2)\alpha AT_{\rm H}^4\, .\eek$$
(The precise Stefan--Boltzmann factor depends on the field species, but we shall
absorb variances in this into $\alpha$.)

Notice that what one has is a black body with a cavity size of the
same order as the dominant wavelength --- both are $\sim R_{\rm Sch}$. 
\it This corresponds to an object
which is very dim in an invariant sense.  \rm
While the spectrum is indeed thermal, it is at such a low temperature
compared to its physical dimensions that it cannot be thought of very
accurately as a classical flux of radiation.
In fact, a measure of the
time between emission of quanta may be given by
$$\hbox{(mean energy per quantum)}/L =
  240\pi ^{-1}\alpha ^{-1} \hbar /(kT_{\rm H})\, ,\eek$$
that is, somewhat longer than the order of the mean period of the quanta.  
The emission of Hawking radiation from the black hole is thus a
process which not only has a quantum origin but is quantum in its
presentation.

Despite this essentially quantum character of the radiation,
it is natural to expect that we should in a
time--averaged sense be able to talk about the rate of mass loss of
the hole due to the process.  This is on its face a very natural
suggestion, and seems to rely only on the assumption that energy is conserved.
Then we have
$${{\d M}\over{\d t}}=-L/c^2\, ;\eek$$
for a Schwarzschild hole
$${{\d M}\over{\d t}}=-(\alpha c^4\hbar /960 G^2)n_{\rm eff}M^{-2}\,
,\eek$$\xdef\lumeq{\the\EEK}%
where $n_{\rm eff}$ is the effective number of radiating species.
This number will depend (weakly) on $M$, since (for example) as the
mass decreases, the temperature rises and linear fields of masses
$\lesssim kT_{\rm H}/c^2$ will contribute significantly to the Hawking
process.  However, ignoring this dependence, we may get a rough
estimate of the time scale over which the mass changes significantly
due to the Hawking process (which will also be an estimate of the
lifetime of the hole) by solving equation (\lumeq ):
$$t_{\rm life}\sim  (320 G^2 /\alpha \hbar c^4 n_{\rm eff})
  M^3\sim (M/M_\odot)^3\cdot 10^{65}\, {\rm y}\, .\eek$$

We are not interested in speculating about physics over scales longer than the
age of the Universe.  Our interest in this equation is rather that it
constrains the possible masses of ``mini'' black holes generated early in the
Universe.  If we set $t_{\rm life}$ to the present age of the Universe, we
obtain a minimum mass such a ``primordial'' black hole must have had (assuming
it Hawking--radiates) to survive to the present day.  This mass is $\sim
10^{15}\, {\rm g}$.

\section{The Hawking process}

In this section, I outline the derivation of the prediction of thermal
radiation.  I have kept the treatment of the essential physical elements of the
mechanism quite close to Hawking's, but I have taken advantage of insights
which have developed in the interim to streamline the presentation and avoid
most of the technical mathematics.\fnote{Actually, Hawking's (1975) paper
contained some important comments on the quantum character of space--time. 
These ideas have fallen out of fashion, and will not be discussed in this
section, but they are close to concerns to be raised in section 11, below.}

In outline, we have a scattering problem to solve:  given an initial
state of the quantum field, before the black hole has formed, find
what the final state of the field will be, after the hole has formed
(and any transients have passed).  The underlying assumption will be
that we can treat space--time classically, and the quantum fields as
propagating on this background space--time. 

\subsection{The space--time}

So let us begin by describing the space--time, $({\cal M}, g_{ab})$.
We will assume it is spherically symmetric, and contains matter
imploding to form a black hole.  We also assume that this is isolated,
and model this mathematically by assuming that the space--time is
asymptotically flat.  (It should be emphasized that in physical terms
``asymptotic'' here means a regime around the collapsing system in
which space--time is suitably flat; at late times, when the matter is
very nearly collapsed, this might be anything from tens of
Schwarzschild radii outwards.  It is a convenient mathematical fiction
to assume that the collapsing object is perfectly isolated, so that
this asymptotically flat regime extends to infinity, and we shall do
so, but one can verify that the physics does not depend on this.)

\epsfysize=3in
\epsfbox{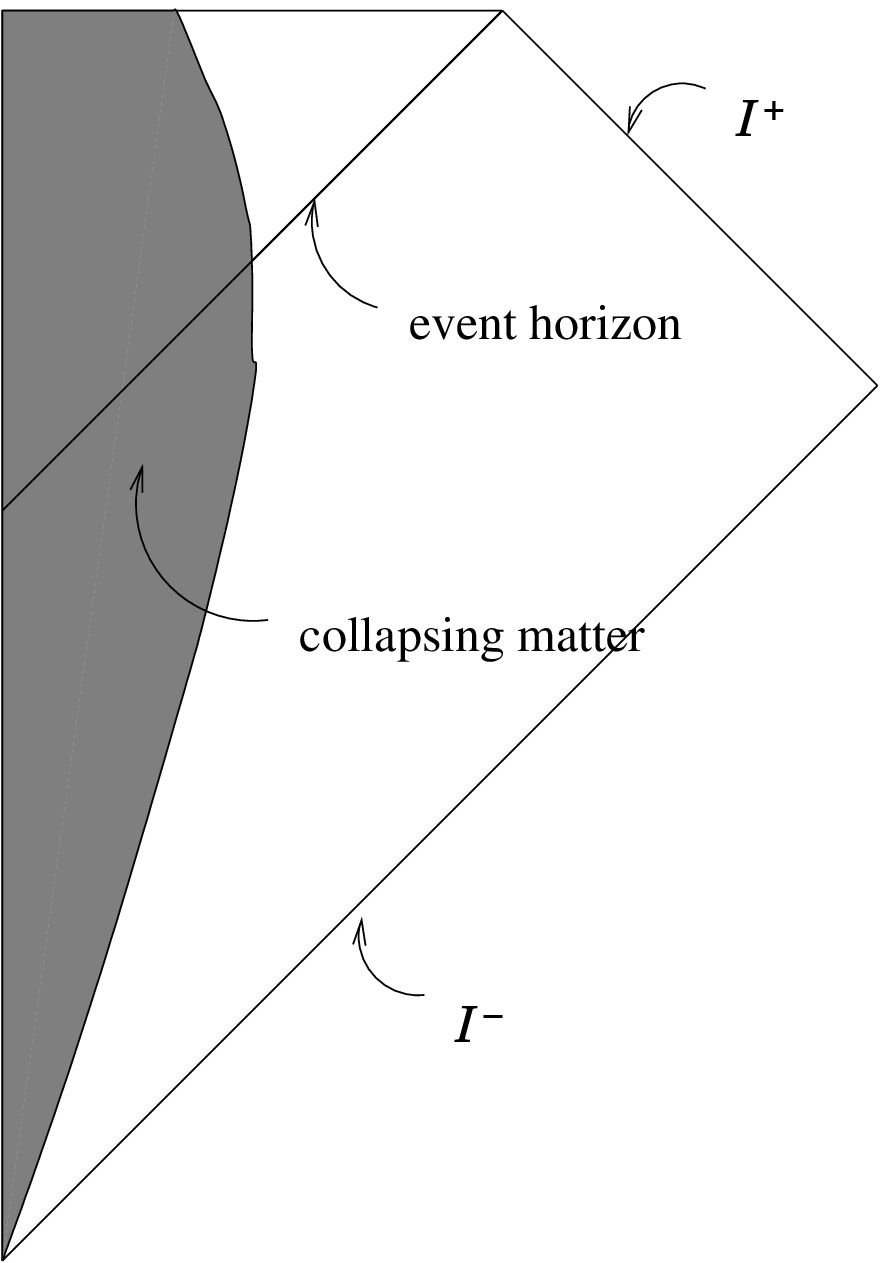}
\figure{The Penrose diagram of a spherically symmetric object collapsing to form
a black hole.  The future and past null infinities $\scri ^\pm$ are shown, as
well as the event horizon.  The black hole is the region at and above the event
horizon.}

The standard Penrose diagram for this space--time is shown in
figure~1, and the reader will find it convenient to refer to this.  In
this diagram, the rotational symmetry has been factored out, so each
point represents a sphere of symmetry.  Scales are distorted in the
diagram, but the causal structure is accurately portrayed.  Thus the
lines at $45$ degrees are null (and correspond to radial null geodesics
in the space--time).  The diagram also allows one to represent various
infinite regimes as finite ones. 

Since by Birkhoff's theorem a vacuum spherically symmetric space--time
must be locally isometric to part of the extended Schwarzschild
solution, we have an explicit understanding of the metric in the
exterior of the matter.  The Schwarzschild metric in standard
coordinates is
$$\fl\d s^2 =\left( 1-(R_{\rm Sch}/r)\right) \d t ^2 -
  \left( 1-(R_{\rm Sch}/r)\right) ^{-1} \d r ^2-r^2\left(\d\theta ^2 +\sin
^2\theta \d\varphi ^2\right)\, ,\eek$$\xdef\Schmet{\the\EEK}%
where $R_{\rm Sch}=2GM/c^2$ is the Schwarzschild radius and $M$ is the mass.
It is also convenient to introduce the ``tortoise'' coordinate
$$r_*=r-R_{\rm Sch} +R_{\rm Sch}\log \left( (r/R_{\rm Sch})-1\right)\, ,\eek$$
as well as the retarded and advanced time coordinates
$$u=t-r_*\, ,\qquad v=t+r_*\, .\eek$$
These are both null coordinates.
The metric (\Schmet ) and the coordinates are valid where both:  (a) the
actual metric is Schwarzschild (that is, exterior to the matter); and
(b) one is outside the Schwarzschild radius.

We shall not need to consider in any detail the region of space--time
within the black hole (that is, within the event horizon).  We will
need, however, some basic information about the portion of space--time
containing matter and exterior to the hole, since the field modes
which are ultimately supposed to give rise to Hawking radiation must
propagate through this region first.  

While the explicit form of the metric
in this region will not be necessary, we can see that the coordinates
$r$, $u$, $v$, $\theta$ and $\varphi$ have well--defined extensions to
this region.
The coordinate $r$ may be defined throughout the
space--time by taking, for any event $p$, the area of the sphere of
symmetry through $p$ to be $4\pi r(p)^2$.
The coordinates $u$ and $v$ may be extended by requiring them to
remain null and spherically symmetric. 
They are good coordinates in the region under consideration except on the axis
of symmetry.
The angular coordinates
$\theta$ and $\varphi$ may be extended unambiguously (except for the
usual spherical coordinate singularities) by flowing in the two--spaces
orthogonal to the spheres of symmetry.

\subsection{Null infinity, the black hole, and time--asymmetry}

A few words are in order at this point about the null asymptotic
structure of the space--time, since this enters both in the
specification of the initial and final data for the massless field and
in the definition of the black hole.  We may attach to space--time
future and past null infinities $\scri ^\pm$, which are null
hypersurfaces.  Each point on $\scri ^+$ is reached by holding
$(u,\theta ,\varphi )$ constant and taking $r\to +\infty$; we take as
usual $(u,\theta ,\varphi )$ as coordinates on $\scri ^+$.  As defined
here, then, future null infinity consists of the future end--points of
the radial null geodesics in the Schwarzschild exterior.  Actually,
every null geodesic which escapes (as one moves into
the future) the gravitational pull of the matter has a well--defined
end--point on $\scri ^+$.  By exchanging ``future'' and ``past,'' and
$u$ and $v$, one has corresponding statements for $\scri ^-$.

The treatment of $\scri ^+$ and $\scri ^-$ has so far been symmetric,
but the discussion of black holes breaks this symmetry.  A black hole
is defined to be a region of space--time from which signals cannot
escape to arbitrarily distant regions.  Formally, escape in this sense
means reaching $\scri ^+$.  Thus the black hole (if it forms) is the
set of events in space--time for which there are \it no \rm
future--directed causal curves reaching $\scri ^+$.\fnote{A 
causal curve is one whose tangent is everywhere null or timelike (if it is
differentiable).}  The boundary of
this set is the future \it event horizon \rm ${\cal H}^+$.  

One could time--reverse the concept of a black hole; the result would
be a \it white hole, \rm a region which into which signals originating
from very great distances could not penetrate, but from which signals
might escape.  Such an object would evidently be quite different from
a black hole.  

Note that this means that if a black--hole space--time is time--symmetric, it
must also be a white--hole space--time.  Such space--times are not models of
gravitational collapse of isolated objects and are not of direct interest in
the Hawking mechanism.  The mechanism requires a time--asymmetric space--time.

\subsection{The mapping of surfaces of constant phase}

A key role will be played by the \it mapping of surfaces of constant
phase.  \rm  A surface $u=$ constant is a spherically symmetric
outgoing null hypersurface.  If the radial null geodesics forming this
surface are traced backwards in time through the spatial origin, they
emerge to form a spherically null hypersurface which (if read forwards
in time) would be incoming, that is, would be a $v=$ constant
surface.  Thus to each retarded time $u$ we may associate an advanced
time $v(u)$.  Since the $u=$ constant and $v=$ constant surfaces are the
spherically symmetric surfaces of constant phase in the
geometric--optics approximation, the function $v(u)$ is the mapping of
surfaces of constant phase.  

(Below, we shall need to refer both to the function $v(u)$ and to the
coordinates $u(p)$, $v(p)$ of an event $p$.  These are quite different
functions, with $v(u)$ an ordinary real--valued function of one real variable,
but $v(p)$ and $u(p)$ each a real--valued function of the event $p$ in
space--time.  One could alternatively write something like $V(u)$ for the
function $v(u)$, but this introduces an undesirable asymmetry between $u$ and
$v$.)

Suppose a spherically symmetric massless wave is sent through the space--time,
and the period is initially $\delta v$.  In the geometric--optics
approximation, it will emerge with period $\delta u$, where $\delta
v/\delta u =\d v(u)/\d u$.  This is the ratio of the frequencies, and
so $\d v(u)/\d u$ is precisely the red--shift factor.

The mapping of surfaces of constant phase is intimately bound up with
the formation of a black hole.  It can be shown on very general
grounds that when matter collapses to form a black hole, one has
$\d v/\d u\to 0$ and the asymptotic relation
$${{\d ^2v}\over{\d u^2}} =-\kappa {{\d v}\over{\d u}} +O((\d v/\d
u)^2)\qquad\hbox{as\ \ }u\to +\infty\, ,\eek$$
where $\kappa$ is the surface gravity of the hole (Helfer 2001).  
Notice that this
is a universal relation, depending only on the surface gravity and not
on any details of the formation of the hole.  It implies
$${{\d v}\over{\d u}}\sim \exp -\kappa u\qquad\hbox{as\ \ }u\to
+\infty\, ,\eek$$\xdef\exprel{\the\EEK}%
so signals from the hole--to--be are exponentially red--shifted.
Also one must have
$$\lim _{u\to+\infty} v(u)=v_0\, ,\eek$$
where $v_0$ is the \it advanced time of formation \rm of the hole, the
advanced time at which the event horizon forms.  We have
$$v_0-v(u)\sim\exp -\kappa u\qquad\hbox{as\ \ }u\to +\infty\,
.\eek$$\xdef\mapform{\the\EEK}%

The details of the formulas (\exprel , \mapform ) will be absolutely 
central to the
arguments for the Hawking process.  \it This should be contrasted with the
significances of the formulas for classical physics.  \rm Classically, the
equations imply that (given that a horizon is about to form), there is
a fast (time scale $\sim c\kappa ^{-1}$) approach to the black--hole
state, 
a state where $v(u)$ is indistinguishable from $v_0$ and $\d v/\d u$ is
indistinguishable from zero.  In distinction, \it the derivation of the
Hawking process will rely on the literal validity of (\exprel , \mapform ) for
arbitrarily late retarded times $u$.  \rm  Thus the exponential increase of
the red--shift, for arbitrarily long times, will be a
central assumption of Hawking's argument.

Another important feature of the mapping of surfaces of constant phase
is that it gives us a way of resolving the singularity of the
coordinate $u$ at the event horizon.  (At the event horizon, we have
$u\to +\infty$.)  If for any event $p$, we take the past--directed
radial null
geodesic inwards, let it pass through the spatial origin and then outwards
to $\scri ^-$, it arrives with an advanced time $\tilde v (p)$.  (The
tilde is to distinguish $\tilde v (p)$ from the coordinate $v(p)$.)
If $p$ lies before the event horizon, so that its retarded time $u(p)$
is well--defined, then $\tilde v (p) =v(u(p))$, where $v(u)$ is the
mapping of surfaces of constant phase and $u(p)$ is the value of the coordinate
$u$ at $p$.  However, the coordinate
$\tilde v$ is easily seen to be a good coordinate throughout
space--time (except at the spatial origin), 
and by its definition is a coordinate constant on radial
outgoing null surfaces.

Thus $v(u(p))=\tilde v(p)$, where $v(u)$ is the mapping of surfaces of constant
phase, provides a good coordinate which extends naturally past the horizon.  In
the case of Schwarzschild space--time, this coordinate would be (up to a
constant factor) the usual null Kruskal coordinate.

\xdef\coorddef{\the\secno{}.\the\subno{}}

\subsection{Propagation of the quantum field}

Now let us turn our attention to the quantum field and its
propagation.  We shall work with a minimally coupled massless scalar
field $\phi$, so the field equation is 
$$\nabla _a\nabla ^a\phi =0\, ,\eek$$
however the essentials of the argument would be the same for
conformally coupled fields, or for fields of non-zero helicity.  

The
basic strategy is this.  We assume the field in the distant past is
specified.  (For definiteness, we take it to be the vacuum, although
the results would be the same for any reasonably quiescent state.)
Thus we understand the expectations of 
combinations of the field operators $\phi (p)$
for events $p$ in the distant past.  We therefore take the field
operators in the distant past as initial data for the field equation.
We may work out the field operators $\phi (q)$ for later events in
terms of these data by solving the field equation, and then see what
the quantum state looks like by forming expectations of these $\phi
(q)$'s.  The expectations will be discussed in the next subsection;
this subsection deals only with the problem of propagation.

Since the field is linear, it can be written as a sum (or integral) of
c--number mode functions times ordinary creation and annihilation
operators with no space--time dependence.  In other words, the
operator character factors through the field equation, and we may
discuss the propagation the field equation 
engenders without distinguishing between 
the mode functions and the corresponding operators.

Taking advantage of the spherical symmetry, we decompose the field
into spherical harmonics:  $\phi =\sum _{l,m}\phi _{l,m}Y_{l,m}$.  It
is also convenient to extract a factor of $r$ from the field: we let
$\phi =\phi ^0/r$ and $\phi _{l,m}=\phi ^0_{l,m}/r$.  Then the fields
$\phi ^0_{l,m}$ satisfy reduced wave equations
$$\partial _u\partial _v\phi ^0_{l,m} +\quad
  \hbox{lower--order terms}\quad =0\, ,\eek$$
where the lower--order terms contain a potential in the vacuum region
and more complicated, time--dependent, terms in the matter region.

An explicit solution of these equations would clearly depend on the
details of the metric within the matter, and thus we cannot reasonably
expect to obtain this in generality.  However, if we restrict our
attention to those modes relevant to the detection of the field at
late retarded times and in moderate frequency regimes, we can get an
exact asymptotically valid formula, as first realized by Hawking.

Consider the observation of some field modes of moderate wavelengths
$\sim\lambda _{\rm char}$ at late retarded times and large spatial
distances, in other words, for large $u$ and near $\scri ^+$.  
Here ``moderate'' means that we hold fixed a finite
interval of wavelengths contributing to the
wave--packet; doing so, we will find certain asymptotic
behavior as $u\to +\infty$, $r\to +\infty$.\fnote{Strictly speaking,
one cannot simultaneously have the wave packet bounded in $u$ and have
its Fourier transforms bounded in wavenumber space.  However one can
arrange for either one to  be bounded and the other to decay
exponentially rapidly, or both to decay rapidly.}  
The important assumption is that the interval of wavelengths
is not allowed to grow
with $u$ or $r$.\fnote{The important restriction here is actually the
infrared one.  To analyze infrared effects one would need to refine
the analysis given here.}
In fact, the wavelengths of
interest will turn out to be those of order $\sim R_{\rm Sch}$.

Let us consider the propagation of these field modes backwards in
time in two stages, the first of which is unaffected by the matter or
its collapse, and the second of which brings in the effects of the
collapse on propagation.  It is helpful to think of propagating the
data from a Cauchy surface (that is, an initial--data surface) $\Sigma
^+$ where it is given, back in time to an intermediate Cauchy surface
$\Sigma$ dividing the two stages, and finally back further to $\Sigma
^-$ (which will essentially be $\scri ^-$).  Thus the propagation of
the data from $\Sigma ^+$ to $\Sigma$ will involve only the exterior
vacuum Schwarzschild geometry, whereas the propagation from $\Sigma$
to $\Sigma ^-$ will show the effects of the collapse.

\epsfbox{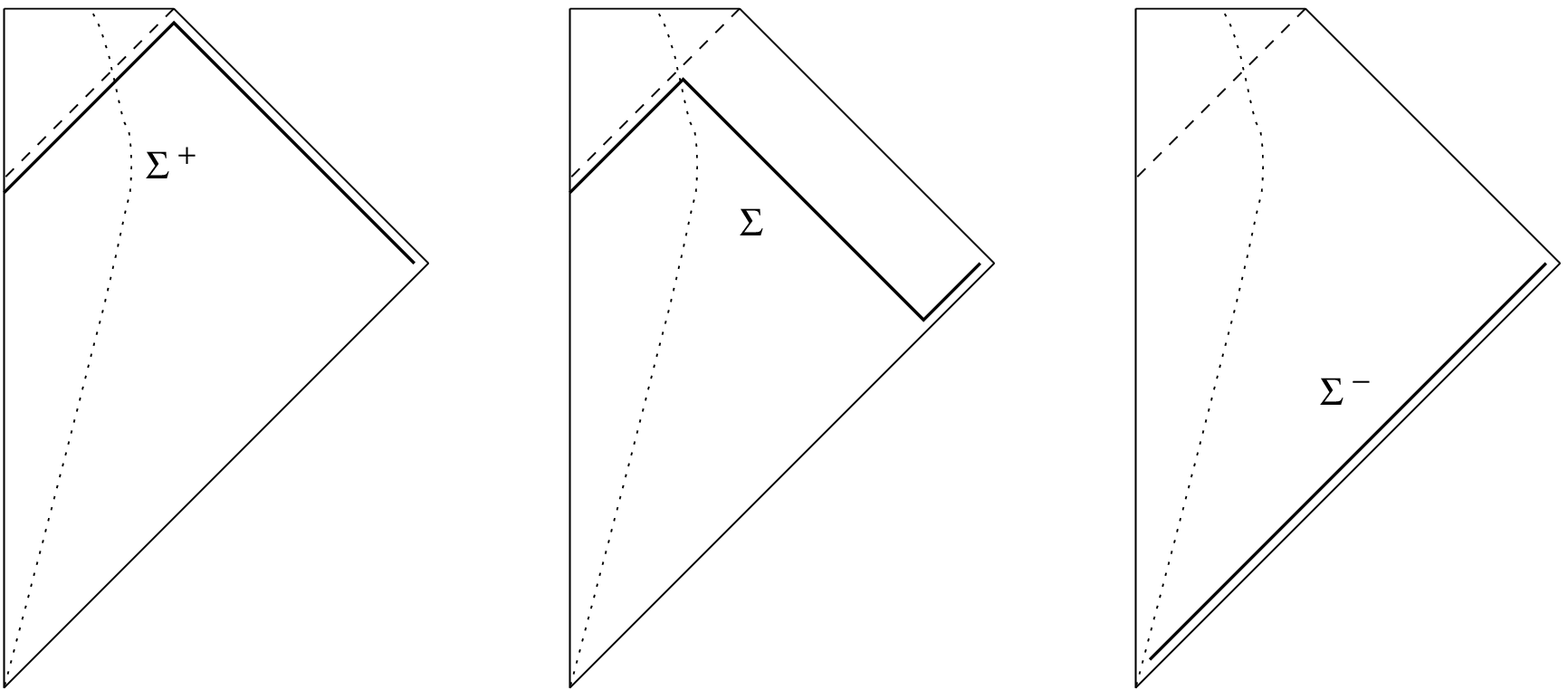}
\figure{The three Cauchy surfaces used in propagating the field modes backwards
in time from $\scri ^+$ to $\scri ^-$.  The portions in the black hole (on and
above the event horizon) are given for conceptual completeness and do not
contribute to the calculation.}

In what follows, it will be convenient to refer to figure~2.  The
field data we are given at late times on $\scri ^+$.  While $\scri ^+$
does not by itself constitute a complete Cauchy surface (roughly
speaking, it contains no data for modes propagating into the hole),
we may take $\Sigma ^+ =\scri ^+ \cup {\cal H}^+$ as a Cauchy surface
on which our data are given (and also one could give data for field
modes propagating into the hole).\fnote{This is a bit of a simplification,
because ${\cal H}^+$ does not actually meet $\scri ^+$.  We should
more properly use a radially--symmetric null hypersurface slightly to
the past of ${\cal H}^+$.  
There is a similar potential technicality involving the identification
of $\Sigma ^-$ with $\scri ^-$.
These issues will not be important here.}
The inclusion  of ${\cal H}^+$ in the definition of $\Sigma ^+$ is
really only for completeness --- a Cauchy surface must be large enough
to accommodate all possible data for the field.
The data we are considering are identically zero on ${\cal H}^+$,
because we are at present interested in what distant observers
perceive, rather than in what happens at the event horizon.  

As mentioned above, the precise definition of the Cauchy surfaces in the
black--hole interior will not be important.  It will be convenient to define
$\Sigma$ in a sort of zig--zag, as follows.  The boundary of the collapsing
matter crosses the event horizon at some advanced time $v_1$.  Let $\Sigma$
consist of three parts:  the portion of the event horizon to the past of
$v=v_1$; the portion of $v=v_1$ in the exterior of the hole; and the portion of
$\scri ^-$ to the future of $v=v_1$.
By the causality
of propagation, any data of the sort we are interested in, which were
initially supported on $\scri ^+$, will give rise to non--zero data
only on that portion of $\Sigma$ external to the hole.  Indeed, the
propagation of such signals from $\scri ^+$ to $\Sigma$ depends only
on a portion of the exterior vacuum Schwarzschild geometry.
\xdef\sigsec{\the\secno{}.\the\subno}%

The precise propagation is determined by the reduced wave equation
governing the $\phi ^0_{l,m}$'s.  It has the form (in the vacuum region)
$$\partial _u\partial _v \phi ^0_{l,m} +V_l(r)\phi ^0_{l,m}=0\,
,\eek$$\xdef\zereq{\the\EEK}%
where $V_l$ is a positive potential falling to zero as $r\to 2M$ and
$r\to +\infty$, and including a centrifugal term.  As the wave packet
propagates backwards from $\scri ^+$ to $\Sigma$, it is partly
dispersed and partly reflected by the potential.  The reflection is a
time--symmetric process (since reflected modes never enter the
time--dependent region), and thus does not contribute directly to the
Hawking process; it will be ignored.  The dispersion means that the
portion of the wave--packet that does propagate through the potential
will arrive at $v=v_1$ somewhat distorted, and with a tail (falling
off by a power law).  

Now we come to a very important point.  A given wave packet with wave
profile $\phi ^0_{l,m}(u)\Bigr| _{\scri ^+}$ at $\scri ^+$ will give rise
to some profile $\phi ^0_{l,m}(u)\Bigr| _{v=v_1}$ at $v=v_1$.  We are
interested in asymptotics for late times:  the effect of translating
by a time $T$ is to replace $u$ by $u+T$ in both of these profiles.
Now recall that $u$ is \it not \rm a good coordinate near the event horizon,
but rather $v(u)\simeq v_0-C\exp -\kappa u$ is a good coordinate
(cf. section \coorddef ).
This means that the profile $\phi ^0_{l,m}(u)\Bigr| _{v=v_1}$ is, as
$T\to +\infty$, squeezed into an exponentially small portion of the
$v=v_1$ surface, just before the horizon.  In other words, in terms of
any fixed local frame in the neighborhood of the point where the
matter crosses the horizon, the relevant wave profiles become
compressed into very tiny intervals; correspondingly, the frequencies
of their components become exponentially blue--shifted.

We may thus, when we follow the waves back in time from $\Sigma$,
apply the geometric--optics approximation.  This approximation remains
mathematically valid throughout this second stage of propagation,
because there is nothing in this stage to undo the divergent
blue--shift.  The waves here propagate backwards in time, inwards
through the matter, through the spatial origin, outwards through the
matter again, and finally out to $\Sigma ^-=\scri ^-$,
but this portion of the
trip only involves bounded red--shifts.  Thus asymptotically (as $T\to
+\infty$) the geometric--optics approximation becomes \it exactly \rm
valid for this second portion of the trip.  One has simply
$$\phi ^0_{l,m}(v(u))\Bigr| _{\scri ^-}=(-1)^{l+1}\phi
^0_{l,m}(u)\Bigr| _{v=v_1}\, ,\eek$$\xdef\firsthalf{\the\EEK}%
where the factor $(-1)^{l+1}$ is due to reflection through the origin.

Notice that this result is universal (for the class of isolated
spherically symmetric black holes of a given mass); it is independent
of the details of the formation of the hole.  It is undoubtably a very
beautiful picture.

In order to compute the specifics of the field propagation, then, it
only remains to work out $\phi ^0_{l,m}\Bigr| _{v=v_1}$ from $\phi
^0_{l,m}\Bigr| _{\scri ^+}$, a problem in vacuum Schwarzschild
geometry.  The result would be of the form
$$\phi ^0_{l,m}(u)\Bigr|_{\scri ^+} =\int _{-\infty}^u
   K_l (u-u')\, \phi ^0_{l,m}(u')\Bigr|_{v=v_1}\, \d u'\,
   ,\eek$$\xdef\secondhalf{\the\EEK}%
where $K_l$ is a suitable Green's function.  (This is more commonly
expressed in Fourier--transformed terms; here $K_l$ is essentially the
Fourier transform of the transmission coefficients.)  We shall
not need the precise forms of the $K_l$'s here.  A very rough understanding
of the propagation will be adequate for our needs.

It is evident from the foregoing that the only scale which enters the
wave propagation is set by the Schwarzschild radius.  Thus it is only
to be expected that the wavelengths predicted for the Hawking quanta
must be of the order of this radius.  As these are propagated
backwards from $\scri ^+$, the modes with $l\geq 1$ are almost entirely
reflected from the potential barrier.
This reflection occurs entirely in the vacuum Schwarzschild region,
which is time--symmetric, and so contributes little to the
Hawking process.  The Hawking process is 
therefore almost entirely an s--wave process.  

Since the wavelengths involved are of the same scale as the
Schwarzschild radius, one may expect that comparable fractions of the
s--wave are reflected off the potential barrier and penetrate it.
Similarly, the kernel $K_0$ must fall off beyond a length scale of the
order of the Schwarzschild radius.  It must have a $\delta (u-u')$
contribution (because of the $\partial _u\partial _v$ term in 
equation (\zereq )),
but apart from this it cannot have structure on scales much smaller
than the Schwarzschild radius.  

It is therefore not surprising that one can get a zeroth--order
(order--of--magnitude)  approximation to the physics by simply taking
$K_0=\delta (u-u')$.  This is the same as using the geometric--optics
approximation for the propagation from $\Sigma ^+$ to $\Sigma$.  
(We emphasize that the geometric--optics model of propagation from $\Sigma$
back to $\Sigma ^-$ is expected to become rapidly
asymptotically exact at late
times.  It is using geometric optics to model the propagation from
$\Sigma ^+$ to $\Sigma$ which is a rough--and--ready approximation.)
Then
one has simply
$$\phi ^0_{0,0}(u)\Bigr| _{\scri ^+}=-\phi ^0_{0,0}(v(u))\Bigr|
_{\scri ^-}\eek$$\xdef\simplescat{\the\EEK}%
for the propagation of the fields, that is, the scattering.  A more
precise computation, including the kernel $K_0$, would involve only a
small, exponentially compressed, smearing of the right--hand side.

The very simple scattering formula (\simplescat ) is the same as that for a
field on one side of a moving mirror (perfect reflector) in
two--dimensional Minkowski space--time, with $v=v(u)$ the trajectory
of the mirror (and $u=t-x$, $v=t+x$ Minkowski null coordinates).  In
fact, moving mirrors were extensively investigated \it after \rm
Hawking's work, largely with the aim of clarifying aspects of it.  We
shall return to this point later. 

To summarize, then.  The propagation of the quantum fields is given very nearly
exactly by (\firsthalf ) and (\secondhalf ),  and to zeroth order (that is,
accurately incorporating the main physical ideas but ignoring factors of order
unity) simply by (\zereq ), where $v(u)$ is the mapping of surfaces of constant
phase. We have the asymptotic behavior
$$v(u)\simeq v_0-C\exp -\kappa u\eek$$\xdef\forrm{\the\EEK}%
as $u\to +\infty$, that is, as the horizon is approached.  

Note that the formula (\forrm ) depends on the black hole forming at a
finite time (precisely, at the advanced time $v_0$).  This assumption
is used in propagating the field backwards in time out from the
vicinity of the collapsing object and towards $\scri ^-$.  For a hole
formed at the very origin of the Universe, such an analysis would not
apply, and one would have to consider the propagation of the field
back in the vicinity of the initial singularity (the big bang).  The
treatment of such holes will not be considered here.

In Hawking's model, the
behavior of the field for a range of $u$--values
$\Delta u$ around $u$ will arise from initial field data from an
exponentially compressed range of $v$--values, with $\Delta v\simeq v'(u)\Delta
u$ and $v'(u)\simeq \exp -\kappa u$.  This dependence of the predictions on the
initial field values on exponentially fine scales is the \it trans--Planckian
problem.  \rm  It means that the Hawking quanta arise from the structure of the
initial state at exponentially increasing frequencies and wave--numbers.

\subsection{The quantum state and the two--point functions}

We have so far discussed the evolution of the field operators; we now turn to
the quantum state on which they act.  This is specified as being a reasonably
quiescent state in the distant past.  We shall see in a little while that the
precise choice is not too important; for simplicity, we shall take the state to
be the in--vacuum, $|0_{\rm in}\rangle$.  

While it is natural to ask about the particle content of the state, it is for
many purposes better to concentrate on the $n$--point functions, especially the
two--point function, $\langle 0_{\rm in}|\phi (p)\phi (q)|0_{\rm in}\rangle$. 
This is because the definition of ``particle'' is ambiguous here, as indeed it
is in any problem where there is a non--stationary external potential (here, the
time--dependent, collapse, phase of the space--time).  Particles are defined in
terms of creation and annihilation operators, which are in turn defined by
splitting the field into positive-- and negative--frequency components.  When
the field propagates through a non--stationary regime, this splitting is not
preserved.  Thus while the particle content of the theory is interesting, its
analysis involves an extra issue, the splitting by frequencies, which is often
best dealt with separately.

In fact, our primary interest is in whether the mathematical
quantities appearing in Hawking's
analysis (especially the field at very fine scales) can be regarded as credible
representations of the real physics.  This does not depend on whether one works
in the particle representation or the field representation, and so analyzing the
particle content of the theory is not really necessary for us.  A brief sketch
of this will be given, however, for completeness, in section 3.5.3.

We begin, then, by considering the two--point functions.

\subsubsection{The two--point function in Minkowski space}

The state is supposed to be the in--vacuum.  Since space--time is 
asymptotically
flat, this means that near $\scri ^-$ the two--point function has its
Minkowskian form.  So we shall begin by reviewing the two--point 
function of the
massless scalar field in Minkowski space.
This two--point function is
$$\langle 0_{\rm M}|\phi (p)\phi (q)|0_{\rm M}\rangle =
  -{1\over{4\pi ^2}}\cdot {1\over{(p-q-\i\epsilon )^2}}\,
  ,\eek$$\xdef\minkfun{\the\EEK}%
where $\epsilon$ is an infinitesimal future--directed timelike vector.  
Of course, strictly speaking it is not an ordinary function, but a 
distribution,
and it must be averaged against test functions to be given meaning.

We can see a great deal of the physical content of the two--point function, and
the part averages play, by considering a simple computation of vacuum
fluctuations.  Suppose we consider the average 
$$\Phi (a) =(4\pi a^3/3)^{-1}\int _{x^2+y^2+z^2\leq a^2}\phi (0,x,y,z)\, \d x\,
 \d
y\,\d z\, ,\eek$$
of a field operator over a sphere of radius $a$.  The expected squared
fluctuation in this is
$$\eqalign{\langle 0_{\rm M}|\Phi (a)\Phi (a)|0_{\rm M}\rangle
  &=(4\pi a^3/3)^{-2}(4\pi ^2)^{-1} \int _{|{\bf x}|\leq a}\d ^3{\bf x}
     \int _{|{\bf y}|\leq a}\d ^3{\bf y}\, |{\bf x}-{\bf y}|^{-2}\cr
   &=\const a^{-2}\cr}\eek$$
on dimensional grounds.  (One can check that the integral is convergent.)  Thus
attempts to measure the field on a scale of size $a$ (equivalently, using modes
with wavenumbers up to $\sim a^{-1}$) result in fluctuations of the order of
$a^{-1}$.  These are precisely the vacuum fluctuations.   That these diverge as
$a\to 0$ shows that questions of localizing fields are very delicate ones. By
differentiating (\minkfun ), and then integrating, one can also find
fluctuations in the  derivatives of the fields.  

While the very simple form of (\minkfun ) is a consequence of both the choice 
of state (vacuum) and the field equation (free massless in Minkowski space), 
its asymptotic behavior as $p\to q$ is expected to hold much more generally. 
Essentially, this is because the asymptotic regime $p\to q$ corresponds to
ultra--high energies.  For any fixed, ``reasonable,'' state, near any event,
there should be an energy scale beyond which the field modes are essentially
unexcited, and also much higher than any masses appearing in the theory.  In
this case, one would expect the two--point function to approximate that of a
massless field in vacuo.  Such states are called \it Hadamard\rm ,  and are
generally regarded as the candidates for  physically realistic
states.\fnote{The discussion in this paragraph is a bit of a simplification,
because the energy of a quantum field propagating on a background cannot be
analyzed simply in terms of the frequencies of the modes.  A more accurate
statement would be that the Hadamard condition is preserved by propagation.}

Thus one expects asymptotic behavior like that of (\minkfun ) as $p\to q$  to
hold even in curved space--time, as the averaging scale $a\to 0$.  We shall
discuss the Hawking process in detail, below, but for the moment  let us recall
that we saw above that the Hawking quanta are supposed to arise from initial
field data averaged over scales $a\sim R_{\rm Sch}\exp -\kappa u$.   This means
that the Hawking quanta are supposed to arise from exponentially extreme field
fluctuations at exponentially tiny scales in the in--state.

To connect these results with the calculations in the Hawking model, we
decompose the two--point function into spherical harmonics.  I shall only give
the result for $l=0$ here, as that is the only one I shall use explicitly. 
This is got by averaging $p$ and $q$ over their spheres.  Remembering that a
factor of $r$ is absorbed in the expansion $\phi =r^{-1}\sum _{l,m}\phi
^0_{l,m}Y_{l,m}$, one  finds
$$\fl\eqalign{
 \langle 0_{\rm M}|&\phi ^0_{0,0}(u_1,v_1)\phi ^0_{0,0}(u_2,v_2)
    |0_{\rm M}\rangle\cr
  &=-r_1r_2 (4\pi )^{-1}\int\sin\theta _1\d\theta _1\d\varphi
_1\int\sin\theta _2\d\theta _2\d\varphi _2\left( (t_1-t_2-\i\epsilon
)^2
  -({\bf r}_1-{\bf r}_2)^2\right) ^{-1}\cr
  &=-(16\pi ^2 )^{-1}\log{{(u_1-u_2-\i\epsilon
)(v_1-v_2-\i\epsilon)}\over {(v_1-u_2-\i\epsilon )(u_1-v_2-\i\epsilon )}}\cr
  }\eek$$\xdef\redfun{\the\EEK}%
where $u=t-r$ and $v=t+r$.  This is then the two--point function of
a dimensionally--reduced field theory, a field which is a function of
the coordinates $u$ and $v$ alone.

This $l=0$ two--point function can be understood in a very convenient
way by thinking of the field in terms of its initial datum, that is,
its value at $\scri ^-$.  To see this, recall the classical result
that for the $l=0$ solutions of the wave equation, the
geometric--optics approximation is exact.  (That is, any $l=0$
solution has the form $(f(u)+g(v))/r$.)  When we require regularity of
such solutions at the spatial origin, the solution is forced to depend
on one function only.  (The spatial origin is $u=v$; regularity there
implies $f(u)=-g( v)$.  One can consider $g(v)$ as the initial datum
for the field.\fnote{In the classical context, one is free to add a
constant to $f$ and subtract the same constant from $g$, so there is
an ambiguity in this choice of initial datum, which does not affect
the field.  There is a similar freedom in the quantum case, which will
not be important here.})  Thus we may write
$$\phi ^0_{0,0}(u,v)=\phi ^0_{0,0}\Bigr| _{\scri ^-}(v)-\phi
^0_{0,0}\Bigr| _{\scri -}(u)\, .\eek$$
Comparing this with (\redfun ), we see that we can think of the initial
datum for the field as a quantum field on $\scri ^-$ with two--point
function
$$\langle 0_{\rm M}|\phi ^0_{0,0}\Bigr| _{\scri ^-}(v_1)
   \phi ^0_{0,0}\Bigr| _{\scri ^-}(v_2)|0_{\rm M}\rangle
  =-(16\pi ^2)^{-1}\log (v_1-v_2-\i\epsilon )\, .\eek$$
This logarithmic behavior is typical of quantum fields in two dimensions.  The
four factors in the logarithm in (\redfun ) arise from the different possible
coincidences as the ingoing and outgoing radial null surfaces from $(u_1,v_1)$,
$(u_2,v_2)$ are extended back to $\scri ^-$, either directly or after passing
through the origin.

\subsubsection{The two--point function in the Hawking model}

We may apply the foregoing results directly to the Hawking model,
since the assumption there is that the in--vacuum has the Minkowskian
form in the asymptotic past.  If we use the zeroth--order
approximation (only $l=0$ modes contribute to the Hawking process, and
those propagate by  geometric optics), then we have
$$\fl\eqalign{\langle 0_{\rm in}|&\phi ^0_{0,0}(u_1,v_1)\phi
^0_{0,0}(u_2,v_2) |0_{\rm in}\rangle\cr
  &=\langle 0_{\rm in}|\left(\phi ^0_{0,0}\Bigr| _{\scri ^-}(v_1)
   -\phi ^0_{0,0}\Bigr| _{\scri ^-}(v(u_1))\right)
\left(\phi ^0_{0,0}\Bigr| _{\scri ^-}(v_2)
   -\phi ^0_{0,0}\Bigr| _{\scri ^-}(v(u_2))\right)|0_{\rm in}\rangle\cr
  &=-(16\pi ^2)^{-1}\log{{(v(u_1)-v(u_2)-\i\epsilon
  )(v_1-v_2-\i\epsilon )}\over{(v(u_1)-v_2-\i\epsilon
  )(v_1-v(u_2)-\i\epsilon )}}\, .\cr}\eek$$
Had we used the exact treatment, the only difference would be a mild smearing
of the $v(u)$ terms, over ranges of size $\Delta u\lesssim R_{\rm Sch}/c$
(and the inclusion of similar formulas, but much smaller ones, for higher $l$
values).

We shall consider only local field measurements, so $(u_1,v_1)$, $(u_2,v_2)$ are
nearby.  Then the two--point function is large.  
The
important factors in the logarithm are those nearing zero,
$(v(u_1)-v(u_2)-\i\epsilon )$ and $v_1-v_2-\i\epsilon$.  

Of these, the second
has the form it does in Minkowski space, and so represents the same vacuum
fluctuations one would have there.  The interesting effects are those which
differ from the Minkowskian ones, that is, the renormalized two--point function
$$\eqalign{\langle 0_{\rm in}|&\phi ^0_{0,0}(u_1,v_1)\phi
^0_{0,0}(u_2,v_2) |0_{\rm in}\rangle -\langle 0_{\rm M}|\phi ^0_{0,0}(u_1,v_1)
  \phi ^0_{0,0}(u_2,v_2)|0_{\rm M}\rangle\cr
  &=-(16\pi ^2)^{-1}\log {{ v(u_1)-v(u_2)-\i\epsilon }\over{u_1-u_2-\i\epsilon}}
 +\hbox{less significant terms}\, .\cr}\eek$$\xdef\renfun{\the\EEK}%
For $|u_1-u_2|\ll R_{\rm Sch}$, the function $v(u)\simeq v_0-C\exp -\kappa u$
may be approximated linearly, and we have
$$v(u_1)-v(u_2)\simeq v'(u)(u_1-u_2)\, .\eek$$
Thus the divergent parts of the Hawking and Minkowski two--point functions
cancel for $|u_1-u_2|\ll R_{\rm Sch}$.  This is the Hadamard condition discussed
above, that on sufficiently fine scales the two--point function should approach
the Minkowskian one.

If we consider points which are a bit more separated, say $|u_1-u_2|\sim R_{\rm
Sch}$, the nonlinearities of $v(u)$ become significant over the range of $u$ in
question, and thus the Hawking analysis predicts significant excitations at
these scales.  This means that the production of quanta becomes significant for
wavelengths $\sim R_{\rm Sch}$; this is Hawking's prediction.  The fact that
the spectrum is thermal with temperature $T_{\rm H}=\hbar\kappa /(2\pi ck)$ is
essentially due to the fact that expressions like (\renfun ) are periodic in
imaginary time with period $2\pi c/\kappa$ (KMS condition).\fnote{The
Kubo--Martin--Schwinger (KMS) condition characterizes thermal states.  If a
system has a Hamiltonian $H$ and is in thermal equilibrium with inverse
temperature $\beta$, then it will be described by a density matrix $\rho =\exp
-\beta H$.  For any two observables $A(t)$, $B(t)$ (where the time--dependence
is given by the Hamiltonian evolution), we should have formally $\langle
A(t_1)B(t_2)\rangle ={\rm tr}\, (\exp -\beta H)A(t_1)B(t_2) ={\rm tr}\,
B(t_2)(\exp -\beta H)A(t_1) ={\rm tr}\, B(t_2)A(t_1+\i\beta )(\exp -\beta H)
={\rm tr}\, (\exp -\beta H)B(t_2)A(t_1+\i\beta )=\langle B(t_2)A(t_1+\i\beta
)\rangle$. The equality of the left--most and right--most terms is the KMS
condition.  It includes periodicity in imaginary time as as the special case
$B=1$.  So the argument given above for the Hawking process is  not sufficient
for establishing the KMS condition, but it is necessary.  A full, careful,
treatment reveals no surprises.} Because the zeroth--order approximation is
reasonably good, the frequency--dependence of the transmission of the relevant
modes through the space--time is weak, and the spectrum is close to a
Planckian, black--body, one.  Notice that the argument for thermality uses
essentially the exponential form of $v(u)$ at late retarded times.

We can see here that the Hawking quanta, which correspond to disturbances near
$\scri ^+$ extended over $\Delta u\sim R_{\rm Sch}$, do indeed arise
from vacuum
fluctuations in the past over scales 
$\sim R_{\rm Sch}v'(u)\sim R_{\rm Sch}C\exp
-\kappa u$.  
This exponential compression means that (after a finite passage of
retarded time), one is probing deeply enough the ultraviolet
asymptotics of the in--state's two point function, that any such state
which is Hadamard will produce the same result.  In other words, it is
not necessary to assume that the in--state is the vacuum, but only
that it is Hadamard.

On the other hand, taking this unbounded exponential compression of
the scales at face value may fairly be called fantastic.  The energies
of the in--modes in question very quickly surpass, not just the Planck
energy, but the entire estimated energy of the Universe.

To emphasize the essential way these high--frequency modes enter, suppose we had
initially imposed an ultraviolet cut--off $\Lambda$ on the in--modes.  Then we
should have found no Hawking quanta at late times, for the out--modes' maximum
frequency would be $\sim v'(u)\Lambda$, which goes to zero rapidly.  
(It is worth pointing out that this procedure is within what may be fairly
described as text--book quantum field theory:  start with a cut--off, do the
calculation, and at the very end take the cut--off to infinity.  That this
results in no Hawking quanta emphasizes the delicacy of the issues.  In this
sense, the trans--Planckian problem may be thought of as a
renormalization--ambiguity problem.)

\subsubsection{Particles}

As mentioned earlier, our real concerns are with Hawking's use of \it field
modes \rm of exponentially increasing frequencies, not whether we express the
physics of those modes in terms of particles or fields, two alternate quantum
representations.  However, for completeness we include a brief outline of the
analysis in particle terms.

Particles are defined in terms of the creation and annihilation operators of the
field, and those in turn are determined by splitting the field into negative--
and positive--frequency parts.  Suppose a physicist carries a device (like a
photomultiplier) whose output he interprets as particle counts.  That device
functions according to its own internal physics (responding to whatever fields
are around); in particular, it presumably uses its own local notion of proper
time to distinguish positive from negative frequencies.  Thus a device in the
``in'' region in the gravitational collapse problem will essentially be
distinguishing positive from negative frequencies on the basis of Fourier
transforms with respect to $v$ (near $\scri ^-$), whereas one in the ``out''
region will make the distinction based on Fourier transforms with respect to $u$
(near $\scri ^+$).  

Precisely because the space--time is not stationary, a field
mode which is $v$--positive frequency will propagate through the space--time
into a mixture of $u$--positive and $u$--negative frequencies.  The in--vacuum
is characterized as the state annihilated by all $v$--positive--frequency field
operators.  However, owing to the mixing, this will not be annihilated by all
$u$--positive--frequency operators.  It will not be the out--vacuum.

This is most commonly expressed in terms of \it Bogoliubov transformations.  \rm
We write schematically
$$\eqalign{\phi _{\rm out}^+&=\alpha\phi _{\rm in}^++\beta\phi _{\rm in}^-\cr
    \phi _{\rm out}^-&=\alpha ^*\phi _{\rm in}^-+\beta ^*\phi _{\rm in}^+\,
    ,\cr}\eek$$
where the mode indices have been suppressed.  (So really $\alpha$, $\beta$ are
infinite continuous or discrete matrices.)  The $\alpha$'s and the $\beta$'s are
the \it Bogoliubov coefficients.  \rm  By using the commutation relations, one
can show that the expected number of particles created is simply
$${\rm tr}\, \beta ^*\beta\, .\eek$$
Since the analysis is mode--by--mode, the number of quanta in a given out--mode
is simply the sum of $\beta ^*\beta$ for this out mode (the matrix
multiplication effecting the sum over all the in--modes).

In fact this was how Hawking did his analysis.  One can find the $\alpha$'s and
$\beta$'s by taking the positive-- and negative--frequency parts of the
equations for the propagation of the field from $\scri ^-$ to $\scri ^+$,
equations (\firsthalf ), (\secondhalf ).  Then one can explicitly compute the
expected spectrum as a function of frequency.

\subsection{Stress--Energy}

One would like to understand the energy--momentum budget of the Hawking
process.  This has not been fully achieved, on account of fundamental
difficulties in constructing (Wald 1994)
and interpreting (Helfer 1996, 1998) the stress--energy operator for
quantum fields in curved space--time. 

In order to get a finite stress--energy operator, one needs to renormalize, and
there are ambiguities in how to do this.  In Minkowski space, these are
resolved by appealing to Poincar\'e invariance, but in curved space--time, in
general circumstances, it is not known how to fix the ``finite part'' of the
stress--energy operator (Wald 1994).  

It should also be remarked that,
while, up to the problem of getting the finite part right, the
stress--energy operator is well--defined, the \it sense \rm in which it is
well--defined is that it becomes an operator--valued distribution.
That is, in general it must be averaged against a tensor
test--function over a space--time volume in order to really be a
self--adjoint operator.  If one tries to average against a tensor test
function over, for example, a Cauchy surface, one does not in generic
circumstances get a self--adjoint operator, but an apparently
pathological object (Helfer 1996, 1998).  These issues are not wholly
understood and will not be discussed further here, however, they
certainly raise questions about whether we have an adequate
understanding of the energetics of the Hawking process.

In an asymptotically flat regime (for example, near $\scri ^+$), one can use
the Minkowskian structure to identify the correct
renormalization.  This enables us to conclude that the stress--energy
of the state predicted by Hawking, measured by distant observers, should indeed
be that of a quasi--thermal state.  
The calculations involved even in this case are not trivial, for the
renormalization must be done carefully.  However, using the fact that
our ``zero--order'' approximation is mathematically identical to a
moving--mirror model, we can make use of general formulas in the
moving--mirror case (Fulling and Davies 1976, Davies and Fulling 1977) 
to conclude that in this
approximation
$$\eqalign{\langle T_{ab}\rangle &=(12\pi r^2)^{-1}
  \hbar\left( {3\over 4}\left(
{{\ddot v}\over{\dot v}}\right) -{1\over 2}{{v^{(3)}}\over{\dot v}}\right) 
  l_al_b
  \quad\hbox{as}\quad r\to \infty  \cr
                 &=(48\pi r^2)^{-1}\hbar\kappa ^2 l_al_b
	\quad\hbox{as}\quad r\to \infty  	 \cr}\eek$$
where $l^a$ is the outgoing null vector.
This is indeed the stress--energy of a radiative flux of energy; it is
in fact the correct asymptotic form up to a numerical factor 
(attributable to the neglect
of higher--angular momentum modes and transmission coefficients).
But this is only what one would expect. 
The interesting question is what the stress--energy is at finite values of $r$,
and particularly near the horizon.

While, as explained above, it is not known precisely how to compute the
stress--energy at finite values of $r$, there are two important statements
about it one can make.  The first is that the \it only \rm difficulties
associated with renormalizing it are in fixing the finite part and the
interpretational issues raised earlier.  Thus (unless somehow the correct
choice of c--number becomes divergent at the horizon --- something which could
not happen if the choice is fixed by local physics), the stress--energy must be
as regular at the horizon as it is anywhere else.  The second is that given any
reasonable, locally conserved, candidate definition for the effective
stress--energy which is stationary at late times, the flux of energy outwards
across $\scri ^+$ must equal minus the flux across the event horizon.  This
means that the energy flux \it into \rm the black hole is negative.

To see this, let us suppose that there exists a well--defined
classical effective stress--energy $T_{ab}^{\rm eff}$.  We shall not be
concerned with precisely how this is computed; the argument will be
independent of that.  We shall however require the tensor to be
\it conserved, regular \rm (of class $C^1$), and
\it stationary \rm in the quasistationary regime under
consideration, \it including the future horizon. \rm
Let $\Sigma$ be a hypersurface in this
regime from a cut $S_{{\cal H}^+}$ of the
event horizon to a cut $S_{\scri ^+}$.  We may compute the effective
energy on $\Sigma$,
$$\int _\Sigma T_{ab}^{\rm eff}\, \xi ^a\, \d\Sigma ^b\, ,\eek$$
where $\xi ^a=\partial /\partial t$ is the Schwarzschild Killing
vector.
Now imagine flowing $\Sigma$ forward along $\xi ^a$, to some $\Sigma
_t$.  The energies on $\Sigma$ and on $\Sigma _t$ will be identical,
because $T_{ab}^{\rm eff}$ and the Schwarzschild geometry are
stationary.  However, since $T_{ab}^{\rm eff}$ is conserved, this
means that the energy flux across the portion of $\scri ^+$ from
$S_{\scri ^+}$ to its image $S_{\scri ^+}^t$ under the flow must
be compensated by a negative flux across the horizon from
$S_{{\cal H}^+}$ to its image $S_{{\cal H}^+}^t$.

This is
just what one would expect on the basis of the Area Theorem:  if Hawking
radiation carries energy away from the hole, the hole will lose mass, hence
area, and this must mean a negative energy flux is crossing the horizon.

The picture that one has, then, is that the in--state (vacuum) is
being split by 
its passage through the collapsing geometry into positive-- and
negative--energy 
fluxes.  The negative--energy fluxes fall into the black hole, while the
positive--energy ones escape to infinity.  

\subsection{The analysis of Fredenhagen and Haag}

Hawking's original analysis was in terms of Bogoliubov coefficients,
with the modes taken to be Fourier ones.  An alternative computation
based more directly on the two--point functions was given by
Fredenhagen and Haag (1990).  Their results reproduced Hawking's, but
clarified certain issues.

First, because the singularities of the two--point functions are local, 
Fredenhagen and Haag
were able to bring out very clearly the dependence of the computation on the
Hadamard form of the initial state and (in the notation of section \sigsec )
its propagation between the surfaces $\Sigma$ and $\Sigma ^+$.\fnote{Actually,
Fredenhagen and Haag did not use precisely our surface $\Sigma$, but a similar
one.} Fredenhagen and Haag emphasized that, given the Hadamard form of the
two--point function on $\Sigma$, one would reproduce Hawking's results. This
seems to have been misinterpreted by some workers as meaning that the analysis
resolved the trans--Planckian problem.  However, this is not the case.  The
Fredenhagen--Haag analysis confirms Hawking's, and, as we shall see in detail
in the next section, that propagation involves trans--Planckian problems.

Second, their analysis gave better control of infrared issues than did
Hawking's.  Fredenhagen and Haag were able to show that the Hawking
process is asymptotically stationary.  This is plausible but not quite
clear in Hawking's analysis.  This is because mild infrared
divergences (which are hard to rule out in Hawking's approach) could
give rise to secular growth in the $n$--point functions.  That this
does not occur requires rather careful estimates, which are outlined
in Fredenhagen and Haag's paper.

\subsection{Almost--black holes}

What happens if a gravitating object collapses to the verge of forming a black
hole, but never quite does so?  Suppose, say, a star collapses to
$R=(1+\epsilon )R_{\rm Sch}$ but then becomes stationary? It is useful to
examine this, both to deepen our understanding of Hawking's model and  to be
able to compare its predictions with those of others.

In the Hawking model, while some transient quanta may be formed, after a while
(perhaps a long while), the modes detected by distant observers arise  almost
entirely by propagation through a stationary region, and so there are no
Hawking quanta.  

The interval of transience is determined as follows.  Consider the usual
argument about tracing modes backwards in time through the collapsing object. 
The mode encounters the limb of the object first at a radius which for late
times will be $(1+\epsilon )R_{\rm Sch}$.  It propagates backwards in time
through the object, emerging at a radius say $r$.  The question is whether $r$
is equal to $(1+\epsilon )R_{\rm Sch}$ or greater than it.  At sufficiently
late times, this part of the limb has contracted to $(1+\epsilon )R_{\rm Sch}$
and (by hypothesis) goes no further.  This means the propagation of the mode
inwards to the object and outwards from the object are symmetric, there is no
red--shift and there are no Hawking quanta.  On the other hand, in the
transient regime, the radius $r$ is still somewhat larger than $(1+\epsilon
)R_{\rm Sch}$, and thus the blue--shift of the mode in from $\scri ^+$ to
$(1+\epsilon )R_{\rm Sch}$ is larger than the red--shift out from $r$ to $\scri
^-$, and one has the possibility of production of quanta.

In other words, the period of transience ends when a distant observer, looking
\it through \rm the collapsing object, sees its trailing edge reach
$(1+\epsilon )R_{\rm Sch}$. After this, no quanta are expected.

\section{The trans--Planckian problem}

We have seen that the quanta which are supposed to be produced
by the Hawking process at late retarded times $u$, that is, the
physically dominant modes of the quantum field, have their origins
near $\scri ^-$ in vacuum fluctuations of frequencies $\sim (c/R_{\rm
Sch})(v'(u))^{-1}$, where
$$v(u)\simeq v_0-C\exp -\kappa u\, .\eek$$\xdef\ekrel{\the\EEK}%
We also saw that the exponential form of this mapping of surfaces of
constant phase was integral to the thermal character of the final state.

Exponential relations like (\ekrel ) are never accepted
uncritically in physics.  They are never supposed to hold for
arbitrarily long times.  There always comes a point, after some number
of $\e$--foldings, when one has passed the scales at which the
mathematical models used are valid, and new physical effects must be
considered.  

In our case, the frequencies of the original modes quickly pass (not
only all conventional quantum field--theoretic scales, but) the Planck scale.
This is called the \it trans--Planckian problem.  \rm
It is clear that any analysis that relies on assumptions about physics
at the Planck scale is speculative, and that an analysis which relies
on the application of conventional physics beyond the Planck scale is
questionable.  

The foregoing comments are negative ones, and one would like to
approach the trans--Planckian issue constructively.  In order to do
this, we must try to be as precise as possible about how the
trans--Planckian modes are bound up with the physics of the collapse.

The main aim of this section is to show that the trans--Planckian
problem can be localized to two regimes.
Roughly speaking, these are a neighborhood of the event
horizon, and a neighborhood of the surface of advanced time $v=v_0$ at
which the event horizon forms.  This means that if somehow one could
find alternative, cis--Planckian, 
physics to that of the Hawking model which operated
in those regimes, but reproduced Hawking's results elsewhere, one
would have overcome the trans--Planckian problem.  This will be
discussed in section 8.

Strictly speaking, we cannot expect a localization of the
trans--Planckian problem in space--time, but must pass to the frame
bundle.  This is because frequency is a frame--dependent quantity, and
so the trans--Planckian regime really consists of all those frames, at
different events, for which the characteristic field modes giving rise
to Hawking quanta have trans--Planckian frequencies.  This will be
analyzed carefully below, and it will be shown that for certain
important cases the trans--Planckian regime extends substantially away
from the event horizon.  These have implications for models where
black holes are formed by sending in massless particles or dust.

The most important frames, physically, are those defined by the matter
whose collapse drives the formation of the black hole.  (At any given
event in the matter, there may be a family of natural frames.  These
will all differ from each other by bounded boosts.)  Field modes which
give rise to Hawking quanta and become trans--Planckian in such frames
are the problematic ones.  Knowledge of precisely what happens to
these modes requires some knowledge of Planck--scale physics, which
presumably means quantum gravity.  In some sense, one would expect the
modes to become entangled with quantum--gravitational fluctuations in
the collapsing matter.  The Hawking model, which relies on
``painting'' the field modes on a fixed classical space--time, would
not be valid.

As mentioned above, there are two trans--Planckian regimes:  one near
the event horizon; and the other near the surface of constant advanced
time $v=v_0$ at which the event horizon forms.  Only the former will
require detailed analysis here.  The latter is equally important, but
for our purposes it will be enough to note that the frequencies of the
field modes diverge there with respect to the asymptotic rest frame.

\subsection{Localization of the problem}

Here we study how the trans--Planckian problem may be localized.
We will need
to take into account the fact that
frequency is not a scalar,
but an observer--dependent quantity, so that different observers will
have different notions of when a given wave--vector passes the Planck
scale.  Even a visible photon would, according to special relativity,
appear to have Planck energy to sufficiently boosted observers.  This
means that a full treatment really localizes the trans--Planckian
problem, not in space--time, but in the bundle of frames.  The
localization consists of those frames of observers for whom the
field modes giving rise to Hawking quanta are trans--Planckian.
For every event in space--time, there are \it some \rm
trans--Planckian frames, corresponding to sufficiently boosted
observers.

However, not all mathematically constructible frames are relevant to
the physics of the situation.  In fact, we distinguish two main
classes of
frames of interest:

\itemitem{} (CH) Those of observers whose world--lines 
cross the horizon;

\itemitem{} (SO) Those of stationary observers (that is, observers
moving along the timelike Killing field $\partial /\partial t$).

\noindent We will develop a general framework for analyzing the
problem, and then specialize it to these cases.

We use the standard Schwarzschild coordinates, and put
$$f^2=1-{{R_{\rm Sch}}\over r}\, .\eek$$
Then an orthonormal frame (in the $(t,r)$ space) is given by
$$T^a=f^{-1}\partial _t\, ,\qquad R^a=f\partial _r\, .\eek$$
It will also be convenient to introduce the associated null frame
$$L^a=2^{-1/2}(T^a+R^a)\, ,\qquad N^a=2^{-1/2}(T^a-R^a)\, .\eek$$
We have 
$$\eqalign{L^a\nabla _a v&=L^a\nabla _a (t+r_*)\cr
  &=2^{1/2} f^{-1}\, ,\cr}\eek$$
and similarly for $N^a\nabla _au$, so in $(u,v)$ coordinates we have
$$L^a=2^{1/2}f^{-1}\partial _v\, ,\qquad N^a=2^{1/2}f^{-1}\partial
_u\, .\eek$$
If an observer is boosted to a velocity $c\tanh\xi$ radially outward
from this frame, her frame will be
$$T^a_\xi =\cosh\xi T^a+\sinh\xi R^a\, ,\qquad R^a_\xi =\sinh\xi
T^a+\cosh\xi R^a\, ,\eek$$
or
$$L^a_\xi =e^\xi L^a\, ,\qquad N^a_\xi =e^{-\xi}N^a\,
.\eek$$\xdef\xiforms{\the\EEK}%

The wave covector for a field mode producing a Hawking quantum (of
characteristic wavenumber $R_{\rm Sch}^{-1}$) is 
$$K_a=R_{\rm Sch}^{-1}\d u\, ,\eek$$
and the corresponding vector is
$$K^a=2^{-1/2}f^{-1}R_{\rm Sch}^{-1}L^a\, .\eek$$
Thus the trans--Planckian problem occurs for
$$\eqalign{l_{\rm Pl}^{-1}&\lesssim g_{ab}T^a_\xi K^b\cr
  &=\e ^{-\xi} f^{-1}R_{\rm Sch}^{-1}\, ,\cr}\eek$$
or equivalently
$$ \e ^\xi f\lesssim l_{\rm Pl}/R_{\rm Sch}\, .\eek$$\xdef\tpcond{\the\EEK}%

The formula just given is the frame--bundle form of the localization of the
trans--Planckian problem.  Interestingly, it does not depend
explicitly on the retarded time $u$, although for a given family of
observers both $r$ (and hence $f=f(r)$) and $\xi$ may depend on $u$.  
Unless $r$ is close to $R_{\rm Sch}$, the function $f$ is of
order unity, and thus the trans--Planckian modes arise only for
$$\e ^{-\xi} \gtrsim R_{\rm Sch}/l_{\rm Pl}\, .\eek$$
Thus, for observers not close to $r=R_{\rm Sch}$, the
trans--Planckian issue arises only for high inward boosts.  
For observers close to $r=R_{\rm Sch}$, it is important to discuss the
physical definition of the frame in question
before interpreting the formulas.  Thus we examine our two families
of observers.

\subsubsection{The family (CH)}

The family (CH) of observers crossing the horizon is the most
important one, because particular cases of it correspond to frames of
the matter whose collapse drives the formation of the black hole.
We shall find, rather remarkably, that the onset of the
trans--Planckian problem occurs for such observers at a local time
$\sim t_{\rm Pl}$ before they cross the horizon, \it independent of
the velocity at which they cross.  \rm   The implications of this will
be discussed in section 4.2.

We will specify the families of observers in the vacuum region; the
behavior of observers tracking the limb of the collapsing object may
be obtained from this.  (We shall not consider the interior of
the collapsing matter.  What happens there must connect continuously
with the case we analyze, and that will be a strong enough result for
us.)

To parameterize the observers in a physically meaningful way, let us
start from the standard null frame $L^a$, $N^a$ at $\scri ^-$, and
parallel transport it along the inward direction $N^a$ to the event
horizon.  One finds that the inwardly--transported null frame is
$fL^a$, $f^{-1}N^a$.  If we consider a frame boosted outwards by
$c\tanh\zeta$ from this, its unit timelike vector is
$$2^{-1/2}\left( f\e ^\zeta L^a +f^{-1}\e ^{-\zeta}N^a\right)\, .\eek$$
The normalized null vectors in the frame are evidently $f\e
^{\zeta}L^a$, $f^{-1}\e ^{-\zeta }N^a$.  One can interpret
$c\tanh\zeta$ as a measure of the observer's velocity as it crosses
the horizon.  Larger positive values correspond to shallower
crossings; more negative values correspond to sharper ones.

To connect with the
notation of the previous subsection, we have $\e ^\xi =f\e ^\zeta$,
and the condition for the characteristic field modes to appear
trans--Planckian is (from (\tpcond ))
$$\e ^\xi f=\e ^\zeta f^2\lesssim l_{\rm Pl}/R_{\rm Sch}\,
.\eek$$\xdef\tpcondor{\the\EEK}%
Thus for any fixed $\zeta$, the trans--Planckian regime consists of
events with coordinate values of $r$ sufficiently close to $R_{\rm
Sch}$.  For shallower crossings of the horizon (larger $\zeta$), the
regime is smaller, whereas for sharper crossings it is larger.

To get a better understanding of this, let us ask how the function
$f^2$ appears to depend on our observer's local time.  We have
$$\eqalign{T^a_\xi \nabla _a f^2&=2^{-1/2}\left( T^a\cosh\xi
+R^a\sinh\xi \right)\nabla _a f^2\cr
  &=2^{-1/2}\sinh\xi f\partial _rf^2\cr
  &=2^{-3/2} (f\e ^\zeta -f^{-1}\e ^{-\zeta})fR_{\rm Sch}/r^2\cr
  &\simeq -2^{-3/2}\e ^{-\zeta}R_{\rm Sch}^{-1}\cr}\eek$$
near the horizon.  This means that if we choose the zero of the local
time $t_{\rm loc}$ to be when the observer crosses the horizon, we
have
$$f^2\simeq -2^{-3/2}\e ^{-\zeta}R_{\rm Sch}^{-1}t_{\rm loc}\, .\eek$$
Using this, we may recast the condition (\tpcondor ) 
for the characteristic modes
to become trans--Planckian as
$$\e ^\zeta f^2\simeq 2^{-3/2}R_{\rm Sch}^{-1} |t_{\rm loc}|\lesssim
l_{\rm Pl}R_{\rm Sch}^{-1}\, ,\eek$$
which is simply (dropping the $2^{-3/2}$)
$$|t_{\rm loc}|\lesssim t_{\rm Pl}\, .\eek$$
This result is, curiously, independent of $\zeta$.

In other words, for any observer whose world--line crosses the event
horizon, the onset of the trans--Planckian regime will be at a local
time $\sim t_{\rm Pl}$ before the crossing.  This result does not
depend on whether the observer is freely falling or not; it depends
only on the world--line being of class $C^1$.

The case of observers crossing the horizon ultrarelativistically
inwards is of special interest.  For these observers, the increment of
the world--line of length $t_{\rm Pl}$ before the event horizon
extends far from the horizon.  In particular, if we were to consider
an ultrarelativistic collapse of matter to form the black hole, then
the onset of the trans--Planckian problem could occur well before the
event horizon (in the frames of observers who are not
ultrarelativistically boosted).  This issue should be kept in mind
when evaluating models where black holes are formed by
ultrarelativistic (or null) collapse.

\subsubsection{The family (SO)}

We consider here the family of stationary observers.
For these, the
timelike unit tangent is simply $T^a$, and so this is the case $\xi
=0$.  We find that the trans--Planckian problem is manifest for
$$f\lesssim l_{\rm Pl}/R_{\rm Sch}\, .\eek$$\xdef\socond{\the\EEK}%
This regime is properly contained in that for the horizon--crossing
observers, since $f^2<f$.  Since the proper radial distance of an
event near the horizon to the horizon is $\simeq 2R_{\rm Sch}f$,
another way of expressing (\socond ) is to say that the proper radial
distance of the observer from the horizon should be $\lesssim l_{\rm
Pl}$.

It turns out that the trans--Planckian problem for the Hawking modes
for stationary observers manifests itself at the same scale as another
difficulty, the point where the accelerations
of the observers become so great they
cannot be considered meaningful classically.  This latter occurs when
the local acceleration experienced by the observer becomes of the
order $l_{\rm Pl}/t_{\rm Pl}^2$.  (One can easily check this by
computing the local acceleration; it turns out to be
$GMf^{-1}r^{-2}$.)

\subsection{Discussion}

We have seen that the trans--Planckian problem is essentially a local one,
except for the case of ultrarelativistic collapse.  This means that if one
could somehow get the modes through these problematic regions by alternative,
cis--Planckian, one would have a solution to the trans--Planckian problem.  

Discussions of this are sometimes phrased in terms of ``getting the right
vacuum'' outside of the event horizon.  This is picturesque but can be
misleading.  What is really meant is getting the state, and in particular, the
$n$--point functions, right.  However, we may accurately say that the problem
is  to understand how to propagate the in--vacuum to the neighborhood of the
event horizon.

\section{Connection with the Unruh process}

One might hope that, given the very beautiful and simple form of Hawking's
predictions, they could be recovered from alternative physical arguments.  In
particular, if the existence of Hawking radiation really did not depend on
ultra--high energy physics or quantum--gravitational hypotheses, one would hope
for alternative arguments within the realm of conventional physics.

Despite an enormous amount of work on the Hawking process, no such arguments
exist, and indeed very little of the work has confronted this problem directly.
The aim of most research has been not so much to address the foundational
difficulties of Hawking's analysis as to show that there are connections
between that analysis and  known physics.   In other words, most of the work
really simply \it assumes \rm Hawking's analysis is correct, and, on that
basis, looks for connections with other physics.  Such work may nevertheless
lead to circumstantial evidence for the Hawking effect, as will be discussed in
section 7.

In one sense, it is clear that it will be impossible to dodge the
trans--Planckian problem, or the issue of quantum--gravitational hypotheses,
and present a thoroughly conventional derivation of the Hawking process.  This
is because any conventional argument must reproduce the propagation of the
field assumed by Hawking, and this involves trans--Planckian physics.  In other
words, in any ``conventional'' picture, the propagation of the field will be by
the same equation as that used by Hawking, and this means that the  Hawking
quanta \it would \rm come from field modes propagating arbitrarily close to the
event horizon and subject to divergent red--shifts.  

There could in principle be, however, an alternative physical argument for the
Hawking process, one which reproduced the trans--Planckian difficulties as an
unfortunate by--product but in which one could somehow argue these were not
essential.  While no argument like this is presently known, the best approach
to one goes back to DeWitt (1979) and Unruh (1976), and has ben
outlined most explicitly by Jacobson (1996).

According to Unruh, a uniformly accelerated detector in Minkowski space will
respond as if it is in a thermal bath.  By the principle of equivalence, an
observer held fixed in a uniform gravitational field will find physics locally
the same as for a uniformly accelerating observer.  Thus (one would think) an
observer held fixed in a uniform gravitational field will perceive Planckian
radiation.  This is tantalizingly close to Hawking's prediction.  

\subsection{The Unruh process}

A quick sketch of the Unruh process is in order. While the result is generally
accepted, it has not been experimentally verified and has been contested by
some workers.  (See e.g. Fedotov et al 1999.)  I give a treatment in a form
which I believe is convincing, but it has been phrased carefully to avoid some
of the trickier issues.

Let us assume we have some sort of detector which responds to the quantum field
$\phi$.  We shall also assume that we can ignore the spatial extent of the
detector.\fnote{Since particles have finite spatial extent, this means that the
detector cannot be in the strict sense a \it particle \rm detector.  It will be
a field strength detector.}  (This is not an entirely trivial assumption, but
it is a reasonable first approximation.\fnote{When the spatial extent of an
accelerated detector must be considered, it is necessary to consider how the
acceleration varies over the detector. The Unruh analysis (and the
Bisognano--Wichmann theorem to be discussed in the next subsection) does not
apply to an extended detector all of whose elements suffer the same
acceleration, but to those where the acceleration varies in such a way that the
spatial separation between adjacent elements is preserved.})  Then if the
detector's world line is $\gamma (s)$ (with $s$ the proper time), it responds
to $\phi (\gamma (s))$.  In particular, by comparing the two--point function
$$\langle 0_{\rm M}|\phi (\gamma (s_1))\phi (\gamma (s_2))|0_{\rm
M}\rangle\eek$$\xdef\minkfunner{\the\EEK}%
with that of an inertial one
$$\langle 0_{\rm M}|\phi (t=s_1,x=0,y=0,z=0)\phi
(t=s_2,x=0,y=0,z=0)|0_{\rm M}\rangle\, ,\eek$$
we get a measure of whether the state would appear to an observer to
be excited beyond the expected vacuum fluctuations. 

A world--line with uniform acceleration $a$ in the $x$--direction is
$$t=a^{-1}\sinh as\, ,\quad x=a^{-1}\cosh as\, ,\quad y=0\, ,\quad
z=0\, ,\eek$$\xdef\world{\the\EEK}%
and, on substituting this into the formula (\minkfun ) for the Minkwoskian
two--point function, we find
$$\fl\langle 0_{\rm M}|\phi (\gamma (s_1))\phi (\gamma (s_2))|0_{\rm
M}\rangle\cr
   =-(4\pi ^2)^{-1}a^2 ((\sinh as_1-\sinh as_2-\i\epsilon )^2
  -(\cosh as_1-\cosh as_2)^2)^{-1}\, .\cr\eek$$
This is clearly periodic in imaginary time with period $2\pi /a$, and
one can check that the full KMS condition holds
for measurements of the field on the world--line.  Thus the state is a
thermal one with temperature $T_{\rm U}=\hbar a/(2\pi ck )$.
The Planckian nature of the spectrum can be established by direct calculation,
but it is no surprise because the only dimensionful quantity is the
acceleration.  Note that the characteristic time to detect a quantum is $\sim
c/a$, the acceleration time.

Of course, a world line representing a detector which accelerates uniformly for
all time is unrealistic.  It would be more natural to consider (say) a detector
which was inertial prior to some time, and then was smoothly brought into a
state of uniform acceleration, and then eventually smoothly returned to an
inertial trajectory.  This would not have led to any result differing from
Unruh's in the interval of constant acceleration, since what enters is the
two--point function (\minkfunner ).   Only if the detector were used to measure
quanta of such low frequencies that their wave packets extended back to the
time prior to the uniform acceleration would discrepancies with Unruh's
predictions arise.  So for practical purposes, unless very sensitive detections
of infrared effects are important, once the uniform acceleration has been
sustained over a few acceleration--times' worth of proper time, the spectrum of
expected excitations is Planckian.

\subsection{The argument from the equivalence principle}

The argument most nearly connecting the Unruh and Hawking effects
goes back to DeWitt (1979) and Unruh (1976) and has more recently been
explicitly outlined by Jacobson (1996).   Consider an observer
hovering outside
of a black hole at Schwarzschild coordinate $r$.  That observer will 
experience a local acceleration, which can be computed,
$$a=f^{-1}(2M/r)^2\kappa\, ,\eek$$
where $f=\left( 1-R_{\rm Sch}/r\right)^{1/2}$ as before.
Now on small enough scales near this observer, the two--point function should
approach its Minkowskian form, and therefore (applying the principle of
equivalence and the Unruh argument) 
the observer will perceive the vacuum to be at a
temperature $T(r)=\hbar a /(2\pi )$.  On account of the red--shift, an observer
at infinity would perceive this as $fT(r)$.  
If we assume that the agreement of
the two--point function with its Minkowskian form becomes better as 
$r\to R_{\rm
Sch}$, then the temperature of the black hole, as measured at infinity, should
be
$$\lim _{r\downarrow R_{\rm Sch}}fT(r)=\kappa /(2\pi )=T_{\rm H}\, .\eek$$

Jacobson 
outlined this argument with the goal that it would provide an
explanation for the Hawking effect without having to invoke physics at the
Planck scale (or beyond).  The idea was that the trans--Planckian physics would
only have to be invoked for observers within the Planck scale of the
Schwarzschild radius.  However, one would hopefully get a very good
approximation to the Hawking picture for observers somewhat farther away.  So if
the argument held for $r$ separated from $R_{\rm Sch}$ by a proper distance of
$1$ nm (say), one could recover essentially Hawking's prediction while only
needing to invoke energies $\sim \hbar c/(1$ nm$)\simeq 200$ eV.  The
trans--Planckian regime would still be necessary to describe physics very close
to the hole, but hopefully this would simply decouple from the Hawking
radiation.  Thus one could take the limiting form of the two--point function for
$r$ close (but not in Planck terms) to $R_{\rm Sch}$ as a sort of boundary
condition which would give rise to the Hawking effect.

It is clear that this argument turns on the sense in which the two--point
function approaches its Minkowskian form.  As we shall see quantitatively
shortly, the approach does not in fact have quite the character
hypothesized.  First, though, there is larger issue to raise.

The argument is apparently time--symmetric, whereas the Hawking process
is not.  The only place a time asymmetry might possibly come into the
argument would be if the limit which has been written simply as $r\downarrow
R_{\rm Sch}$ must actually be taken in some time--asymmetric way in order to
get the correct limiting two--point behavior. If this is not the case,
then the argument by appeal to the equivalence principle 
and Hawking's analysis must speak to different physics.   However,
even such time--asymmetry in the approach could not be a complete explanation
of the discrepancy. If we consider, for example, an object which collapses not
quite to a black--hole state but so that its limb is at $r=(1+\epsilon )R_{\rm
Sch}$ and it is stationary, then Hawking's analysis would predict that (after
some transients) there would be no production of quanta.  However, the 
argument by appeal to the equivalence principle
would predict a Planckian (``Hawking'') spectrum.

In order to understand what really is happening, let us consider the situation
quantitatively.  To make contact with the previous analysis of the Hawking
prediction, let us consider a spherically symmetric family of accelerated
observers with detectors are arranged to respond only to the $l=0$
modes.  Then the Minkowskian null coordinates of the world line of one of these
would be $u=-a^{-1}\exp -as$, $v=a^{-1}\exp as$, and the $l=0$ two--point
function for the detectors would be (cf. (\redfun ))
$$-(16\pi ^2)^{-1}\log {{\left( \e ^{as_1}-\e ^{as_2}-\i\epsilon\right)
           \left( -\e ^{-as_1}+\e ^{-as_2}-\i\epsilon\right)}\over
	   {\left( \e ^{as_1}+\e ^{-as_2}-\i\epsilon\right)
	   \left( -\e ^{-as_1}-\e ^{as_2}-\i\epsilon\right)}}\, ,\eek$$
whereas the detectors in the Hawking process would have\fnote{Note
that we expect the geometric--optics approximation to become exact in
this case, as we have $r\downarrow R_{\rm Sch}$.}
$$-(16\pi ^2)^{-1}\log{{ (v_1-v_2-\i\epsilon )(-\e ^{-\kappa u_1}+\e
^{-\kappa u_2}-\i\epsilon )}\over{(v_1+\e ^{-\kappa u_2}-\i\epsilon
)(-\e ^{-\kappa u_1}-v_2-\i\epsilon )}}\, .\eek$$

In each of these, the important terms are the numerators, since we are
interested in what happens when the two events are close and the times are
late.  If the argument's hypothesis about the two--point asymptotics
were correct, we should obtain the second from
the first by substituting $\kappa f^{-1}$ for $a$ and $fu$ for $s$.   We see
that this correctly reproduces the second factor, but not the first.   The two
forms become asymptotically close (up to constant prefactors) only for $\Delta
v\ll a^{-1}$, which is to say $f^{-1}\Delta s\ll a^{-1}$.  However, this is
just the condition that we are passing the frequencies at which the Unruh
quanta occur.  This condition is not any more accurately fulfilled as $r\to
R_{\rm Sch}$.  Thus the limiting behavior necessary for the argument,
that the two--point function should approach the Minkowskian form sufficiently
accurately to locally represent the Unruh process as $r\downarrow R_{\rm Sch}$,
does not hold.

Let us now step back and consider the problem generally. In a curved
space--time, in a  small enough neighborhood of any event, one expects the
two--point function to have the Hadamard form, and this means that in small
enough neighborhoods the Unruh effect should apply to accelerating observers
(those whose acceleration times are small compared to the size of the
neighborhood). But what sets the scale at which the Hadamard asymptotics take
over?  It is the choice of quantum state, for different quantum states have
different two--point functions.  Thus an argument like DeWitt's really
depends on the choice of quantum state (and the state used in the Hawking
process does not quite have the required properties).

While the behavior hypothesized in the argument 
does not hold, the actual behavior
is rather close, agreeing in the terms coming from the null coordinate $u$ but
disagreeing in those coming from $v$.  This is closely related to the
possibility of constructing what is called the Hartle--Hawking vacuum.

For us, the lesson to be drawn is that this approach has not succeeded in
weakening the dependence of the argument for Hawking radiation on the details
of the propagation of the quantum field and the two--point function.  We need
to know the scale at which the Hadamard asymptotics take over, and this can
only be determined by analyzing the propagation of the two--point function.
	   
\subsection{The Bisognano--Wichmann Theorem}

Remarkably, at about the same time Unruh was studying the effects of
uniform acceleration on detectors responding to free massless fields,
a far more general theorem was being proved in the context of axiomatic
field theory.  The authors of this theorem, Bisognano and Wichmann (1976),
did not mention its interpretation in terms of accelerated observers;
they were concerned with it as establishing a duality between certain
algebras of observables.  The connection with the work of Unruh was
pointed out much later, by Sewell (1982).  

The Bisognano--Wichmann theorem essentially says that the Unruh effect remains
valid for arbitrary, \it interacting, \rm Poincar\'e--invariant field theories,
in the sense that such a theory with Hamiltonian $H$ will be perceived by a
uniformly accelerating observer as if it were in a thermal state with density
matrix $\exp -H/(kT)$ where $T=\hbar a/(2\pi ck)$.

Since the Unruh effect does not explain the Hawking effect, we cannot expect
the Bisognano--Wichmann theorem to explain the Hawking effect, either. 
Nevertheless, the Bisognano--Wichmann theorem is of great interest.  It shows
that the Unruh effect is profound and not just a peculiarity of an artificially
simple field equation; in this way, it also gives hope that if the difficulties
with the Hawking effect could be circumvented for linear fields, they might
also be circumvented for realistic, interacting, field theories.  (See Sewell
1982.)

It is important to point out that there is one, potentially very serious,
obstacle to applying arguments like those of Bisognano and Wichmann in curved
space--time.  This is that the arguments rely essentially on the existence of a
self--adjoint semibounded Hamiltonian operator.  In a non--stationary
space--time, such Hamiltonian operators do not exist (Helfer 1996).  Since
non--stationarity  plays a key role in the Hawking process, such concerns must
be kept in mind.

\section{Lessons from moving--mirror models}

By a moving mirror model, one generally means a linear massless quantum field
responding to a perfect reflector in two--dimensional Minkowski space.  Such
models are closely related to the Hawking process: we noted earlier in the
geometric--optics, s--wave approximation, the evolution of quantum fields in
the Hawking model is \it precisely \rm given by a moving--mirror model (where
the mirror's trajectory is given by $v=v(u)$, with $u$ and $v$ interpreted as
the Minkowski null coordinates $t-x$ and $t+x$). Relations like this one have
been used over the years to clarify various aspects of the Hawking process. 

While it is sometimes suggested that moving mirror models can explain
the Hawking process, this is not really the case.  
There is  a mathematical identity of the scattering of the quantum fields
within the two models.  The task facing us is to decide in what regimes the
models themselves can be justified, and how the physics driving the scattering
in one model might be identified with that in another.

What we shall find is that the moving--mirror models suggest that a classical
treatment of space--time is inadequate for understanding at least one important
aspect of the propagation of quantum fields through gravitationally--collapsing
regions, the energy budget of the system.  In the moving--mirror models, the
energies associated with the quantum fields are typically smaller than the
errors made in neglecting the quantum character of the mirror and its driving
engine.  If similar results hold in the gravitational case, one must question
whether the neglect of quantum--gravitational corrections is justified. 

\subsection{The trajectories}

Let us begin by seeing what
it is which gives rise to the radiation in the moving--mirror case, that is,
what the trajectory of the mirror is.  (It
is \it not \rm one of uniform acceleration.)  To
understand its structure, let us first work out the proper time along it:
$$\eqalign{s&=\int\sqrt{\d u\d v}\cr
  &=\int\sqrt{v'(u)}\d u\cr
  &\simeq C\exp -\kappa u/2\cr}\eek$$
In what follows, the value of $C$ is irrelevant and we shall take $C=\kappa
^{-1}$ to give the simplest form with the correct dimension.
Then $s\simeq -\kappa ^{-1}(v'(u))^{1/2}$, and $s\uparrow 0$ as $u\to +\infty$.
This means that the mirror moves off to infinity in finite proper time.

The unit timelike tangent to the trajectory is
$$\eqalign{T^a&=(v'(u))^{-1/2}\left(\partial _u+v'(u)\partial _v\right)\cr
  &\simeq -(s\kappa )^{-1}\partial _u -(s\kappa )\partial _v\cr}\eek$$
The unit normal towards the right is
$$R^a\simeq -(s\kappa )^{-1}\partial _u+(s\kappa )\partial _v\, .\eek$$
and the acceleration is
$$\eqalign{A^a&={{\d T^a}\over{ds}}\cr
  &\simeq s^{-2}\kappa ^{-1}\partial _u -\kappa\partial _v\cr
  &=-s^{-1}R^a\, .\cr}\eek$$
Thus as $u\to +\infty$ and $s\uparrow 0$, the mirror moves to the
left, not with 
constant acceleration, but with acceleration increasing in magnitude
unboundedly.  In fact, since $T^a$ grows unboundedly as $s\uparrow 0$, an
infinite amount of energy--momentum must be supplied to the mirror to accelerate
it on this trajectory.\fnote{While $T^a$ remains a unit vector, on
account of the Lorentzian signature it can still increase unboundedly,
moving out to infinity along the unit hyperbola.}

At this point, it would not be unreasonable to dismiss the moving--mirror model
with this trajectory as hopelessly artificial.  Certainly it contrasts grossly
with a gravitationally collapsing object, where no external energy at all,
certainly not an unbounded one, need be supplied.  

There are other concerns about the moving--mirror models as well.   Any real
mirror is not perfectly reflective, but (roughly speaking) reflects modes only
below some plasma frequency $\omega _{\rm p}$.  When the acceleration grows so
large that $a\gtrsim c\omega _{\rm p}$, the velocity is changing appreciably
over a time smaller than the response time of the plasmons and so one does not
have in any simple sense a mirror.  And when $a/c\gtrsim t_{\rm Pl}^{-1}$, of
course, the neglect of quantum gravity is not credible even in these models.

\subsection{The energy budget}
\xdef\mirsec{\the\secno{}.\the\subno{}}%

While we have seen that there are specific concerns about the physical
reasonableness of the mirror trajectory used to recover a Hawking--type
prediction of Planckian radiation, there is an important lesson to be learned
from moving--mirror models as a class, including those with physically
reasonable trajectories. This lesson concerns the adequacy of modeling the
mirror's trajectory classically for understanding the system's energetics.

If we try to understand the energy budget of a moving--mirror problem, that is,
how the mirror can produce quanta by accelerating vacuum fluctuations, we find
that it is not adequate to model the mirror classically.  The energies of the
quanta produced are so low that the uncertainty in the mirror's state due
simply to its finite Compton wavelength must be taken into account.  In fact,
one finds that when the mirror accelerates, its state (and that of the engine
driving it) become entangled in a significant way with the vacuum fluctuations
in question, and so the mirror (and its driving engine) are in a superposition
of energy states whose width is of the order of the energy of the quanta
created from the vacuum fluctuations (Parentani 1996, Helfer 2001a).  

The main idea of the argument can be given as follows.  Consider a
(for simplicity, non--relativistic) mirror of mass $m$, moving in a
potential $V(x)$.  Then the Hamiltonian for the system (of mirror and
field) has the form
$$H_{\rm mirror}+H_{\rm field,\ left}+H_{\rm field,\ right}\, .\eek$$
Here 
$$H_{\rm mirror}={{p^2}\over{2m}}+V(x)\, ,\eek$$
and $H_{\rm field,\ left}$, $H_{\rm field,\ right}$ are the
Hamiltonians of the field on either side of the mirror.  Each of the
field Hamiltonians has a zero--point term of magnitude
$$(12\pi )^{-1}(\hbar /mc) V'(x)\, ,\eek$$
and represents the excitation of the field by the mirror even for a
vacuum in--state (the case of interest here).

Note that the zero--point term could be interpreted as arising from a
displacement of the coordinate in the potential term $V(x)$ by an amount
$\Delta x =(12\pi )^{-1}(\hbar /mc)$.  Now $\hbar /mc$ is the mirror's Compton
wavelength, that is, the scale at which the mirror cannot be considered as
(even) a (quantum) point particle, but must be given a relativistic quantum
field--theoretic treatment.  (This smearing--out of the point on which a
potential acts is a general feature of relativistic quantum field theories,
usually discussed under the heading ``Zitterbewegung.'')

A few comments are in order.  First, while the argument has been phrased simply
in terms of a ``mirror,'' it really refers to the reflective agents, presumably
plasmons in a more realistic treatment. Second, it is rather a surprise that
one needs to pass to a relativistic quantum treatment of the mirror.  That one
must do so is a measure of the delicacy of the energetics of the system.  

In the parallel with the Hawking process, the mirror corresponds to the
gravitationally collapsing space--time.  Thus if the analogy between the two
systems holds, in order to have a treatment of the Hawking system adequate for
the accounting of energy transfers between the collapsing system and the
quantum field, we will need a relativistic quantum treatment of the collapsing
geometry:  a quantum--gravitational treatment.

\section{Connections with thermodynamics}

Before Hawking's prediction of black--hole evaporation, a close formal
similarity between the classical general--relativistic theory of black holes
and thermodynamics had been noted.  It was regarded by most workers, though,
as \it only \rm formal (see however Bekenstein 1973).  With Hawking's
prediction that black holes were indeed thermodynamic objects, there seemed to
be good reason to suppose that the connection was very deep indeed.  Subsequent
work has focused on elaborating this connection.  The theory developed is
sometimes cited as the strongest evidence, albeit circumstantial, that the
Hawking effect is real.  The sentiment is that the overall picture is so
attractive, the details will somehow sort themselves out to produce the
evaporation predicted by Hawking.

This sentiment, while perfectly legitimate, is not really of itself an
argument that the Hawking effect is real, but rather an argument for which way
one should guess --- is it real or not?  Such guesses are a reasonable and
often necessary part of theoretical physics, but they are not substitutes for
trying to sort out what really happens.

All of the work to be described in this section will be of this circumstantial
character, and thus will not directly address the trans--Planckian problem or
the question of quantum--gravitational corrections.  I shall only raise these
issues explicitly, though, when they arise in senses not yet encountered. 

How good is this overall picture of the Hawking effect together with
thermodynamics?  The real answer is that the picture must still be considered
to be in a stage of development.  There are good arguments that the Hawking
effect should respect the ``generalized second law.'' On the other hand, we
shall see that there are difficulties in making precise exactly what this
generalized second law says, and some question about the arguments that there
is a need to generalize the second law. We shall also see that there are some
important open issues in the analyses of some of the thought--experiments
proposed to test the generalized second law.  Thus, while there is indeed a
general overall related set of analyses involving the Hawking effect and
thermodynamics integrally, there are significant unresolved issues, in
establishing the results and even in formulating them.

While the work in this area is not at present definitive, it is some of the
most important in the field.  It attempts to build detailed physical pictures
of what happens, and it is from attempts to model real situations that we can
hope to learn the most.

I will close this introductory section with an only half--way tongue--in--cheek
comment of Eddington's, which brings out both the physicist's conviction of the
primacy of the second law\fnote{A notion which should be given some thought.}
and the distinction between whether one has actually analyzed a
conflict in physical theories or simply has a strong belief in which way it
will be resolved.

\itemitem{}
If someone points out to you that your pet theory of the universe is in
disagreement with Maxwell's equations --- then so much the worse for Maxwell's
equations.  If it is found to be contradicted by observation --- well, these
experimentalists do bungle things sometimes.
But if your theory is found to be against the second law of thermodynamics I
can give you no hope; there is nothing for it but to collapse in deepest
humiliation.  (Eddington 1929)

\subsection{Classical black--hole thermodynamics}

We begin by summarizing what is called classical black--hole thermodynamics.  
Whether it really \it is \rm thermodynamics or not is not wholly clear, and
indeed prior to Hawking's prediction of black--hole radiation it was viewed by
most workers as simply an analogy. However, the formal similarities are so
striking that it is reasonable to conjecture that there is a deep connection,
remaining to be understood.

The \it zeroth law \rm is that an isolated, stationary black hole has a
well--defined ``temperature,'' its surface gravity $\kappa$.  That is, the
surface gravity of an isolated, stationary black hole is constant over the
event horizon.  This is not at all a trivial result for non--spherically
symmetric  holes.  Note that, without bringing in Planck's constant, it is not
possible to convert $\kappa$ to a quantity with units temperature (or energy);
dimensionally a temperature would be $\hbar \kappa /(ck)$.

The \it first law \rm is conservation of energy, which can be expressed in
differential form as $8\pi (c^2/G)\kappa\d A=\d Mc^2-\Omega \d J-\Phi \d Q$
(with $\Omega$ and $\Phi$ the angular velocity and electromagnetic potential at
the horizon), giving the response of the hole to changes in charge, mass and
angular momentum; it parallels a thermodynamic $kT\d S=\d U -\Omega \d J-\Phi
\d Q$.

The \it second law \rm is the statement that, if the energy density of matter
crossing the horizon is positive, then the area of the black hole cannot
decrease.  (This is Hawking's celebrated \it Area Theorem.\rm ) Thus one
identifies the area with entropy. Again, classically these have different
units, and to convert area into an entropy one must form $Akc/(\hbar
G)=kA/l_{\rm Pl}^2$.  Note that the factor $l_{\rm Pl}^{-2}$ will give even
black holes of modest size colossal entropies.  

While these results are very beautiful, they have no obvious connection with
ordinary thermodynamics.  In particular, neither the ``temperature'' $\kappa$
nor the ``entropy'' $A$ has any clear thermodynamic interpretation or
significance.

Hawking's prediction gives an obvious thermodynamic interpretation to
$\kappa$ (or $\kappa /2\pi$)
as the temperature.  It does not give quite so obvious a
thermodynamic interpretation to the area as the entropy, although such
an interpretation can be inferred from the classical first law.  We
shall see later in this section that there is a certain amount of
evidence that this interpretation is \it consistent \rm with other
thermodynamic principles.  On the other hand, the problem of
finding a convincing \it explanation \rm of the identification of area
with entropy (for example, by identifying the black hole's area as a
measure of the logarithm of some volume of phase space) is not
addressed by most of these considerations.  It may be addressed by
quantum--gravitational computations, however, as will be discussed in
section 10.

\subsection{General relativity and the second law}

A main motivation for the generalized second law was a concern that the
ordinary second law of thermodynamics might not hold in general relativity.  
The argument runs as follows.

Imagine adiabatically lowering a box of gas (or thermal radiation) towards a
Schwarzschild black hole.  As it is lowered, potential energy is recovered,
this energy corresponding to the red--shift of the box's energy relative to
infinity.  The \it entire \rm energy of the box is red--shifted, including its
heat content.  Thus as one lowers the box towards the black--hole horizon, one
recovers a fraction arbitrarily close to unity of the energy in the box \it as
useful energy at infinity.  \rm  This process seems to contradict the second
law, at least in its ordinary sense, for it allows one to convert thermal
energy to work adiabatically.  This argument is often cited as a demonstration
that classical general relativity is not compatible with the second law.
It is therefore taken as a sign that the second law will have to be modified if
it is to hold when black holes are present.\fnote{The process described here,
of lowering a thermodynamic system towards a black hole, has come to be called
by some the \it Geroch process, \rm  and by others the \it Wheeler process. 
\rm  It was apparently described by both men, independently, at Princeton
around 1971 (Bekenstein 1973).  However, no published work by Wheeler or Geroch
on this seems to survive.}

However, on more careful consideration, it is not clear that the second law
really has been violated.   Even if one lowers the box to within a proper
distance $\epsilon R_{\rm Sch}$ of the hole and then releases it, external
observers will never see the box, or the entropy it contains, enter the hole. 
They see the box, very red--shifted, moving towards the Schwarzschild radius
exponentially slowly.\fnote{This analysis assumes the cosmic censorship
conjecture.  Were the conjecture to fail, distant observers might see the box
disappear as it went behind a naked singularity to cross the horizon.} No
entropy has been lost, even as far as simply the region outside of the hole is
concerned.  (On the other hand, one should note that, given that the
box \it will \rm fall into the hole, it is beyond the influence of
distant observers.)

There is another issue.  This is that one would not expect the local
temperature, and hence the local energy, of the box to remain exactly constant
as the box is lowered adiabatically.  This is because the gravitational field
will tend to pull matter and radiation within the box towards its bottom.  This
inhomogeneity would decrease the entropy, and would have to be offset  in an
adiabatic process by an increase in temperature (assuming positive specific
heat) and internal energy.  This means one cannot extract quite as
much energy, for an increment of motion of the box,
as thought naively.  There might
even be a point, somewhat before the Schwarzschild
radius is reached, at which the maximum energy extraction for an adiabatic
process is reached. Lowering the box further would require either abandoning
the adiabatic condition, or actually putting energy into the box.  In other
words, the requirement of adiabaticity might lead to a sort of flotation
point.  A similar but not identical behavior, in the quantum analysis, was
noted by Unruh and Wald (1982) (see also Radzikowki and Unruh 1988, which
contains a correction); see section 7.5.

It should also be pointed out that in these models the black hole has been
treated as being at zero temperature.  The validity of this needs to be
justified, as examples like Feynman's ratchet and pawl (or Maxwell's demon)
show.

In the literature, there is much discussion of ``dropping things (or entropy)
into black holes.''  The assertion is made that in this process entropy is
hidden from the outside world, and that for this reason the second law appears
to be violated.  However, this assertion really comes from combining
selectively chosen elements of the physics as perceived by  different observers
in a way which cannot be expected to provide an accurate accounting for the
entropy of the system.  On the one had, the authors want to consider a family
of observers, some of whom fall into the hole, in order to be able to assert
that the box \it has \rm fallen in; on the other, they want to only consider
the entropy measured by those observers in their family who stay outside the
hole.\fnote{The question of whether a box can be considered ``for all practical
purposes'' to be in the black hole is bound up with the choice of
coarse--graining scale. A choice of coarse--graining scale presumably defines a
thickened ``effective''  horizon such that any object which comes within this
thickened region is considered for practical purposes to be part of the
horizon.  However, this scale must also be finer than the scale on which one
wants to consider the degrees of freedom of the thermal system in the box. 
Thus one cannot bring the entire box, or even a significant fraction of it,
within the thickened effective horizon.}

At a deeper level, it is not wholly clear how one should formulate the second
law in even the simplest black--hole space--times:  whether it should refer
only to region external to the hole, or take into account the hole as well. 
One must also bear in mind that inside the hole (and, for rotating black holes,
inside the ``ergosphere''), \it no \rm observer sees space--time to be
stationary.  This means that what is usually taken as a basic underpinning of
thermodynamics is absent. 

In sum, the situation is that there is no clear violation of the second law,
but also some concern about what the precise formulation of the second law
should be.  

\subsection{The generalized second law}

As just pointed out, the argument that ``dropping thermal systems into black
holes leads to violations of the second law'' is probably not really correct.
However, this does not seem to go to the root of the concern about entropy and
event horizons.

The root issue is that (minus) the  entropy represents a measure of
information, and information is  in some sense lost to the external world when
it crosses an event horizon.  Thus there does seem to be an important and
perhaps profound physical problem:  to find an extension of the second law
which provides an accounting of entropy for families of observers up to \it and
including \rm those on an event horizon.  The \it generalized second law \rm is
a hypothesized extension of this sort, first proposed by Bekenstein (1973):

\itemitem{} The common entropy in the black--hole exterior plus the black--hole
entropy never decreases.

\noindent (With the understanding that the black--hole entropy is a constant
multiple of the area.)\fnote{If one wants to retain the original motivation for
the generalized second law (loss of entropy from the external world, etc.), and
accepts the comments I have made earlier to the effect that violations of the
ordinary second law are only possible when naked singularities appear, then one
would presumably conjecture that the common entropy in the exterior of a
singularity, plus the area of the singularity (however that might be defined)
never decreases.}  

It is worthwhile discussing what the precise mathematical and physical
formulation of this ought to be. It seems reasonable that we should measure the
entropy on some ``surface of constant time,'' a Cauchy surface from which the
interiors of any black holes have been excised.  ``Never decreasing'' has a
clear meaning, then, for there is a clear sense in which one such surface is to
the future of another.  

It is less clear what ``common entropy'' should be, because of the difficulty
in defining entropy for very general systems.  It is also not wholly clear, in
the most general circumstances, what it means to simply consider the entropy
``in the exterior,'' for there may be correlations which relate interior and
exterior modes.  

Would the Geroch--Wheeler process obey the generalized second law? At first
sight, it is not at all clear that it could.  If we ignore the buoyancy effect
I described earlier, then one could lower a box of entropy arbitrarily close to
the Schwarzschild radius of a hole.  This would render the box's contribution
to the hole's mass (when the box is finally dropped in) arbitrarily small, so
it would not increase the hole's entropy.  Yet the entropy of the box would
cross the horizon, violating the generalized second law.

This issue was recognized by Bekenstein (1973) from the first, and he proposed
that the generalized second law would be enforced by previously unsuspected \it
entropy bounds, \rm of the form
$$ S\leq 2\pi R E /(\hbar c)\, ,\eek$$
where $E$ and $R$ are the energy and characteristic dimension of the box, and
$S$ is the entropy it contains.  Since then, there has been much thought given
to the problem of establishing entropy bounds of various sorts.  (See Bousso
2002 for a review.)  At present there is not agreement, however, that strong
enough bounds have been proved to enforce the generalized second law.  However,
the buoyancy concern that I raised above, and one raised earlier by Unruh and
Wald (1983), also raise the possibility that the generalized second
law could be 
enforced by physics other than entropy bounds.  (See Bekenstein 1982,
1994b, 2002, and Wald 2001 for discussion of the debate over entropy bounds.)

In any event, it should be noted that when  ordinary macroscopic objects cross
the event horizon, the gain in black--hole entropy is typically colossal
compared to any common entropy involved.  For example, a $1$ kg mass falling
into a solar--mass black hole would increase the black--hole entropy by $\sim
10^{47}$, whereas the common entropy of the mass would be (very roughly) the
number of baryons, $\sim 10^{27}$, many orders of magnitude smaller.  It is
hard to think of ways of violating the generalized second law with ordinary
sorts of matter.

A related point is that if indeed it were possible to violate the
ordinary second law (whatever that means in a general--relativistic
context) up to the limits imposed by the generalized second law, those
violations would be enormous.  For the loss in ordinary entropy would
be the gain in black--hole entropy.

Note however that if the area of the black hole could decrease (this
would correspond to the black hole gaining negative energy), the
requirement that the generalized second law hold might be much more
difficult to meet.  If somehow a $-1$ kg mass could be dropped into a
solar--mass black hole, the generalized second law would require this
to be accompanied by a huge ($\sim 10^{47}$) increase in the entropy
of the exterior.  Of course, we do not have any $-1$ kg masses.  But
even with the tiny negative energies supposed to be absorbed by the
black hole in the Hawking process, we have a non--trivial test of the
generalized second law, as we now discuss.

\subsection{The generalized second law and the Hawking process}

Does the Hawking process respect the generalized second law?  There is
good (although not quite conclusive) evidence that it does.  And the
picture that emerges is in some respects quite appealing, and is
therefore taken to be circumstantial evidence for the Hawking process.

There are two competing effects which need to be considered in order
to check whether the Hawking process respects the generalized second
law.  On the one hand, the state of the quantum field changes, and in
particular thermal radiation is emitted to infinity.  While precisely
how the field's state, and entropy, change need to be thought about
carefully, one would expect an overall increase in ``common'' entropy
from this.  On the other hand, since (presumably) energy is conserved,
the black hole itself must be losing mass in the Hawking process.
This means its area, and hence the black--hole entropy, should be
decreasing.

\subsubsection{Quantum--gravitational issues}

It should be noted that any assumption about how the black hole loses mass
implicitly involves quantum--gravitational assumptions, if only in a negative
way.   (By negative quantum--gravitational assumptions, I mean that one assumes
that classical general relativity is adequate when dimensional or other
arguments show that there are some potentially quantum--gravitational effects.)
Since the emission of Hawking radiation is a highly non--classical process, the
question of how the quanta carry away energy is potentially one, too.  
However, if Hawking radiation does exist, it is probably legitimate to
approximate the time--averaged energy loss by a semiclassical source term
$\langle T_{ab}\rangle$.  (Here time--averaged means over scales $\gtrsim
GM/c^3$, the characteristic time to emit a Hawking quantum.) This would lead to
a mass loss of the hole due precisely to the expected Hawking flux.  This is
the assumption generally made.

It turns out that one must consider still longer averaging times, though, to
avoid potential quantum--gravitational effects. Neglecting numerical factors,
the mass loss of the hole in a time $\sim GM/c^3$ (the characteristic time to
emit a Hawking quantum) will be $\sim kT_{\rm H}/c^2\sim m_{\rm Pl}^2/M$.  This
means that the change in Schwarzschild radius should be $\sim l_{\rm Pl}m_{\rm
Pl}/M$, a tiny  fraction of the Planck length.  It is unrealistic to assume
that changes like this have any meaning --- it is probably unrealistic to \it
believe \rm they have any meaning.  A classical model of space--time is thus
not clearly credible for the analysis of such processes.  Only if we average
over times $\gtrsim (M/m_{\rm Pl})(GM/c^3)$ does the change in Schwarzschild
radius approach the Planck length and a classical general--relativistic picture
of the change in the horizon become credible.\fnote{Over a time $GM/c^3$, the
area changes by $\sim l_{\rm Pl}^2$, and so one could argue that the change in
area has classical meaning over such times.  However, the point made here is
that one does not have a full classical general--relativistic picture over such
time scales.  Thus we do not really know that we are justified in modeling
space--time classically for these purposes.}  For a solar--mass black hole,
this longer averaging time would be $\sim 10^{25}$ y, but for a $10^{15}$ g
``mini'' black hole, it would be $\sim 10^{-4}$ s. 

\subsubsection{Definition of entropy}

In principle, one would like to compute the field entropy on a partial Cauchy
surface in the exterior of the hole, and see how this  changes as the surface
moves forwards in time.  Some progress has been made towards this goal, by
taking the field entropy to be the entanglement entropy of the modes outside
the black hole relative to those inside (Bombelli et al. 1986, Bekenstein
1994a).  However, the definition of this involves some sort of
quantum--field--theoretic regularization, and a full understanding of this is
yet to come.  In particular, it is not clear if one will be able to avoid the
use of trans--Planckian cutoffs.

\subsubsection{Rate of entropy production}

Another line of approach is to compute the average rate of entropy production
in the radiation.  This would be simply be given by the entropy for blackbody
radiation, were the spectrum exactly thermal. On account of deviations from the
Planckian spectrum, however, there are modifications to the blackbody formula. 
Remarkably, however, there is a general argument that the rate of entropy
production on account of radiation exceeds the rate of area loss (Bekenstein
1975, Panangaden and Wald 1977).  Page (1976)  has found numerically that
(depending on the species) the excess is typically by a factor $\sim 1.5$. 
This line of argument can be extended to take into account non--vacuum
in--states; see Frolov and Page (1993) (as well as the related paper of Hawking
1976).

In the form just given, the radiative--entropy calculations are a little
removed from verifying the generalized second law in the form I have given it. 
This is because the entropy of the radiation is not the entire entropy (on any
partial Cauchy surface) of the region exterior to the black hole. However, it
is possible to use these calculations to get a result which is close to
verifying the second law in this strong sense. This is not quite trivial,
because one has to deal with the problem of defining entropy on the partial
Cauchy surface, as well as the fact that correlations between interior modes
and radiative modes potentially render entropy not an extensive quantity.

However, it is possible to phrase things so as to largely, although not
completely, dodge these difficulties.  The idea is this.  Consider a partial
Cauchy surface $\Sigma$ consisting of two parts, one a spacelike surface
$\Gamma$ extending from the event horizon to $\scri ^+$, and the second the
portion ${\cal N}$ of $\scri ^+$ to the past of $\Gamma$.  We assume that
$\Sigma =\Gamma \cup {\cal N}$ lies to the future of the collapsing matter, so
that it is in the stationary portion of the Schwarzschild solution, and we let
$\Sigma _t =\Gamma _t\cup {\cal N}_t$ be the surface got by flowing forward by
an amount $t>0$ along the vector field $\partial /\partial t$.  (See figure 3.)

\epsfysize=3in
\epsfbox{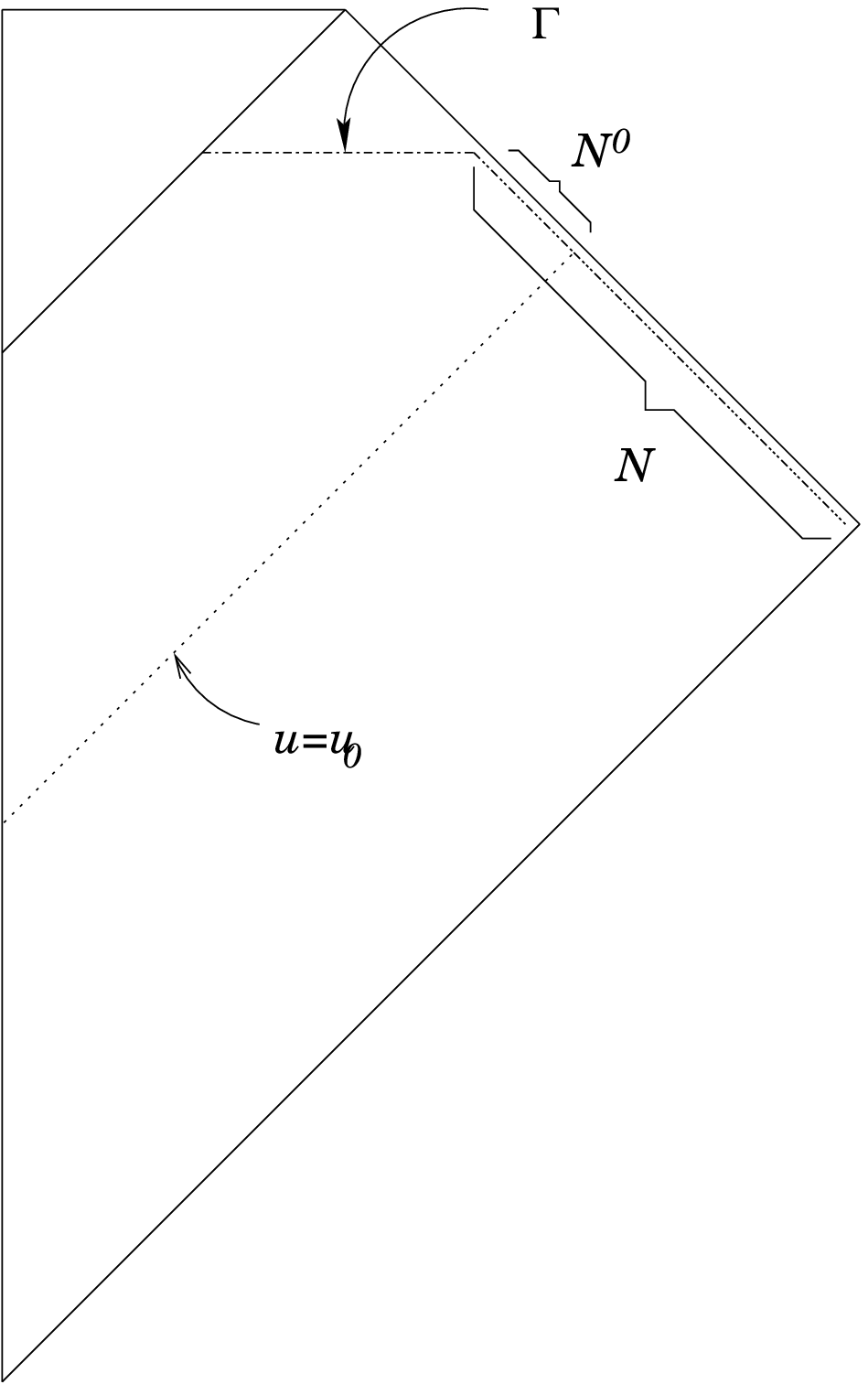}
\figure{The geometry of the partial Cauchy surfaces used in the argument that
the rate--of--entropy--production calculations support the strong form of the
generalized second law.  The partial Cauchy surface $\Sigma$ consists of two
parts, an ``internal'' $\Gamma$, and a portion ${\cal N}$ sensitive to the
radiative data.  Then ${\cal N}_0$ is the portion of ${\cal N}$ to the future of
the retarded time $u=u_0$ at which the Hawking radiation nominally starts.  The
definitions for $\Sigma _t$ are got by flowing all quantities except $u=u_0$
forward along the Killing field.}

There will be some nominal retarded time $u_0$ at which the Hawking process may
be said to have begun (and transient behavior passed).  We shall assume that we
can neglect correlations between the field for the portions of $\Sigma$ (or
$\Sigma _t$) before and after $u=u_0$ in calculating the entropy.  Then to
check the generalized second law, we must compute the difference in field
entropies $$S(\Gamma _t\cup {\cal N}^0_t)-S(\Gamma\cup {\cal N}^0)\, ,\eek$$
where ${\cal N}_t^0=\{ (u,\theta ,\varphi )\in {\cal N}_t\mid u\geq u_0\}$, and
similarly for ${\cal N}^0$.  On the other hand, since the Hawking process is
stationary, the entropies will be the same for that portion of $\Gamma _t\cup
{\cal N}_t^0$ which is the image of $\Gamma _t\cup {\cal N}^0$.  Now since the
radiation is thermal, it is a good approximation to treat the entropy within
the portion of $\scri ^+$ receiving the radiation as an extensive quantity.  
It is also, we assume, a good approximation to treat the field on $$A_t=\{
(u,\theta ,\varphi )\mid u_0\leq u\leq u_0+t\}\eek$$ as uncorrelated with that
on $\Gamma _t$ --- indeed, it seems likely that this assumption becomes better
for large $t$, since $A_t$ becomes larger but $\Gamma _t$ stays the same size. 
Under this assumption, we have
$$S(\Gamma _t\cup {\cal N}^0_t)-S(\Gamma\cup {\cal N}^0) =S(A_t)\, .\eek$$
This is exactly the radiative entropy.

A number of comments on this argument:  (a) The assumptions made about
correlations seem plausible physically. (b) It dodges the technically difficult
issue of defining entropy over $\Sigma$ and relies on the stationarity of the
space--time to cancel out ``internal'' (on $\Gamma$ and $\Gamma _t$) portions
of the entropy.  (c) It is  not clear that the space--time really is stationary
to the accuracy required. 

\subsection{The Geroch--Wheeler process revisited}

The Geroch--Wheeler process was reconsidered by Unruh and Wald (1982) (see also
the correction, Radzikowski and Unruh 1988) in the light of quantum theory. 
They discovered a number of very interesting phenomena which suggested a
coherent picture of the generalized second law, the Hawking process, and Unruh
radiation.  Their aim of developing an explicit physical picture serves as one
of the most important models in the area.  

What Unruh and Wald found was that it was necessary to take into account a
variety of factors involving zero--point fluctuations. It was necessary to
consider not just the Hawking radiation, but also moving--mirror effects
induced by the walls of the box and (at least for observers comoving with the
box) Unruh radiation.  While there are some questions concerning the hypotheses
of the analysis (and hence its conclusions), it is clear that any future work
will have to face up to the issues these authors uncovered.  If
anything, the problem may be still more delicate.

Unruh and Wald actually considered, not quite the Hawking process, but a
Hawking--radiating black hole in equilibrium with thermal radiation at
temperature $T_{\rm H}$; this is known as the \it Hartle--Hawking \rm state. 
And they actually analyzed the problem from two different points of view:  that
of an observer comoving with the box (hence accelerating); and that of an
inertial observer.  

From the point of view of the comoving observer, they found that the Unruh
radiation impinging on the box exerted a buoyant force, which became so strong
as to cause the box to float at a certain point. Thus there was never any
question of dropping the box into the black hole, and in considering various
possibilities for opening and shutting and raising and lowering the box Unruh
and Wald found no violation of the generalized second law.  For inertial
observers, the analysis is more difficult, and Unruh and Wald relied on
two--dimensional model calculations.  These suggested that the buoyant force
could be thought of as arising from radiation induced by the motion of the
reflecting walls of the box.  

There are two, related, concerns about the Unruh--Wald analysis.  The first is
that it is semiclassical.  For example, the  radiation pressure on a wall of
the box is taken to be a c--number, whereas actually it is a
distribution--valued operator sensitive to the scale on which it  is measured. 
This simply means that a small box cannot detect the contributions of the
long--wavelength field modes to the pressure. As Bekenstein (1999) has pointed
out, the scales involved are such that this may be a serious concern.
(See also Bekenstein 2002.)

It is also a bit dubious to give a semiclassical analysis which involves
fictional Unruh observers (that is, accelerated observers who do not actually
make measurements of the quantum field but merely aim to describe it
mathematically).  This is because the energy for the Unruh process, when it is
a real process (that is, quanta are really detected), must presumably be
supplied by the engine which accelerates the observers.  The process inevitably
entangles the field's state with that of the motive engine, and so it is not
clear whether a semiclassical analysis is adequate.

The second concern is best introduced by pointing out that Unruh and Wald
themselves deduced from their analysis that an empty box, once accelerated to a
certain point, would maintain its acceleration, by mining energy from the
quantum vacuum.  Clearly, this is a provocative result and one which needs to
be investigated carefully.  Recently, Marolf and Sorkin (2002) have argued that
the result is unrealistic, and that a realistic treatment would require a
quantization of the internal mirror modes (and does not seem to lead to
self--acceleration).

The Marolf--Sorkin analysis is consonant with findings in other moving--mirror
models (Parentani 1996, Helfer 2001a) (cf. section \mirsec ).   It seems that
in order to have an accurate enough treatment of these to analyze second--law
issues arising from vacuum fluctuations, one needs to treat the mirrors as
quantum objects, and indeed to second--quantize the reflective mirror modes
(the plasmons).  In Helfer (2001a), it was found that these limitations
invalidated some attempts to defeat the (ordinary) second law.

At present, a convincing treatment of the Geroch--Wheeler process at this
level has not been achieved. As Marolf and Sorkin  charmingly put it, when
they  analyzed the process in light of their conclusions,

\itemitem{} [O]ne is forced to consider temperatures at which the ``thermal
radiation'' is dominated by box--antibox pairs.  This clouds the picture
somewhat.

\section{Nonstandard propagation}

Describing the appearance of trans--Planckian modes in the usual Hawking
calculation as ``certainly wrong,''  Unruh (1995) considered what might be a
substitute and discovered one of the most curious results in the area. 
Motivated by the physics of sound waves at very short wavelengths, he
investigated a model with where there was a limiting ultraviolet frequency
$\omega _0$, and dispersive propagation at very high wave numbers.  He found
significant evidence for thermal creation of particles in this model, at the
model's analog of the Hawking temperature.  Subsequent investigations of this
and related models have found similar results. 

This area is very much a matter for investigation.  Some aspects of the physics
are now well--understood, but other essential points are not yet clear.  Also
various models with different features have been considered, and there is at
present no consensus about which (if any) is preferred. Finally, the successes
of these models in dealing with the trans--Planckian problem are unclear.  Most
require ultrahigh wave--numbers although they have bounded frequencies.  Some
of the models of Corley and Jacobson (1996) avoid both ultrahigh wave--numbers
and ultrahigh frequencies, but these have apparently unphysical features. 
There also exist models where the field is confined to a lattice in a
black--hole space--time (Corley and Jacobson 1998, Jacobson and Mattingly
2000), but so far these seem to involve either trans--Planckian modes or
physically unacceptable distortions of the lattice.

Before discussing the Unruh--type models, I shall review a related proposal of
Jacobson (1993).  This was an attempt to modify the
propagation by introducing a boundary condition which would do away with
trans--Planckian problems.  Then
I shall briefly discuss two of the Unruh--type
models, one due to Corley and Jacobson
(1996) and the other to Brout, Massar, Parentani and Spindel (1995b) (BMPS).  
There is a large overlap between these two, and between them and Unruh's
original work, so substantially the same comments apply to all.\fnote{Unruh's
model is not discussed directly here because it involves a spatial periodicity. 
While this periodicity is not really essential to the issues dealt with here,
explaining just why this is, when the asymptotic structures of black--hole
space--times are so much in the foreground, would require a lengthy and perhaps
distracting treatment.}

All models considered so far are in two space--time dimensions, and presumably
connect with the Hawking process by describing the s--wave sector.

\subsection{Jacobson's cut--off model}

Jacobson (1993) introduced a model which can be viewed in several
different ways:  it gives an illustration of the
localization of the trans--Planckian problem discussed in section 4.1; it can
be thought of as an alternative hypothesis for reproducing Hawking's
predictions; it overlaps to a degree with Unruh's ideas.  It is based
on the introduction of a non--standard boundary condition, and so ---
propagation being determined by a field equation and
boundary conditions --- is also really an example of non--standard
propagation.

For our purposes, it is best to present this model from a perspective
a little different from the original one; comments on the
original will be given in the course of the discussion.

Let us
fix a radius $r_{\rm bc}=
(1+\epsilon )R_{\rm Sch}$, with $\epsilon$ small but
not so small that observers freely falling across the horizon (from
zero velocity at infinity) would perceive trans--Planckian frequencies
in the propagation of Hawking quanta for $r\geq r_{\rm bc}$.
If one then simply takes
as data at this boundary the Hawking state (properly, its $n$--point
functions and field operators), one can regard the region $r\geq
r_{\rm bc}$, with this boundary condition, as a
Hawking--radiating space--time for which no trans--Planckian problem
arises.  (One can therefore introduce a cis--Planckian cut--off without
affecting the predictions, so this is sometimes called a cut--off model.)

So far, we have simply cut out from space--time the trans--Planckian
regime near the event horizon; one can think of this as just an
illustration of the localization of the trans--Planckian problem
there.  However, Jacobson suggests we take the boundary condition at
$r=r_{\rm bc}$ as a new law of physics.  

This is reasonable, especially if one believes that Hawking radiation
really does occur but that there are problems with the
trans--Planckian physics in its derivation.  It cannot be counted as
a final theory, however, for it does not really address what the
physics is for $r\leq r_{\rm bc}$, and also one would really like some
sort of an explanation of what deeper physics might give rise to this
boundary condition.
While these concerns were certainly known to Jacobson, it is
appropriate to
spell them out, and raise some related points.

Since the horizon is
determined by global data, one would like to know how the space--time
``knows'' to implement the boundary condition.  Does the boundary
condition really only apply when an actual black hole is to form, or
would it also apply for ``almost--black holes,'' that is, for objects
collapsing to $r=(1+\epsilon ')R_{\rm Sch}$?

The boundary condition is imposed on a timelike surface.  In general,
such boundary conditions lead to sensitive dependence on initial data.
(A simple example of this in Minkowski space, for the massless field
and the surface $z=0$, is
$$\phi =a\e ^{\i\omega t -\i k x}\cosh (\kappa z)\eek$$
with
$$\omega ^2-k^2+\kappa ^2=0\, .\eek$$
By choosing $|a|$ small, the data $(\phi ,\partial _z\phi )\Bigr|
_{z=0}$ can be chosen uniformly small.  However, one can arrange for
$\kappa$ to be arbitrarily large, which makes $\phi$ arbitrarily large
in any given neighborhood of $z=0$.)  This issue has not yet
been investigated.  On the other hand, much of Jacobson's paper is
concerned with a careful argument to the effect that his boundary
condition has an \it internal \rm self-consistency:  it can perpetuate
itself without any reference to ultra--high frequency physics.
Perhaps developments of these ideas could address the sensitivity issue.

There is also the question of the trans--Planckian frequencies
associated with the advanced time of formation 
$v=v_0$ of the hole (that is,
with the extension of the generators of the event horizon backwards
through space--time).  In other words, the model does not yet seem to
address the fact that the Hawking quanta have their origins in
trans--Planckian vacuum fluctuations near $v=v_0$.

Finally, it should be mentioned that Jacobson considered it quite
possible that his boundary condition would \it not \rm be fulfilled,
and that there might therefore be modifications to the Hawking flux.

\subsection{The model of Corley and Jacobson}

The main features of the two models to be discussed (that of Corley and
Jacobson, and that of BMPS.) are similar.  I shall give a sketch of the
Corley--Jacobson model here, and in the next subsection only describe the
essential differences of the BMPS model.

\subsubsection{The space--times}

The models considered are space--times with metrics $$\d s^2=\d t^2 -(\d x
-v(x)\d t)^2\, .\eek$$ (The reader is cautioned that the symbols used for the
coordinates by different authors are not compatible.)  The Schwarzschild metric
(modulo its angular parts) is included in these, the case $v(x)=-\sqrt{ R_{\rm
Sch}/x}$.  We shall assume $v(x)$ is an increasing, continuously
differentiable, function of $x$, which tends to a constant value  (less than
unity) as $x\to +\infty$.  Then the metric is regular everywhere throughout the
$(t,x)$ coordinate system.  The vector field $\partial _x$ is everywhere
space--like, but $\partial _t$ is timelike iff $|v|<1$.  The temporal
orientation  is taken by defining $\partial _t$ to be future--pointing in the
region $|v|<1$; elsewhere, it is determined by continuity.  We shall mostly be
interested in the region $v\geq -1$.

The curve $v=-1$ is null and is a Killing horizon.  (That is, the Killing
vector $\partial _t$ is tangent to it.)  It is also, in a suitable sense, an
event horizon.  Precisely, we define a null curve to escape to \it future right
infinity \rm $\scri ^+ _{\rm r}$ if $x$ increases without bound towards the
future.  (Because of the assumptions made on $v(x)$, this will agree with any
other standard definition of the ``right half'' of $\scri ^+$ in this
two--dimensional space--time.)  Then $v=-1$ is the boundary of the set of
events from which there are causal curves to $\scri ^+_{\rm r}$, as is easily
checked from the differential equation for the null curves.  We may thus call
it the \it future right event horizon.  \rm  It will be the analog, in these
models, of the event horizon of the black hole.

The curves $\d x=v\d t$ are the trajectories of freely--falling observers,
everywhere orthogonal to the surfaces $t=\const$, and crossing the horizon
transversely.  They play an important role in the models, as the sense of what
the ``in--vacuum'' is taken to be will be defined in terms of these.   Note
that this is (at least on its face) a highly coordinate--dependent concept. 
And even if $v$ tends to zero asymptotically, the $(t,x)$ coordinates need not
be inertial.  For example, in the Schwarzschild case, we have
$v(x)=-\sqrt{R_{\rm Sch}/x}$, and so the trajectories are
$-(2/3)\sqrt{x^3/R_{\rm Sch}} =ct+\const$.

It should be emphasized that in the Corley--Jacobson (and in the Unruh) models,
the space--time region that is examined is \it exactly stationary.  \rm  It is
supposed to model space--time after gravitational collapse has occurred, and no
input about the collapse, or the propagation of the quantum fields through the
collapsing region, is used.  In this way it differs essentially from the
Hawking model. 

\subsubsection{The dispersive propagation}

Even in Minkowski space, dispersive propagation is not Lorentz--invariant. 
Physically, this is due to the fact that the dispersive medium defines a
preferred frame.  In order to define the propagation in the Unruh--type models,
one must break local Lorentz invariance.  The authors do this by writing their
propagation equation as
$$(\partial _t +\partial _xv)(\partial _t+v\partial _x)\phi ={\hat F}^2
(\partial _x)\phi \, .\eek$$\xdef\disprop{\the\EEK}%
where $F(k)=-\i {\hat F}(\i k)$ is the dispersion function.  
(The ordinary wave equation is the case $F(k)=k$.)
The right--hand side of (\disprop ) is defined by
Fourier--transforming in $x$; it
is thus a highly coordinate--dependent quantity, since defining it depends on
integrating over the $x=\const$ curves.  It can be written somewhat
more invariantly as 
$$\nabla ^2\phi =\left( {\hat F}^2(X) -X^2\right)\phi \, ,\eek$$
where $X=\partial _x$.  This makes clear that 
(as far as the propagation goes) the local Lorentz
symmetry is broken by the choice of the vector field $X$, and nothing more.

Various choices have been investigated for the dispersion relation.  The
simplest one is the \it hyperbolic tangent form \rm 
$$F_{\rm hyp}(k)=k_0\tanh (k/k_0)\, ,\eek$$
which has $F_{\rm hyp}(k)\simeq k$ for $|k|\lesssim k_0$ and  $F_{\rm hyp}(k)
\simeq
k_0\sgn (k)$ for $|k|\gtrsim k_0$.  This is the one we shall consider, except
for a few comments. The effect of the dispersion relation is to replace the
partial derivative $\partial _x^2$ at a single event  by a weighted average
over $x$--distances of coordinate size $\sim k_0^{-1}$.

A rough approximation to the propagation can be gotten by WKB
methods.  Approximate solutions are then of the form
$$\phi\sim \exp \i (\omega t+k(x)x)\, ,\eek$$
where 
$$(\omega -kv(x))^2=F^2(k)\, .\eek$$
Consider either branch
$$\omega -kv(x)=\pm F(k)\eek$$
of this, for fixed $x$ and $\omega$, as an equality between functions of $k$. 
The left--hand side is a straight line, but the right--hand side, for
non--trivial dispersion, is a curve.  There will thus be several points of
intersection, typically three  three on the positive branch and one on the
negative, in the models considered.  It is the extra solutions on the positive
branch which allow the Unruh--type behavior.

Fix $\omega$, and imagine a wave--packet which, in the distant future
consists of the spatial mode of moderate wave--number, $k\sim \omega
/(1+v)$.  As this is propagated back towards the past, it moves
inwards towards smaller coordinate values, and $v(x)$ decreases
towards $-1$.  This means that the slope of the line $\omega -kv(x)$
increases, and the line becomes more nearly tangent to $F(k)$.  This
gives a chance for the solution to develop a term corresponding to the
right--most intersection point, and indeed Unruh has given an argument
that at this point the packet turns around (its group velocity is zero
at the point of tangency) and moves outwards at the group velocity
defined by the right--most point.

In fact, there is another feature that occurs simultaneously, as pointed out by
Corley and Jacobson, ``mode conversion:'' the left--most mode is excited as
well.\fnote{Possibly this can be understood in terms of Stokes's phenomenon.} 
It too turns out to contribute to the wave packet moving away (as we go
backwards in time) from the hole.

The upshot of the analysis so far is that the late--time mode we started with,
which was approximately that of a field satisfying the ordinary wave equation,
has been propagated back in time to a wave packet moving away from the hole
with group speed a bit slower than that of light.  The packet involved consists
of wave--numbers corresponding to the two extreme solutions of the dispersion
relation, rather than the intermediate one in late time with which we started.
By the time the modes have gotten to this point, their propagation is quite
different from what one would have for the ordinary wave equation.

In order to outline the significance of this for particle creation, we must
discuss the quantization of the theory.

\subsubsection{Quantization}

We recall that particles are defined in terms of the decomposition of a field
into positive and negative frequencies.  In a general curved space--time, this
decomposition is frame--dependent, and so is the notion of particles.  However,
if there is in a region of interest a naturally--determined frame, we have a
natural candidate for the definitions of particles.

The particle content of the out--state is supposed to be analyzed in the
asymptotically Minkowskian region $x\to +\infty$, and here there is a good
definition of the positive--negative frequency decomposition.\fnote{Although
this requires a little care to define, given the dispersive character of the
propagation.}  However, this is \it not \rm the way the particle content of the
\it in--state \rm is defined. (Had we used the same definition for the
in--state, we should find no particle production.)  Rather, the
particle--content of the in--state is determined by Fourier--analyzing with
respect to time \it along the trajectories of the preferred class of freely
falling observers, \rm that is, along the curves $\d x =v\d t$.  

The motivation for this may be described as the hypothesis that the in--state
should be vacuum as measured by these observers.

It has been verified numerically that these definitions of in--vacuum and
out--particle content do result in a (nearly) thermal spectrum at the model's
analog of the Hawking temperature.  Just \it why \rm this occurs is not really
understood, however.

\subsection{The model of Brout, Massar, Parentani and Spindel}

These authors investigated a very similar situation to that of Corley and
Jacobson above,\fnote{And reached many similar conclusions, which will not be
recapitulated here.} with hyperbolic--tangent dispersion, but with the
after--collapse portion of the space--time glued onto an idealized collapse
portion. Specifically, their model consists of an early Minkowski--space
region, followed by collapse of a pulse of incoming null matter to form a black
hole, and then an Unruh--type region.  

I shall not go into any of the technical analysis of this model (which follows
the same spirit as that given above), but list two contributions that it makes:

In this model, the definition of the in--state is given by evolving from the
Minkowski vacuum.  (Recall that in the Corley--Jacobson model this depended on
the freely--falling observers in the distant past.)  Since the two models
produce similar results, this is evidence that the Unruh--type models might not
be artificially sensitive to the choice of in--state.

Brout et al. gave a WKB treatment, and were able (within the accuracy of the
calculation) to give an analytic argument for the emergence of thermal
radiation in this model.  This must be counted as an important result.
Subsequent refined and somewhat
generalized treatments were given by Himemoto and Tanaka
(2000) and Saida and Sakagami (2000).

\subsection{Summary and discussion}

Jacobson's (1993) paper effectively gave hope that one might 
circumvent the trans--Planckian problem by 
exploiting its
localization to the event horizon.  The paper showed that if somehow an
effective boundary condition near the horizon could result from (unspecified)
deeper physics, Hawking's predictions could be recovered without
trans--Planckian modes.

The Unruh--type models very remarkably reproduce results quite close to
Hawking's.  Although superficially they involve only a small tinkering with the
field equation, in fact there is another key input, that is, the definition of
the in--vacuum, which does not follow the standard one but is at least on its
face a coordinate--dependent concept.  That the same results are produced in
the Corley--Jacobson and BMPS models is evidence that this dependence is not
too severe.

The modification of the field equation, that is, the dispersive propagation, is
regular  at the horizon.  It is defined by modification of the spatial portion
of the wave equation along a vector field $X=\partial _x$ which is perfectly
regular and space--like at the horizon.   One should note that the \it
stationarity \rm of this modification is used very strongly, and most
essentially in the neighborhood of the horizon.

While these models do substantially reproduce the final radiation from black
holes, the physical origin of that radiation is quite different.  Instead of
coming from the propagation of vacuum fluctuations through the collapsing
space--time and out to infinity, the Unruh--type models only consider that
portion of space--time after the collapse phase; the radiation arises from a
hypothesized character of the quantum in--state.

In the Corley--Jacobson model, the character of the in--state is set by a
preferred family of freely falling observers.  They are not observers in ``the
distant past,''  since the entire analysis takes place in the portion of the
space--time after the black hole has formed.  In the BMPS model, the in--state
is determined more invariantly, by propagation from an initially Minkowskian
regime.

To what degree do these models resolve the trans--Planckian problem?
There is some debate over this, partly on account of questions of how 
essential elements of the models are,
and partly on account questions of the significance of
trans--Planckian wave--numbers versus trans--Planckian frequencies.

The model with hyperbolic--tangent dispersion relation requires the use of
unboundedly
high wavenumbers as we follow the wave--packet back further into the
past.\fnote{This issue does not come up in Unruh's original model, on account of
the periodic boundary conditions he used.}  One would like to know whether these
are really essential, or whether, by settling for only a finite propagation
backwards, one could avoid them.  There is no definitive answer to this at
present, but another result in the Corley--Jacobson paper suggests
that the ultra--high wavenumbers
\it are \rm essential.  Corley and Jacobson investigated
a quartic dispersion
relation, with bounded frequencies and 
bounded wavenumbers, but were unable to find a
completely satisfactory construction of the theory in that case.

While the elimination of trans--Planckian frequencies is arguably progress, the
presence of trans--Planckian wavenumbers is still disturbing.  It
means that one is (at least implicitly) invoking a continuum model of space on
arbitrarily fine scales, something which one should at least be hesitant about. 
Also, the statement that trans--Planckian energies have been eliminated is
significantly frame--dependent (if trans--Planckian wavenumbers have not been
eliminated).  For example, if $\omega\d t +k\d x$ is a wave covector with
$\omega$ cis--Planckian but $k$ trans--Planckian, then an observer with tangent
$\cosh\xi \partial _t +\sinh\xi \partial _x$ will see a frequency 
$\omega\cosh\xi
+k\sinh\xi$, which will be trans--Planckian even for moderate $\xi$.

The most important issue to be resolved within these Unruh--type models is the
question of \it why \rm the spectrum they produce is nearly thermal and
(generalizing that) understanding how nearly the corresponding Bogoliubov
coefficients match those of Hawking's predictions.   
(We have numerical and analytical arguments, but it seems there ought to be a
brief basic physical argument.)
When we do this, we may
hope to have a better understanding of how sensitive the Unruh--type models are
to:  (a) the choice of coordinates; (b) the definition of the in--vacuum.

Finally, it should be noted that the possibility of dispersive propagation of
high--frequency modes has been an important theme not just in work on Hawking's
model, but in quantum gravity generally.  See Himemoto and Tanaka (2000),
Amelino--Camelia (2000), B\l{}aut et al (2001); and Amelino et al (1998) for the
possibility of experimental verification of such propagation.

\section{'t Hooft's $S$--matrix black hole theory}

It was 't Hooft who coined the term ``trans--Planckian.''
Dissatisfied with the appearance of ultra--high frequencies and with
some other aspects of the Hawking model, he pursued a program to
develop an alternative theory.  

In fact, 't Hooft (1985, 1993, 1996, 1997, 1999)
has put forward a host of interesting ideas, both in
developing his theory and in elaborating the concerns which motivated
it.  Behind this variety of ideas is a set of intuitions about what 
should be the guiding
principles in developing a fundamental theory:  the ideas all aim to
contribute to a unified cohesive view of the quantum character of black 
holes, and of space--time more generally.  In important respects, this
view differs radically (if sometimes subtly) from the conventional one.

One should 
have two, distinct, questions in mind when evaluating 
contributions like these.  The first is rather severe:  to what
degree does the program as a unit
really give a convincing treatment of the
quantum character of a black hole?  The second:  what light do the
individual ideas 't Hooft has introduced throw on the problem of
understanding this quantum character?

Given the current state of the program (and the profoundly difficult
problems it attacks), it would be premature to hold it to any very
stringent standard as far as the first question goes.  As to the second
question, the program has already proved its value.  It has given
models for ways in which quantum--gravitational effects might enter
essentially in the Hawking process, it has provided new physical
perspectives on ideas from string theory, and, perhaps most
importantly, it has provided a vivid reminder that there may be 
profound modifications to conventional quantum field theory in the
presence of black holes. 

While the goal of the
program is a full quantum treatment (including quantum--gravitational
considerations) of black holes, this has not yet been achieved.  And it
is not yet really clear the extent to which the program can be said to 
support the assertion that black holes radiate.
In its present stage of development, the 
program \it assumes \rm that only rather slight modifications of the 
Hawking model, insofar as the radiation it predicts, are necessary. 

It is not really possible to discuss the program's treatment of the
Hawking process without describing the program as a whole.
And to do this will require touching on some interesting aspects of
Hawking's predictions (loss of quantum coherence, eventual explosion
of black holes) which could themselves be discussed at length.
However, only enough of a treatment of these for considering 't Hooft's
program will be given.

I shall give an overview of the program, omitting the computations but
focusing on the ideas involved.  I shall along the way comment on the 
open issues; however, it will be evident that
many of these could be far more fully discussed 
than space here allows.  
Unless otherwise noted, all developments of the program
discussed here can be found in the long review paper 't Hooft (1996).

\subsection{Overview of the program}

The point of departure for the program is the idea
that black holes should
constitute a species of particle.  These particles may have internal
degrees of freedom, and so they may exist in various quantum states.  
The first question to be asked about this particle species is, What is
its spectrum?  In particular, if the particle is placed in a box, is
the spectrum continuous or discrete?

't Hooft argues that Hawking's theory would lead to a continuous
spectrum.  (This reasoning will be sketched below.)
This in turn, it is argued, would make it impossible for
black holes to evaporate completely; there would always be remnants.
This is regarded as unacceptable, because no other particle species
behaves in this way.  It also raises the question of how \it virtual
\rm black--hole remnants might modify other aspects of quantum theory.

As 't Hooft himself points out, this argument rests on a number of
assumptions which are debatable.  Is it really legitimate to apply
conventional quantum physics at the Planck scale, as needed to make
assertions about remnants?
Can we really talk about stationary states (eigenstates) of a
Hamiltonian for black holes, since black--hole
space-times contain not only an exterior
stationary region but an interior time--dependent region?  Is it
really legitimate, to the level of accuracy required, to talk about
putting a black hole in a perfectly reflecting box?  Is it legitimate
to regard the box itself as fixed (infinitely massive)?  How much
trouble would virtual black--hole remnants really cause?  If black
holes can only form from fairly massive objects in the first place,
couldn't it be that virtual remnants were always accompanied by
higher--energy virtual excitations, and so suppressed?

Another concern which motivates 't Hooft is the apparent loss of
quantum coherence associated with the Hawking process.  That is, the
Hawking process appears to convert an initially pure state (the
in--vacuum or other in--state) 
to a mixed state (the quasi--thermal radiative state).
As 't Hooft himself recognizes, it is far from clear that there really
is any loss of coherence in the Hawking process.  Accepting the
conventional quantum--field--theoretic view (and leaving aside
essentially quantum--gravitational questions about what happens at the
very end of the black hole's life), the mixing arises simply because
one is neglecting the modes inside the hole.  Yet 't Hooft is
dissatisfied with this conventional
picture, because it is the infinitely many
modes within the hole which are responsible for the continuum of
black--hole states.

\subsubsection{The spectrum}

As mentioned above, one of the main goals of 't Hooft's program is to
derive the spectrum of excitations of the black hole.  This is done by
\it assuming \rm that Hawking's model is essentially correct, and
appealing to time--reversal invariance.  The use of time--reversal
invariance here is, as we shall see, at least questionable.

To derive the density of states (up to a multiplicative
factor), he 
compares the Hawking
emission probability with the capture cross--section.  
This argument
rests on time--reversal invariance (or more properly, $CPT$
invariance).  There is some concern about this.  The two amplitudes
that must be related are
$${\cal T}_{\rm in}={}_{\rm BH}\langle M+E/c^2|\,|M\rangle _{\rm BH}
|E\rangle _{\rm in}\eek$$
and 
$${\cal T}_{\rm out}={}_{\rm BH}\langle M|\, {}_{\rm out}\langle E|\,
|M+E/c^2\rangle _{\rm BH}\, ,\eek$$
where $|E\rangle$ represents a quantum of energy $E$, and ``BH'' of
course stands for ``black hole.''  The concern is that the
time--reverse of a black hole is not a black hole, but a white hole.
While the black hole, once it forms, 
\it is \rm approximately static in some sense, it
is not clear that this sense is strong enough to justify the use of
time--reversal invariance, since the Hawking process depends on the
global history, before the quasistatic regime, of the space--time.

As emphasized by 't Hooft (see also Bekenstein 1972a,b), it
is difficult to reconcile the Hawking model as it stands with \it any
\rm finite density of states.  (That is, the multiplicative factor
undetermined in 't Hooft's calculation might be infinite.)
If, for example, one drops baryons into
a hole, and lets it radiate, one can increase the baryon number
indefinitely without increasing the mass.  Thus either baryon number
must cease to be a quantum number in a theory with black holes, or the
theory must contain an infinite density of energy levels.
Thus in order to achieve the discrete spectrum desired, some sort of
modification of Hawking's theory is necessary.

\subsubsection{The holographic principle}

Partly to render the density of states finite, and partly to deal with
concerns about quantum coherence, 't Hooft (1993)
made the radical suggestion
that conventional quantum field theory seriously overcounted the
internal degrees of freedom of a black hole.  He proposed that there
should be a \it holographic principle, \rm which should relate these
degrees of freedom to the boundary.  He was led in fact to conjecture
that such principles applied still more generally (to cases other than
black holes); there have since been many attempts to formulate
``holographic principles.''

I shall make no attempt to discuss
alternatives beyond those necessary to give a sense of 't Hooft's
program; see Smolin (2001), Bousso (2002) for fuller accounts and reviews.
For our purposes, it will be
convenient to distinguish \it weak \rm from \it strong \rm forms, as
well as those which are meant to 
apply to \it black holes \rm from those which
are meant to apply \it generally.  \rm  (This terminology is not
standard, although the weak/strong distinction is similar to one
used by Smolin .)

Let $\Sigma$ be an acausal hypersurface\fnote{An \it acausal \rm
hypersurface is
one for which there is no causal path connecting two points.  The
condition is somewhat stronger than requiring it to be locally
spacelike.
For our purposes, one could equally well require $\Sigma$ to be \it
achronal, \rm that is, to have no points connected by timelike paths.}
with boundary $\partial \Sigma$ of
area $A$.  The \it black--hole \rm case will be when $\Sigma$ lies in
the interior of the hole and $\partial\Sigma$ on the event horizon;
the \it general \rm case will be when $\Sigma$ is unrestricted.

\itemitem{}
The \it weak holographic principle \rm holds that the information 
in $\Sigma$  should
amount to no more than $A/(4l_{\rm Pl}^2)$ bits.

\itemitem{} 
The \it strong holographic principle \rm holds that the information in
$\Sigma$ is actually contained in
(or can be recovered from) at most $A/(4l_{\rm Pl}^2)$ bits of
information on the
boundary.

\noindent The term ``information'' is of course vague; an important
issue is how to make it precise.  However, we shall accept it at an
intuitive level.

A \it general \rm holographic principle, even a weak one, would be
very exciting, for it would imply non--local constraints on quantum
field theories even in Minkowski space.  This is because the number of
field modes in a given volume would be constrained; presumably, there
would be ultraviolet cut--offs arising from non--local
physics.
Depending on the scales and the circumstances in which these cut--offs
became important, there could be profound alterations in physics.  One
should in particular keep this in mind for cosmological models.

While there have been attempts to formulate rigorous versions of
general holographic principles, no really definitive one exists,
and there is a serious concern
about whether such a formulation will be possible.  Consider
a finite perturbation of the sphere $t=0$, $r=r_0$ 
Minkowski space to $t=f(\theta
,\phi )$, $r=r_0$.  The area of the perturbed sphere is
$$A=\int \sqrt{1-r_0^{-2}\|\nabla f\| ^2}\, 
  r_0^2\sin\theta\d\theta\d\phi\,
,\eek$$\xdef\arbh{\the\EEK}%
where $\| \nabla f\|$ is the ordinary two--sphere norm of the gradient
of $f$.  By choosing $f$ small but oscillatory one can arrange for the
area (\arbh ) to be arbitrarily close to zero while keeping $|f|$
arbitrarily close to zero.  
Thus there are arbitrarily small
perturbations of the sphere making its area arbitrarily small.
(Had we perturbed in space--like directions, we could have made the
area arbitrarily large.)  

This lack of stability of the area is especially troubling
from a physical point of view.  One should not have to specify to
arbitrary precision the surface $\partial\Sigma$ in order to say how much
information it surrounds.
It is
very difficult to see how, in such circumstances, one can take
$A/(4l_{\rm Pl}^2)$ as a measure of the information in
$\Sigma$.  (See however Bousso 2002 for other forms of general
holographic principles which many workers consider more 
likely to be viable.)

For \it black holes\rm , the area of the section $\partial\Sigma$ of
the event horizon has a stability not enjoyed by the \it general \rm
case, and so the sort of pathology uncovered above
cannot occur.  We shall from now on consider only the black--hole case.

While it would seem that even the weak form of the black--hole
holographic principle would meet 't Hooft's concerns about a continuum
of black--hole states, in fact 't Hooft seems to seek a theory in
which the strong form holds.  He is not alone in this; many workers do
not like the idea that information can actually be lost to a black
hole, and would prefer some structure where the information can be
recovered from the boundary of the hole.  

It is rather difficult to see how the strong form of the black--hole
holographic principle could
be compatible with any classical description of physics within the
hole, since it
seems to be at odds with the principle of freely specifying Cauchy
data on the interior of $\Sigma$.  
(Note that this point applies even when we take into
account limitations in precision of specification of the data.)  
Indeed, attempts to explicate the physics of this strong form have led to
the notion of \it
black hole complementarity \rm  ('t Hooft 1993, Susskind et al 1993), which 
asserts that quantum measurements within the hole \it interfere \rm with those
at late times outside of it, in a manner contrary to that required by a
classical causal structure of the hole.  
(In section 9.1.4, we shall see that 't Hooft is led to explicitly
quantize the causal structure.)
This is in accord with the
``no quantum photocopy machine'' principle:  if at the quantum level the
information contained in the infalling matter \it is \rm
recoverable from the boundary, it \it cannot \rm be encoded in the interior.

While highly speculative, and apparently at odds with classical
general relativity, the notion of black hole complementarity deserves
serious thought.  Since both black holes (through their causal
properties) and quantum theory (through complementarity
and measurements) profoundly
affect the transmission of information, it is worth considering
the possibility that they modify each other.  
Not just phenomena, but the laws of physics may be deeply altered in
the presence of black holes.

To summarize:  
\it General \rm holographic principles, that is, ones 
extending beyond the black--hole case, would potentially be of
far--reaching importance, but it is not yet clear that they can even be
stably formulated.  If the \it strong black--hole \rm
holographic principle applies, as envisioned in 't Hooft's treatment,
the interior of the hole seems to acquire an essentially quantum
structure and may not even be approximable by a classical theory.

\subsubsection{The trans--Planckian problem}

There is not yet a clear resolution of the trans--Planckian problem
within 't Hooft's theory, although there have been some interesting
suggestions.

The \it brick wall model \rm corresponds to quantizing the field in a
regime $r\geq (1+\epsilon) R_{\rm Sch}$, where $\epsilon$ is chosen
large enough that no trans--Planckian modes appear.  It is thus quite
similar to Jacobson's (1993) timelike boundary condition (section 8.1).  
It should
be noted that the actual quantization used by 't Hooft corresponds not
to the Hawking model, but to the \it Boulware \rm (1975) vacuum (a
time--symmetric state defined in the exterior region only of the
Schwarzschild black hole).
As in the case of Jacobson's timelike boundary, the brick wall represents an
ad--hoc model and is not meant to be a final theory.

It has also been suggested that the gravitational interactions of
quanta in the vicinity of the horizon are strong enough to modify the
theory and provide an effective ``brick wall'' which serves as a
trans--Planckian cut--off.  
One might say the quanta are supposed to build their own brick
wall.
This idea, as presently conceived, seems
to require a radical alteration of the conventional 
Hawking picture.  This
alteration appears in a model calculation as a large,
unconventional stress--tensor near the horizon ('t Hooft 1997).  
An appeal to black--hole complementarity is made to justify this
stress--energy.

It is
possible to give a more direct understanding of the issue.
First note that,
in the conventional picture, the excitations that
appear to distant observers as Hawking quanta do not appear as
excitations at all near the horizon, but as combinations of vacuum
fluctuations. 
This is simply because the Hawking process after all converts any
state which is Hadamard at the horizon to thermal radiation.
As one follows a Hawking quantum backwards in time, there comes a
point where the curved--space
difficulties in defining what a ``particle'' is become
significant, and the quantum would look to observers not like a single
particle, but like a more complicated field excitation.  On
dimensional grounds, this must happen for $r\sim 3GM/c^2$.  Following
the quantum backwards much further, it must become simply a
combination of vacuum fluctuations. 

The proposal  that gravitationally--interacting
quanta should ``build their own brick wall'' requires energetic 
real quanta in
the neighborhood of the horizon.  If these are to be the Hawking
quanta, then one must modify the
conventional theory so that the quanta are real particles near the
horizon.  The modified stress--tensor suggested in 't Hooft (1997) can
be viewed as the result of just such a change.

On the other hand, it should be noted that up to this point the
discussion has left aside what actually happens when quantum
measurements of Hawking quanta are made.  This neglect is customary in
analyses of the Hawking effect but is not really justified.  When a
measurement of Hawking quanta is made, the state vector is reduced,
and near the horizon this does result in a complicated excitation of
ultra--high frequency modes.  However, the fact that the notion of
what a ``particle'' is alters as one goes from a neighborhood of
$\scri ^+$ to a neighborhood of the horizon means that this excitation
is not simply the presence of a high--energy quantum.  These points
will be discussed more fully in Helfer (2003).

\subsubsection{The $S$--matrix theory}

One of the goals of 't Hooft's program is a theory which explicitly
preserves purity.  As we saw above, the apparent destruction of purity
comes about in the conventional picture by ignoring the internal
black--hole modes.  In 't Hooft's theory, it is anticipated that this
picture will be modified by the strong form of the holographic
principle; the details of this have yet to be worked out.
However, 't Hooft develops the
theory by \it assuming \rm that it can be written in $S$--matrix form,
and then deducing properties of the $S$--matrix.

The main idea may be outlined as follows.  In order to avoid
difficulties with bound states, it is assumed that the in-- and
out--states contain only asymptotically free particles.  Thus a
fiducial in--state $|$in${}_0 \rangle$ might look, in the distant
past, like a set of particles whose trajectories will lead them
eventually to implode and form a black hole.  A fiducial out--state
$|$out${}_0 \rangle$ would contain all the particles emitted after the
final evaporation of the black hole via Hawking radiation.  Note that
this implicitly involves quantum--gravitational assumptions, since it
necessarily involves the Planck--scale physics in the very last,
explosive, stage of the evaporation.

Evidently, actually developing the theory in this form would 
require studying the black hole over its entire existence, and also
addressing difficult quantum--gravitational issues.  This would be too
hard, at least initially, and what has been done so far corresponds rather
to studying the Hawking process over periods of time in which at least
up to quantum corrections the black hole can be taken to be static.

The idea is to start from some (unknown, but finite) amplitude
$\langle$out${}_0|$in${}_0\rangle$ and perturb both the in-- and the
out--states by adding or deleting particles from them in order to work
out an $S$--matrix (up to a phase, presumed irrelevant).  In practice,
this is done by assuming the matrix (operator, really) factors as
$$S=S_{\rm out}S_{\rm hor}S_{\rm in}\, ,\eek$$\xdef\sfact{\the\EEK}%
where $S_{\rm in}$ represents the propagation of particles inwards
from infinity towards the horizon, the factor $S_{\rm hor}$ is the
scattering near the horizon, and $S_{\rm out}$ represents propagation
outwards from the horizon to infinity.

Of course, the factorization (\sfact ) can at best only be
expected to be approximate and only to apply to one sector of the
theory.  Making these limitations precise is highly non--trivial, and
is related to some deep questions, to be taken up below, about whether
any of these factors should in fact be expected to be unitary.

Each of the three factors should ultimately be important to 't Hooft's
program.  The factor $S_{\rm in}$ contains the information of the
collapsing matter, and so presumably much of the treatment of the
quantum--coherence issues should be bound up with understanding it.
However, some ingoing particles can be considered in the factor
$S_{\rm hor}$, and so one can get a sort of perturbative handle on the
issue by studying $S_{\rm hor}$.  Finally, a key step in the Hawking
process is the propagation of field modes outwards from a neighborhood
of the horizon, accomplished here by $S_{\rm out}$.

Of the three factors, the program has so far been concerned with the
middle one, $S_{\rm hor}$.  't Hooft
considers the addition or deletion of a number of particles near
the horizon, and the way that an incoming particle might distort the
gravitational field and so alter the trajectory of an outgoing
particle.\fnote{In this computation, the particles are approximated as
pointlike; more properly, their finite Compton wavelengths are
neglected.}  He is able to deduce a great deal of the form of $S_{\rm
hor}$ based on this physics.  This is of
considerable interest as a model
of possible quantum--gravitational corrections.

The way in which the computation of $S_{\rm hor}$ really speaks to the physics
of the Hawking process needs to  be elaborated.   As discussed above, the
Hawking quanta arise, not from ordinary particles near the horizon, but from
combinations of high--frequency vacuum fluctuations there.   These, when
propagated outwards,  give rise to the observed Hawking particles.  On
dimensional grounds, one would expect the concept of a Hawking quantum as a
quantum--mechanical particle to become valid for $r\gtrsim 3GM/c^2$; much
closer to $r=R_{\rm Sch}$, it should  be a combination of vacuum
fluctuations.  

Whether the Hawking quanta appear as particles of as  combinations of
vacuum fluctuations (or as some intermediate) to $S_{\rm hor}$ depends
then
on at what distance from the horizon the transition from $S_{\rm hor}$
to $S_{\rm out}$ is made.  The analysis which has been made in 't
Hooft's program
presumes the Aichelburg--Sexl metric form is valid, which essentially
means one is much closer to the horizon than to $r=3GM/c^2$.  Thus it
seems that the transition must be made close to $r=R_{\rm Sch}$, and
this suggests that the Hawking quanta are actually created, at least
within the conventional picture, by the factor $S_{\rm out}$.  It would
simply be the Hadamard form of the state in the $S_{\rm hor}$--sector
which is the seed for their creation.  

To resolve this concern about exactly how
$S_{\rm hor}$ contributes to the
Hawking process, one would need a fuller treatment of it together with
$S_{\rm out}$, keeping careful track of the precise hypotheses on the
transition from one regime to the other.
Alternatively, perhaps the sorts of modifications to
Hawking's theory made in the ``quanta building their own brick wall''
model discussed above might be invoked.

One final, important, aspect of 't Hooft's $S$--matrix computation
should be brought out.  As mentioned above, the computation explicitly
takes into account the distortion of space--time geometry caused by
incoming particles.  This led 't Hooft to introduce the idea of
quantum operators representing the location of past and future
horizons, and to start to consider the consequences of quantum
complementarity for measurements of these operators.  While these ideas
are very speculative, the profound
importance of addressing such issues --- how
quantum behavior might alter the global causal structure of
space--time --- should be clear.

\subsubsection{Purity and unitarity}

A major goal of 't Hooft's program is a theory which is unitary.  This
is motivated partly by a desire to make black holes look just like any
other quantum--mechanical particles, and partly by a desire to
preserve purity of quantum states.  While it is possible that this
will be achieved, there are subtleties in understanding just
what this would involve.  While unitarity would imply the preservation
of purity, the converse need not be true, and there are questions
about whether (or to what approximation) a unitary theory can be
constructed even for the factored $S$--matrix approach.

We shall adopt the conventional Heisenberg picture, so that the state
vector remains fixed (except when reductions occur) and the fields are
functions (operator--valued distributions, really, on space--time).  As
long as this picture is applicable, preservation of purity is
automatic.  Let us turn to unitarity.

In ordinary quantum mechanics, the evolution of a set of canonical
variables $(p_j,q_j)$ from one time to another \it must \rm be
unitarily implementable, that is, there must be a unitary $U$ such
that
$$U\left[\matrix{p_j\cr q_j\cr}\right] _{t_0}U^{-1}
  = \left[\matrix{p_j\cr q_j\cr}\right] _{t_1}\, .\eek $$
This is a consequence of the Stone-von Neumann theorem:  for finitely
many degrees of freedom, as long as the canonical commutation
relations are preserved, such a unitary $U$ must exist.

However, for field theories, which have infinitely many degrees of
freedom, the situation is more complicated.  Evolution does preserve
the canonical commutation relations, but it need not be unitarily
implementable.  Indeed, if the evolution is represented as a
Bogoliuobov transformation, there is a well--known square--summability
criterion for unitary implementability (Wald 1994).

In certain restricted cases, one can argue on physical grounds that
one expects unitarily implementable evolution.  For example, for \it
any \rm Poincar\'e--invariant quantum field theory, the evolution from
a $t=\const$ to a $t'=\const$ surface (with $t$ and $t'$ inertial time
coordinates) should be given by a Poincar\'e motion, which one would
expect to be represented by a unitary operator.  (Indeed, this is
almost a definition of what one would mean by a Poincar\'e--invariant
quantum field theory.)

It is something of a surprise to find out, though, that unitary
implementability is the exception rather than the rule.  In Helfer (1996), for
quantum fields propagating through a general curved space--time, it was shown
that the ``sum of squares'' to be computed in the Bogoliubov criterion is an
integral of products of Green's functions and two--point functions.  For an
explicit non--pathological class of cases this was computed and found to
contain ultraviolet divergences. These divergences will also be present
generically, on account of the smooth dependence of the Green's and two--point
functions on the structure.

Physically, these divergences come about 
because the two--point functions have different
asymptotics on different Cauchy surfaces.  As
mentioned earlier, reasonable physical states are \it Hadamard, \rm
that is, their ultraviolet two--point asymptotics are modeled on
$-(4\pi )^{-1}(p-q-\i\epsilon )^{-2}$ in Minkowski space.  Now in
curved space--time, this asymptotic form is of course expressed in
terms of the metric distance between two events $p$ and $q$.  In
particular, then, the two--point function restricted to a Cauchy
surface encodes the surface's 
intrinsic metric.  That is, the geometry of the
Cauchy surface can actually be recovered from the ultraviolet
asymptotics of the two--point function.  This means that if the two
Cauchy surfaces are not isometric, the two--point functions on them
will have different ultraviolet asymptotics.  Finally, the
``sum of squares'' in the Bogoliubov condition is essentially an
integral over the square of the difference of the two--point function
between the two surfaces (times a wave--vector space integral).  
Since in general two Cauchy surfaces are not isometric, this
difference cannot be made to vanish.
And it is found that the integral in question diverges
generically.  

Another way of looking at this result is that the two--point function
is a correlation function between vacuum fluctuations.  Since the
ultraviolet asymptotics of this
correlation function, restricted to a Cauchy surface, allow one to
recover the geometry of the surface, the physics represented by this
ultraviolet asymptotic regime is essentially different on one surface
from another (unless they are isometric).  This is why the evolution
is not unitarily implementable except in exceptional cases.

It should be noted that the fact that one has a clear--cut way of
isolating the divergences means that one could introduce a cut--off;
the cut--off theory would be unitarily implementable.  This also
means that the theory is \it formally \rm unitarily implementable; the
difficulty is in taking the cut--off to infinity.
 
This issue of non--unitarity is not really fully understood.
However, there are some hints of how to resolve it.

First, while the non--unitarities are most evident in the case of
quantum fields in curved space--time, they also occur (but with lower
divergences) for quantum fields propagating through general
time--dependent external potentials in Minkowski space (Helfer
1996).\fnote{The explanation given above in terms of intrinsic metrics
does not apply, of course, but it turns out that the subdominant
ultraviolet asymptotics encode the external field.}  
We shall consider charged quantum fields moving in a time--dependent
external electromagnetic potential.

For these charged fields, there
is a reasonable, conservative
speculation as to what the resolution of the non--unitarity problem is.
This is that the
non--unitarities result from the neglect of quantum fluctuations in
the electromagnetic field, and a full quantum--electrodynamic
treatment (if this could be really achieved) would indeed be unitary.

If a similar resolution holds for gravitational fields, then it would
appear a quantization of gravity would be necessary to restore
unitarity.  On the other hand, it might be that quantum gravity cannot
be understood without a deeper treatment of the physics of reduction
(see e.g. Penrose 1986); so perhaps a quantum gravity theory should
not have unitarity as a goal.

In particular, the analysis above suggests that the factors $S_{\rm
in}$, $S_{\rm hor}$, $S_{\rm out}$ in 't Hooft's program
cannot be exactly unitary without a quantization of gravity.  Some
such quantization is implicit in 't Hooft's program, and, as we saw,
progress towards it has been made in the case of $S_{\rm hor}$.

Finally, while a basic tenet of 't Hooft's program is that ultimately a
unitary theory should be attainable, we have seen here that unitarity
is a stronger condition than preservation of purity.

\subsection{Summary}

't Hooft has proposed a radical and ambitious program; to execute it
successfully will require deep modifications of conventional quantum
field theory, and the incorporation of essentially
quantum--gravitational ideas.  The program is still in a stage of
development, and it is not yet clear that it will meet its goals.
It does not at present provide a clear resolution of the
trans--Planckian problem.

The program has however proved valuable in raising deep questions
about the possibility of modifying quantum field theory, and also for
providing models for possible such modifications and for
quantum--gravitational effects.  While these models may ultimately not
turn out to be correct, innovative ideas such as these are
of the greatest importance in tackling the very difficult problems involved.

\section{Evidence from theories of quantum gravity}

In this section, I shall review the evidence for the Hawking effect from
theories of quantum gravity.  Of course, all of this work is speculative, but
it is very important in trying to understand what might happen beyond the
classical treatment.

Virtually all work discussed here is quite technical, and no attempt to present
the details of arguments will be given; the emphasis will be on the basic
physical assumptions and the results.

\subsection{Dilatonic black holes}

These are meant to be model theories of the s--wave sector of quantum gravity
coupled to matter fields.   (Thus in particular the radius of any sphere of
symmetry becomes an operator, and it is essentially this which is the ``dilaton
field.'') They are non--linear field theories. Their forms are motivated by
formally integrating out the angular dependence of Lagrangians which are
supposed to represent the theories.  This involves a number of conceptual and
technical problems, most especially related to the diffeomorphism invariance. 
The results therefore are not unique, but in each case represent the workers'
best opinions of how to resolve these difficulties.  See Grumiller et al.
(2002) for a recent review.

Dilaton models which are semiclassical perturbations of Schwarzschild (coupled
to scalar matter) have been investigated and have been able to reproduce the
Hawking radiation (Kummer and Vassilevich 1999),  but the freedoms in defining
the theories in light of our present ignorance are large enough that not all
admissible theories do produce Hawking radiation. One can even have negative
luminosities (Balbinot and Fabbri 1999).  

Since conventional semiclassical quantum field theory in curved space--time \it
does \rm predict Hawking radiation, how can even some semiclassical dilaton
models fail to predict it?  One way of phrasing the answer is there are
difficulties in ensuring that dimensional reduction and renormalizations (or
well--definition) of the theory commute.   The underlying physical question
seems to be that even though one ``integrates out'' many degrees of freedom,
the vacuum fluctuations from these degrees might leave a non--trivial
modification of the semiclassical theory. Our limited understanding of this
should serve as a caution of the need for confronting the  subtleties involved
a correct treatment of the physics.

\subsection{TTFKASS}

M--theory (the theory formerly known as superstrings) has been able to
reproduce the Bekenstein--Hawking entropy formula $S=A/(4l_{\rm Pl}^2)$ for
``near--extremal'' black holes (see Peet 2000 for a review).   These are a
restricted class of holes, with charge very nearly equal to their mass (and
surface gravity very nearly zero). 

The argument for this is a mode--counting one involving one of the remarkable
dualities discovered in the theory.  It is thus a serious candidate for an
explanation of the origin of  black--hole entropy (at least for near--extremal
holes).   (Recall that the Hawking process does not do this.)

While the mode--counting result is clear, it is difficult to follow the details
of the physics through the duality and understand what a black hole would
really look like in this case.  Would Hawking radiation be predicted only in
collapse situations or for all black holes?   What would be the resolution of
the trans--Planckian problem?  (The near extremality --- that is, nearly zero
$T_{\rm H}$ --- makes understanding this harder.)  These questions are, at
present, unresolved.

\subsection{Ashtekar's approach}

Ashtekar's approach to quantum gravity, a Hamiltonian one based on his ``new
variables,'' is much more obviously connected to general relativity than is
M--theory.  Nevertheless,  one still has both conceptually and technically
difficult tasks in trying to model the Hawking process, even in idealized ways,
in this theory.  

Remarkably, however, workers using this approach were able to push through a
mode--counting argument to reproduce the relation $S\propto A$ (although not to
fix the constant of proportionality).  See Ashtekar et al. (2000), and
references therein.  This result holds not just for extremal black holes, but
for any spherically--symmetric hole large compared to the Planck scale.  

It should be pointed out that in this approach there is a certain ad--hoc
character to the treatment of the horizon.  (This is a Hamiltonian approach,
and so data must be given on a Cauchy surface.  However, the location and even
the existence of an event horizon are in general difficult to determine from
Cauchy data, as they are defined in terms of the global asymptotics of the
space--time. Thus a full treatment would require, at least implicitly,
information about the global evolution of the data --- clearly, such a
treatment would be unrealistically difficult.)

One would hope it is possible to understand the resolution of the
trans--Planckian problem within this approach, but this seems not yet to have
been treated in detail.  At the least,  from the general nature of the
construction, one would expect each Hawking quantum to become entangled with a
quantum perturbation of the horizon.

\subsection{Euclidean quantum gravity}

Possibly the earliest quantum--gravitational approach to verifying the Hawking
mechanism was undertaken within the Euclidean quantum gravity program.  Gibbons
and Hawking (1977) evaluated the partition function
$$Z={\rm tr}\, \exp -\beta H\eek$$\xdef\pfun{\the\EEK}%
in the one--loop approximation, and were able to reproduce the
Bekenstein--Hawking formula for the entropy.

There is a question of the consistency of this scheme, however, which also
brings up a more general problem (Wald 1994).  If black--hole entropy can
indeed be interpreted as the logarithm of the number of available states, then
the density of states must grow spectacularly quickly ($\sim\exp 4\pi (M/m_{\rm
Pl})^2$) with the mass of the hole.  This makes it very hard for thermodynamic
averages like (\pfun ) to converge.

\section{Quantum character of space--time}

In the previous section, I discussed the consequences, for the Hawking analysis,
of assuming particular theories of quantum gravity.  But there is another
approach to understanding quantum--gravitational effects, where rather than
hypothesizing a particular theory of quantum gravity, one considers general
properties of quantum theory and studies how they might be integrated with
general relativity.  While there is not a sharp line between the two approaches,
the latter one does tend to approach questions more directly physically than
mathematically.  Given the highly speculative nature of any specific theory of
quantum gravity, the complementary character of the second approach is
especially valuable.

\subsection{Validity of the semi--classical approximation}

I will begin, not by describing specific possible quantum--gravitational
issues, but by discussing a popular fallacious argument to the effect that one
can neglect such concerns in modeling the Hawking process, except perhaps in
accounting for the change of the hole's mass over very large times.  I'll
recount it, challenging the alert reader to spot the difficulty.

The argument goes like this.  Everyone agrees that the Hawking effect is a very
small perturbation of an essentially classical system, the gravitational
collapse of a large body.  So to first approximation, it ought to be perfectly
legitimate to treat the quantum field as simply ``painted on'' to the classical
background.  Only over times for which the emission of Hawking quanta can
significantly perturb the geometry of the collapsing object should this
treatment break down.

Here is a parallel argument about another physical system.  Consider the
emission of a quantum of radiation by an atom embedded in a crystal.  Since
everyone agrees the crystal is essentially a macroscopic system, it ought to be
sufficient to treat the crystal classically, and model the emitting atom
quantum--mechanically.  Only for very fine measurements of (for instance) the
momentum of the crystal would it be necessary to go beyond this.

The argument about the crystal is an argument against the M\"ossbauer effect,
and so is specious.   It is correct insofar as it asserts that the
back--reaction of the emission on the crystal only makes the tiniest correction
to the macroscopic state of the crystal. It errs crucially in neglecting the
fact that \it the coupling between the emitting atom and the crystal has a
quantum character, \rm and one can only get the physics right by taking into
account the existence of quantized phonon modes in the crystal.  That is, the
coupling of the atom to the crystal \it constrains \rm the allowed transitions
of the atom; it restricts the allowable matrix elements for the transition of
the atom.  Neglecting the coupling, the atom may make transitions to a
continuum of recoil momenta; taking the coupling into account, most of these
transitions are forbidden.  What forbids them is conservation of momentum, of
the atom plus crystal system, at the level of quantum operators.  Thus some
quantum characteristics of the crystal enter essentially.

While the analogy between the Hawking process and the M\"ossbauer effect cannot
be made a very close one, the point being made here is simply that when a large
system (the collapsing object; the crystal) couples to a small quantum one (the
quantum field; the emitting atom), one needs to consider whether the coupling
will lead to constraints on transitions.

\subsection{Quantization of black--hole area}

It is certainly credible that quantum gravity should act to quantize a black
hole's area.  Indeed, such a feature would attractively unite the
interpretation of the black hole's area with entropy with the
information--theoretic interpretation of entropy.  So let us consider the
consequences of such a relation.  This argument follows work of Bekenstein and
Mukhanov (1995), the importance of which was emphasized by Ashtekar (1998).
(For a list of works on area quantization, see Das et al 2002.)

Let us suppose for simplicity that the black hole's area is quantized in units
$\alpha l_{\rm Pl}^2$, where $\alpha$ is a numerical constant.  Now, when the
black hole emits a quantum of energy $\Delta E$, its area changes by $\Delta
A\simeq 32\pi  (G^2/c^6)M\Delta E$, and so we must have
$$n\alpha l_{\rm Pl}^2=32\pi (G^2/c^6)M\Delta E\, ,\eek$$
or
$$\eqalign{\Delta E&\simeq (32\pi )^{-1}n\alpha (l_{\rm Pl}^2c^6/G^2)M^{-1}\cr
  &=n(\alpha /4) kT_{\rm H}\, .\cr}\eek$$
In other words, the energies of the Hawking quanta would be discrete multiples
of $(\alpha /4)kT_{\rm H}$.  If $\alpha$ were around unity, this would lead to a
line spectrum, rather than a continuum.  If $\alpha$ were much smaller than
unity, the spectrum would be quasi--continuous.  But if $\alpha$ were larger
than about $10$, almost all Hawking radiation would be suppressed, since there
would be few Hawking quanta energetic enough to allow the requisite transition
of black--hole states.

Any discretization of the spectrum will tend to increase its
information--content and therefore to decrease its entropy.  However, this does
not affect the Bekenstein--Hawking formula and thus one expects that some
discretizations of the spectrum may lead to violations of the generalized
second law.  This has been confirmed by computations of Hod (2000).

Of course, one need not consider simply a quantization of the area in integral
multiples of $\alpha l_{\rm Pl}^2$;
there are other possibilities.  The real point of the investigation, from our
point of view, is to show that quantum--gravitational corrections which could
very plausibly be present on dimensional grounds could considerably alter
not just Hawking's predictions, but the conjectured connection of it with
thermodynamics.

Finally, it should be noted that these sorts of concerns are \it beyond \rm the
trans--Planckian problem.  The trans--Planckian problem is a difficulty \it
within \rm the semi--classical approximation.  Even if the trans--Planckian
problem were to be resolved by some sort of (say) non--standard field
propagation, there is no obvious way that such a resolution would speak to the
consequences of quantizing black--hole area.

\subsection{Quantum measurement issues}

We can get another perspective on the possible consequences of a quantum
character of space--time, as follows.  

Measurements of geometrical quantities in space--time are presumably subject to
quantum limitations, just as measurements of other quantities are.  This means
that all of the calculations we do with classical general--relativistic
quantities should be regarded as approximations valid for when quantum
fluctuations are negligible.  This approximation would break down when those
fluctuations are significant, and we may try to estimate when this happens.

The key geometrical quantity that enters in the Hawking analysis is $v(u)$, the
mapping of surfaces of constant phase.  For example, the rough approximation to
the propagation we used was simply
$$\phi ^0_{0,0} (u)\Bigr| _{\scri ^+}=-\phi ^0_{0,0}(v(u))\Bigr| _{\scri ^-}\,
.\eek$$\xdef\angeq{\the\EEK}%
We are now going to admit the possibility that it may not be adequate to treat
$v(u)$ as a classical quantity to arbitrary accuracy, that it might be subject
to quantum fluctuations.  This means that $\phi ^0_{0,0}(u)\Bigr| _{\scri ^+}$
entangles the state of the field at $\scri ^-$ with $v(u)$.  
(Note that this is similar to 't Hooft's 1996 idea of quantizing the horizon,
discussed in section~10.1.4.)

Let us consider a measurement of $v'(u)$.  (It is the quantities $v'(u)$ and
$v''(u)/v'(u)$ which enter expectations of the field operators most
directly.)   The most direct way to measure $v'$ would be to send in (massless,
non--interacting) field quanta of known frequencies from $\scri ^-$ and measure
their frequencies at $\scri ^+$.  The red--shift would be $v'(u)$ (and ratios
of successive differences of this would give an estimate for $v''/v'$).  It
should be emphasized that these ``probe'' quanta have nothing direct to do with
any Hawking quanta and could be of a different species than the  Hawking quanta
to be observed.  The probe quanta  are just part of a thought--experiment to
understand what the effects of quantum limitations on accuracies of measurement
might be.

For such a measurement to be relevant to an observation of Hawking quanta over a
time interval $\Delta t$ (near $\scri ^+$), the probe quanta must have
frequencies near $\scri ^+$ of $\omega _+\gtrsim 1/\Delta t$.  However, this
means that their frequencies near $\scri ^-$ must have been $\omega
_-=(v')^{-1}\omega _+\gtrsim (v')^{-1}/\Delta t$.  Since $(v')^{-1}$ increases
exponentially quickly, we quickly pass any finite scale for $\omega _-$.  In
particular, presumably it is impossible to have $\hbar\omega _-> E_{\rm Pl}$,
and thus there is a fundamental limitation to the possible measurements.  If we
take $\Delta t\sim R_{\rm Sch}/c$ (the time scale for the emission of a Hawking
quantum), then this fundamental limitation is reached when
$$v'(u)\sim m_{\rm Pl}/M\, .\eek$$
This of course happens very quickly.  It is in fact the same scale at which the
trans--Planckian problem sets in.

It should be emphasized that the difficulties at this scale are potentially much
more severe than simply an entanglement of the state of the field with the
quantum state of space--time.  The difficulty is that at this point
trans--Planckian physics enters, and it is probably not meaningful to talk about
the value of $v'$ even as a quantum operator.

\section{Experimental prospects}

The real proof or disproof of Hawking's predicted radiation, of
course, would be experimental.  It is not impossible that we shall one
day have experimental results one way or the other.  The main
difficulty is that, at present, there are no known black holes of the
right size that we might observe Hawking radiation from them; also,
there is the difficulty of knowing exactly what one should look for
--- one must go beyond Hawking's rather idealized linear--field model
to have an accurate physical prediction of what a Hawking--radiating
black hole would really look like.

\subsection{What sorts of black holes might exist?}

At present, two sorts of black holes are believed to have been observationally
identified:  ones of a few solar masses, and ``super--massive'' ones (with
masses $\gtrsim 10^6 M_\odot$).  There seems to be no prospect for identifying
Hawking radiation from such objects:  they would be far too cold and far too
dim.

It is very plausible that black holes might have been created in the early
Universe, from density fluctuations (and also possible that they may have formed
by other means).  These are called \it primordial black
holes. \rm  For our purposes one should note that ``primordial'' is not to be
taken absolutely; these holes (if they exist) were formed by gravitational
collapse after the big bang and should, according to Hawking's theory, radiate.

Estimates of the mass function of primordial black holes (number of black holes
of a given mass) are model--dependent.  The main constraints on the mass
function come from the assumption that black holes \it would \rm radiate, and
that too much radiation would either be directly observed or disturb other
features of the cosmological model.\fnote{Radiation from black holes can
potentially disturb nucleosynthesis, alter the cosmic microwave background, and
change the entropy--to--matter ratio.} On the other hand, if black holes do not
radiate, then the only known constraint is that they should not contribute more
than the critical density.  See Liddle and Green (1998) for a summary of
constraints.

No generally--accepted models have been put forward for the formation of  small
(sub--solar mass) black holes later in the Universe.  We should be mindful,
though, that astronomy has a tradition of surprising us.

To summarize:  If we do not assume that black holes radiate, there are few
constraints on them in cosmological models.

\subsection{What would a Hawking--radiating black hole really look
like?}

Hawking's calculations apply directly only to massless linear fields.
No such fields are believed to exist in Nature; all fields are
believed to couple to others.  How important is the assumption of
linearity, and how would the inclusion of nonlinear corrections affect
the predictions?

These are questions to which we do not have wholly convincing
answers, partly due to technical difficulties in treating nonlinear
fields, and partly due to the trans--Planckian problem.
In principle, one would have to work through the steps of Hawking's
derivation, but for a realistic non--linear field theory, and this
would not only be a massive quantum--field--theoretic undertaking, it
would involve hypotheses about what physics is like in the ultrahigh
energy regimes necessary to the Hawking process in the vicinity of the
event horizon.

However, some progress has been made, along with some very nice
physical analyses.  Essentially the starting--point for these is the
assumption that for (say) $r\gtrsim 10 R_{\rm Sch}$ one can hope that
Hawking's predictions would give rise to a conventional field theory
with a central source at the Hawking temperature.
General--relativistic corrections should be negligible, and one has a
(difficult) conventional field--theory problem to solve.

In order to get an idea of the physics and the open questions, let us take the
following, very crude, division of scales (figures are given to
order--of--magnitude only):

\itemitem{}\qquad $kT_{\rm H}\lesssim 1$ MeV.  (Energies below the threshold
for electron--positron production.)  In this case, one expects Hawking
radiation of photons and neutrinos, modeled as free massless particles.  The
corresponding black--hole mass range is $\gtrsim 10^{17}$ g.

\itemitem{}\qquad $1$ Mev $\lesssim kT_{\rm H}\lesssim 100$ MeV.  In this
range, electron--positron production, as well as non--linear
quantum--electrodynamic effects, must be considered, but strong--interaction
physics can be neglected.  The corresponding mass range is $10^{15}$ g
$\lesssim M\lesssim 10^{17}$ g.

\itemitem{}\qquad $100$ MeV $\lesssim kT_{\rm H}\lesssim 1$ GeV.  In this range,
one must take into account the production of unstable hadrons (mesons) which
will decay into leptons and photons.  The corresponding mass range is $10^{14}$
g $\lesssim M\lesssim 10^{15}$ g.

\itemitem{}\qquad $1$ GeV $\lesssim kT_{\rm H}$.  This is the scale at which
proton--antiproton production sets in.  

The first one of these is fairly well understood, but beyond that there are
questions.   The appearance of the radiation depends crucially on precisely
what sorts of self--interactions the Hawking quanta do undergo on their way out
from the hole. For example, it has been suggested (Heckler 1997, Cline et al
1999)  (although disputed:  Kim et al 1999) that for $kT_{\rm H}\gtrsim 200$
MeV quantum--chromodynamic Bremsstrahlung and other effects become important
enough for a photosphere to form.  This would reduce the observed temperature. 
Indeed, \it any \rm mechanism which resulted in (approximately) thermalizing
the final radiation at a radius $r$ would reduce the temperature by $\sim
(R_{\rm Sch}/r)^2$, since the luminosity goes as $T^4r^2$.

While the regimes above only go down to around $10^{14}$ g, this is only
because that much suffices to give a general consideration of the sorts of
physical questions that come up, and our (present) lack of ability to answer
them.  One should certainly consider the possibility of smaller (hence higher
$kT_{\rm H}$) holes.

Thus there is significant uncertainty in what we should expect for the emergent
spectrum for a Hawking--radiating black hole of given mass.  This point needs
to be considered, not only in asking what sorts of holes might be detected
observationally, but in estimates of the effects of Hawking--radiating holes in
the early Universe.  Lower--than--expected temperatures would presumably lead
to smaller effects of black holes on cosmology, and therefore to weaker
constraints of the possible number of black holes in the early Universe.

Whatever the self--interactions of the Hawking quanta are on their way out from
the hole, total luminosity (including all species)
should be given by Hawking's prediction (taking
account of the number of effective degrees of freedom at any temperature).  
These luminosities
are quite low, being
$$L\simeq 10^{20}\, {\rm erg\ s}^{-1}\, (10^{15}\,{\rm g}/M)\, .\eek$$
For comparison, the Sun's luminosity is $\sim 4\cdot 10^{33}$ erg s${}^{-1}$.
Thus one needs a great many black holes, or ones close to the Earth, for the
radiation to be detectable.

Given the uncertainties in the calculations, I will only quote two numerical
estimates, to give the reader a sense of the magnitudes involved.  Heckler
(1997) estimates that a $10^{13}$ g black hole could only be detected as a point
source by the 
GLAST satellite if it were closer than about $1.5\cdot 10^{-3}$ pc $\simeq 30$
astronomical units.  
One can also infer limits on the density of evaporating black holes by requiring
their radiation not exceed the observed gamma--ray spectrum.  
Halzen et al (1991) give such a limit
$$\Omega _{\rm pbh}
\lesssim 7.6(\pm 2.6)\cdot 10^{-9}h_0^{-1.95\pm 0.15}\, ,\eek$$
where $h_0$ is the scaled Hubble parameter (of order unity).
The reader is cautioned that both of these are model--dependent (and different
models are used) and are displayed only to give a rough sense of the scales
involved.

Work aimed at providing realistic models of what radiating would look like is a
subject of active research, with much scope for interesting physics to be
discovered.  Besides the papers already mentioned (Cline et al 1999, Halzen et
al 1991, Heckler 1997), see Daghigh and Kapusta (2002), Kapusta (2000).
A brief summary from the point of view of experimental prospects 
occurs in section II.F of the Snowmass report (Buckley et al 2002).

\subsection{Summary}

To confirm the existence of Hawking radiation, we would need (not only
observations), but reliable models of what radiating black holes would actually
look like.  We are at present limited in our understanding of how to construct
such models, as they require the study of non--linear, realistic, quantum field
theory at high temperatures.  This is a very difficult technical problem, but it
does not require the resolution of foundational issues.

To show that black holes did not radiate, we 
would need to detect the holes, or to
have strong indirect evidence of their existence, as well as to show that they
do not radiate.  At present we are far from doing this.  The possibility should
be taken seriously, however.  Black holes are attractive dark--matter
candidates.  Primordial black holes present as halo objects could conceivably be
detected by microlensing (Minty et al 2001).  If one could get a lower bound on
the number of black holes in this or a similar way, and get an incompatible
upper bound from gamma--ray observations or cosmology, one might be able to rule
out black--hole radiation experimentally.

\section{Conclusions}

The introduction to this paper listed six main conclusions, which are
recapitulated in the appendix following, along with references to the
places in the text which support them.  The present section gives a
more narrative account.

It seems unlikely we will have a convincing theoretical answer to the question
``Do black holes radiate?'' soon.  General dimensional
arguments, like that of Bekenstein and Mukhanov (1995), show that
quantum--gravitational effects may enter essentially, so, without an accepted
theory of quantum gravity (at least in the regimes necessary for treating the
Hawking model), we cannot expect a definitive resolution.

While we may not come to an answer to this question soon, we may expect to
learn a great deal from theoretical investigations, especially those attempting
to model as explicitly as possible physical processes related to the Hawking
mechanism.  The work of Unruh (1976) and of Unruh and Wald (1982) is an example
of how much can be learned by making models of the elements of processes which
cannot, in the present context be taken for granted.  
For such models to contribute to resolving the quantum--gravitational issues,
they will presumably have to take up issues involving limitations on the
measurability of space--time quantities.

More work needs to be done to understand the models with non--standard
propagation of the quantum fields (such as Unruh's dispersive propagation).  We
need to know whether such models could be used to provide a solution of the
trans--Planckian problem (at present, they address the problem of
trans--Planckian frequencies but not that of trans--Planckian wave--numbers). 
We also need to understand just what aspects of these models are essential to
producing thermal spectra.  Also an open question is how quantum--gravitational
concerns (like the one raised by the Bekenstein--Mukhanov analysis) might be
resolved by such models.

The connection between black holes and thermodynamics remains inadequately
understood.   The Hawking process would explain the temperature of a black hole
but not, of itself, explain the nature of black hole entropy. Just what black
hole entropy is, and how it relates to other notions of entropy, remain
matters of speculation.  The main attempt to link the two is
the generalized second law.  There are some questions about the motivation for
this, as well as difficulties in formulating it precisely, and there is not at
present agreement about whether one needs to understand limitations on the
entropy--content of matter in order to have an adequate treatment.

We cannot verify the generalized second law in its strongest form for the
Hawking process, because the accounting of entropy which would be required is
extremely fine. (The entropy typically changes by about one unit when a Hawking
quantum is emitted, yet the absolute entropies involved are very large, and
indeed, divergent in the case of the quantum field before regularization.) With
some apparently reasonable approximations, however, the Hawking effect has been
shown to respect the generalized second law.  There are also reasonably good
indications that the quantum Geroch--Wheeler process will respect the
generalized second law, although this is a case where it seems a more delicate
physical analysis is needed to settle the question.

Quantum--gravitational models have had mixed results in supporting Hawking's
predictions.  Some dilaton models have confirmed them, but others have not. 
The Bekenstein--Hawking entropy formula has been confirmed for near--extremal
black holes by a mode--counting calculation in string theory, but so far that
theory does not provide a detailed model in which one can see that the holes do
radiate.   (Extremal black holes have zero temperature.) Ashtekar's ``loop
quantum gravity'' program has established the general relation $S\propto A$ for
spherically symmetric holes, again by a mode--counting argument.  One would
hope this program could be developed to the point where the emission of quanta
and the resolution of the trans--Planckian problem understood.

A major problem is to understand what a radiating black hole really would look
like (under the assumption that Hawking's mechanism is correct). 
Self--interactions in the quanta could lead to (at least partial)
thermalization at large (compared to $R_{\rm Sch}$) distances from the hole,
lowering its temperature considerably; there is also the question of reliably
computing how many particles of each species are produced. This is a difficult
problem involving the dynamical behavior of non--linear realistic field
theories at high temperature.  But only when we have an idea of what the answer
is will we be able to assess any experimental evidence.  The answer to this
problem is also important insofar as the Hawking mechanism is used to constrain
the numbers of black holes in cosmological models.  

While this review has been limited to black--hole physics, one should bear in
mind that closely related issues arise in inflationary theories (e.g.
Brandenberger 2002); one could hope for some cross--fertilization of ideas.

From an astrophysical point of view, the limitations that have been placed on
cosmological models by assuming that black holes do radiate should be
considered tentative, and the consequences of not making this assumption should
be given comparable weight.  In particular, serious consideration should be
given to the possibility that ``mini'' black holes may contribute to the dark
matter in the Universe.

Astrophysical studies constraining the possible number of ``mini'' black holes
by means independent of the assumption that they radiate would be most welcome.

\ack

I am grateful to many colleagues for their comments on aspects of these ideas. 
I would like particularly to mention Jacob Bekenstein,
Steven Carlip, Robert Geroch, Theodore Jacobson, Don Page, William Unruh and
Robert Wald.  Needless to say, the perspective and judgments in this
paper are my own.

\Appendix{Text passages supporting the conclusions}

The introduction listed six main conclusions of this paper.  This
appendix contains, for each conclusion, a list of the places in the
text where the arguments supporting it are given.

\smallskip
\it (a) All derivations of Hawking radiation involve
speculations of what physics is like at trans--Planckian scales.  \rm

For the Hawking model, see section 3, especially 3.5.2 (how trans--Planckian
field modes enter), and section 4 (the trans--Planckian problem generally).
For the responses to
arguments based on the Unruh process, see section 5.2 (the arguments
need as input the ultraviolet behavior of the Hawking state).  
For models with non--standard propagation, see 8.2.2 and 8.4
(ultra--high wave--numbers enter essentially).

Moving--mirror models and 't Hooft's model do not really \it derive \rm Hawking
radiation; for comments on these, see the introduction to section 6, and 9.1.3.

Of course, arguments bases on theories of quantum gravity (section 10)
all involve, at least implicitly, speculations about Planck--scale
physics.

\smallskip
\it (b) There are equally plausible speculations about physics at such
scales which result in no radiation at all, or in non--thermal
spectra.  \rm

There is some question here of when two speculations are equally
plausible, of course.  What I list here corresponds to ideas which
either have been taken seriously by respected workers, or fall within
a reasonable conservative physical judgement.

See the last paragraph of 3.5.2 (a textbook--style cut--off theory would
lead to no Hawking radiation), the last paragraph of 8.1 (Jacobson's analysis
led him to think there might be modifications to the Hawking spectrum), section
10.1 (some dilaton models reproduce Hawking radiation, some do not), 
section 11.2 (quantization of black--hole area could alter or obviate the
production of Hawking radiation).

\smallskip
\it (c) The various derivations which have been put forward are not all
mutually consistent.  \rm

Two models which could in principle be experimentally distinguished will be
considered inconsistent (although they may agree in some regimes, e.g., they may
both produce the Hawking spectrum at large distances).

The models with non--standard propagation (section 8) evidently differ
essentially from the original Hawking model (section 3).  't Hooft's
model probably differs essentially from Hawking's near the horizon
(section 9.1.3, cf. also section 9.1.4).

\smallskip
\it (d) Quantum--gravitational corrections are very plausibly of a
size to alter or even obviate the prediction of thermal radiation.
\rm

Note that the assertion is not that quantum--gravitational corrections \it do
\rm alter Hawking's prediction, but that they are of a size where they plausibly
\it might.  \rm

See the end of section 6.2 (moving--mirror models
suggest quantum--gravitational corrections might be significant),
section 10.1 (dilatonic models might or might not reproduce Hawking radiation,
depending on how they are tweaked), and section 11, especially 11.2
(quantization of black--hole area could lead to alteration of Hawking's
predictions).

\smallskip
\it (e) A number of arguments which have been put forward in support
of the Hawking mechanism are not really direct evidence for the
existence of thermal radiation, but rather are arguments for
interpreting black holes' areas as entropies.  \rm

See the introduction to section 7 (black--hole thermodynamics generally),  and
sections 10.2--10.4 (mode--counting arguments from quantum gravity theories).

\smallskip
\it (f) The proposed mechanism, at least as conventionally understood,
relies precisely on the assumption that quantum--gravitational effects
can be neglected, and so no deep test of quantum gravity can emerge
from it.  \rm

The second part of this assertion follows from
the first.  For the first part, the Hawking model's classical
treatment of space--time at all scales is emphasized in section
3.5.2.  Discussion of the neglect of quantum--gravitational effects
will be found at the end of section
6.2 (parallels suggested by moving--mirror models)
and especially in sections 11.1 (limits to the validity of semi--classical 
treatments), 11.3 (possible quantum--measurement theoretic limitations).

\references

\refjl{Amelino--Camelia, G 2002}{Int J Mod Phys}{D11}{35--60 gr-qc/0012051}

\refjl{Amelino--Camelia, G, Ellis, J, Mavromatos, N E, Nanopoulos, D V and
Sarkar, S 1998}{Nature}{393}{763--765 astro-ph/9712103}

\refbk{Ashtekar, A 1998}{The geometric universe:  science, geometry and the 
work of Roger Penrose}{(Oxford: University Press), eds. Huggett, S A, Mason, L
J, Tod, K P, Tsou, S T and Woodhouse, N M J, pp~173--194}

\refjl{Ashtekar, A, Baez, J C, and Krasnov, K 2000}{Adv Theor Math
Phys}{4}{1--94}

\refjl{Balbinot, R and Fabbri, A 1999}{Phys Lett}{B459}{112}

\refjl{Bekenstein, J D 1972a}{Phys Rev}{D5}{1239}

\refjl{\dash 1972b}{Phys Rev}{D5}{2403}

\refjl{\dash 1973}{Phys Rev}{D7}{2333--2346}

\refjl{\dash 1975}{Phys Rev}{D12}{3077--3085}

\refjl{\dash 1982}{Phys Rev}{D26}{950}

\refbk{\dash 1994a}{The Seventh Marcel Grossmann Meeting on recent 
developments in theoretical  and experimental general relativity, gravitation,
and relativistic field  theories:  Proceedings}{(World Scientific), eds Robert
T Jantzen, G Mac Keiser, and Remo Ruffini, pp~39--58}

\refjl{\dash 1994b}{Phys Rev}{D49}{1912}

\refjl{\dash 1999}{Phys Rev}{D60}{124010}

\refbk{\dash 2002}{in Advances in the interplay between quantum and gravity
physics}{eds P G Bergmann and V de Sabbata (Kluwer), pp~1--26, gr-qc/0107049}

\refjl{Bekenstein, J D and Mukhanov, F F 1995}{Phys Lett}{B360}{7--12}

\refjl{Belinski, V A 1995}{Phys Lett}{A209}{13--20}

\refbk{Birrell, N D and Davies, P C W 1982}{Quantum fields in curved
space}{(Cambridge:  University Press)}

\refjl{Bisognano, J J and Wichmann, E H 1976}{J Math Phys}{17}{303}

\refjl{B\l{}aut, A, Kowalski--Glikman, J and Nowak--Szczepaniak, D 2001}{Phys
Lett}{B521}{364--370 gr-qc/0108069}

\refjl{Bombelli, L,  Koul, R K, Lee, J and Sorkin, R D 1986}{Phys
Rev}{D34}{373--383}

\refjl{Boulware, D G 1975}{phys Rev}{D11}{1404}

\refjl{Bousso, R 2002}{The holographic principle, submitted to Rev Mod
Phys}{hep-th/0203101}{52 pp}

\refjl{Brandenberger, R H 2002}{Trans--Planckian physics and inflationary
cosmology, hep-th/0210186}{}{}

\refjl{Brout, R, Massar, Parentani, R and Spindel, Ph 1995a}{Phys
Rep}{260}{329--454}

\refjl{\dash 1995b}{Phys
Rev}{D52}{4559--4568, hep-th/9506121}

\refjl{Buckley, J et al 2002}{Gamma--ray summary report (Snowmass
2001)}{}{astrop-ph/0201160}

\refjl{Carlip, S 2001}{Rept Prog Phys}{64}{2001}

\refjl{Cline, J M, Mostoslavsky, M and Servant, G 1999}{Phys Rev}{D59}{063009}

\refjl{Corley, S and Jacobson, T 1996}{Phys Rev}{D53}
{6720--6724, hep-th/9601073}

\refjl{Daghigh, R G and Kapusta, J I 2002}{Phys Rev}{D65}{064028, gr-qc/0109090}

\refjl{Das, S, Ramadevi, P and Yajnik, U A 2002}{Mod Phys Lett}{A17}{993-1000}

\refjl{Davies, P C W and Fulling, S A 1977}{Proc R Soc Lond}{A356}{237}

\refbk{DeWitt, B S 1979}{in General relativity:  an Einstein centenary
survey}{eds S W Hawking and W Israel (Cambridge), pp~680--745}

\refbk{Eddington, A S 1929}{The nature of the physical world}{(New York:
MacMillan and Cambridge:  University Press)}

\refjl{Fedotov, A M, Mur, V D, Narozhny, N B, Belinskii, V A and Karnakov, B M
1000}{Phys Lett}{A254}{126--132}

\refjl{Fredenhagen, K and Haag, R 1990}{Commun Math Phys}{127}{273}

\refjl{Frolov, V P and Page, D N 1993}{Phys Rev Lett}{71}{3902--3905}

\refjl{Fulling, S A and Davies, P C W 1976}{Proc R Soc Lond}{A348}{393}

\refjl{Gibbons, G W 1977}{in Proc. First Marcel Grossman Meeting on General
Relativity, ed. R. Ruffini}{}{449--458}

\refjl{Gibbons, G W and Hawking, S W 1977}{Phys Rev}{D15}{2752}

\refjl{Grumiller, D, Kummer, W and Vassilevich, D V 2002}{Phys Rept}{}{to
appear}

\refjl{Halzen, F, Zas, E, MacGibbon, J H and Weekes, T C
1991}{Nature}{353}{807--815}

\refjl{Hawking, S W 1974}{Nature}{248}{30}

\refjl{\dash 1975}{Commun Math Phys}{43}{199--220}

\refjl{\dash 1976}{Phys Rev}{D13}{191--197}

\refjl{Heckler, A F 1997}{Phys Rev Lett}{78}{3430--3433}

\refjl{Helfer, A D 1996}{Class Q Grav}{13}{L129--L134}

\refjl{\dash 1998}{Class Q Grav}{15}{1169--1183}

\refjl{\dash 2001a}{Phys Rev}{D63}{025016}

\refjl{\dash 2001b}{Class Q Grav}{18}{5413--5428}

\refjl{\dash 2003}{Quantum energetics of the Hawking process, in
preparation}{}{} 

\refjl{Himemoto, Y and Tanaka, T 2000}{Phys Rev}{D61}{064004 gr-qc/9904076}

\refjl{Hod, S 2000}{Phys Rev}{D61}{124016}

\refjl{'t Hooft, G 1985}{Nucl Phys}{B256}{727}

\refbk{\dash 1993}{in Salamfestschrift:  A collection of talks}
{eds. A. Ali, J. Ellis, S. Randjbar-Daemi, World Scientific,
pp~284--296, gr-qc/9310026}

\refjl{\dash 1996}{Int J Mod Phys}{A11}{4623--88, gr-qc/9607022}

\refbk{\dash 1997}{Proceedings:  Twelfth International Congress of Mathematical
Physicists (ICMP 97)}{eds D De Wit, A J Bracken, M D Gould and P A Pearce,
International Press, pp~64--79, gr-qc/9711053}

\refbk{\dash 1999}{in Basics and highlights in fundamental
physics (Erice 1999)}{ed. A Zichichi, World Scientific,
pp~72--86, hep--th/0003004}

\refjl{Itzhaki, N 1996}{Some remarks on 't Hooft's S--matrix for black
holes}{}{hep-th/9603067}

\refbk{Jacobson T 1990}{Brighton 1990:  Relativistic astrophysics,
cosmology and fundamental physics (proc. 15th Texas symposium on relativistic
astrophysics}{104--116}

\refjl{\dash 1991}{Phys Rev}{D44}{1731--1739}

\refjl{\dash 1993}{Phys Rev}{D48}{728--741}

\refbk{\dash 1996}{in Recent developments in gravitation and
mathematical physics: proceedings}{[First Mexican school on gravitation and
methamtical physics], eds. Macias, A, Matos, T, Obregon, O and Quevedo, H,
World Scientific, pp~87--114}

\refjl{\dash 1999}{Prog The Phys Suppl}{136}{1--17, hep-th/0001085}

\refjl{Jacobson, T and Mattingly, D 2000}{Phys Rev}{D61}{024017}

\refbk{Kapusta, J I 2000}{in Phase transitions in the rarly 
universe:  theory and observations (International School of Astrophysics 
D. Chalonge, Erice 2000)}{eds H J de Vega, I Khalatnikov, N Sanchez, Kluwer
Academic, astro-ph/0101515}

\refjl{Kim, H I, Lee, C H and MacGibbon, J H 1999}{Phys Rev}{D59}{063004}

\refjl{Kummer, W and Vassilevich, D V 1999}{Annalen Phys}{8}{801--827}

\refjl{Marolf, D and Sorkin, R D 2002}{Perfect mirrors and the
self--accelerating box paradox}{preprint}{hep-th/0201255}

\refbk{Minty, E M, Heavens, A F, and Hawkins, M R S 2001}{Testing dark matter
with high--redshift supernovae}{astro-ph/0104221}

\refjl{Page, D N 1976}{Phys Rev}{D14}{3260--3273}

\refjl{Panangaden, P and Wald, R M 1977}{Phys Rev}{D16}{929--932}

\refjl{Parentani, R 1996}{Nucl Phys}{B465}{4526}

\refjl{\dash 2001}{Phys Rev}{D63}{041503}

\refbk{Peet, A 2000}{Boulder 1999:  Strings, branes and gravity}{(TASI
lectures), pp~353--433}

\refbk{Penrose, R 1986}{in Quantum concepts in space and time (Oxford,
1984)}{eds R Penrose and C J Isham (Oxford) pp.129--146}

\refbk{Penrose R and Rindler W 1984--6}{Spinors and
Space--Time}{(Cambridge:  University Press)}

\refjl{Radzikowski, M J and Unruh, W G 1988}{Phys Rev}{D37}{3059--3060}

\refjl{Saida, H and Sakagami, M 2000}{Phys Rev}{D61}{084023 gr-qc/9905034}

\refbk{Schweber, S S 1961}{An introduction to relativistic quantum field 
theory}{(Evanston, Illinois:  Row, Peterson and Co.)}

\refjl{Sewell, G 1982}{Ann Phys}{141}{201--224}

\refjl{Smolin, L 2001}{Nucl Phys}{B601}{209 hep-th/0003056}

\refjl{Susskind, L, Thorlacius, L and Uglum, J 1993}{Phys Rev}{D48}{3743--3761,
hep-th/9306069}

\refjl{Unruh, W G 1976}{Phys Rev}{D14}{870--892}

\refjl{\dash 1981}{Phys Rev Lett}{46}{1351--1353}

\refjl{\dash 1995}{Phys Rev}{D51}{2827--2838}

\refjl{Unruh, W G and Wald, R M 1982}{Phys Rev}{D25}{942--958}

\refjl{Visser, M 2001}{hep-th/0106111, Essential and inessential features of
Hawking radiation}{}{}

\refbk{Wald, R M 1994}{Quantum Field Theory in Curved Spacetime and
Black Hole Thermodynamics}{(Chicago:  University Press)}

\refjl{\dash 2001}{Living Rev Rel}{4}{6}

\bye